\documentclass[twocolumn,traditabstract]{aa}  

\usepackage[hyperindex,breaklinks=true, colorlinks, citecolor=blue]{hyperref}  
\usepackage{graphicx}
\usepackage{array}
\usepackage{txfonts}
\usepackage{caption}
\usepackage{lipsum}
\usepackage{hyperref}
\usepackage{url}
\usepackage{natbib}
\usepackage{breqn}
\usepackage{diagbox}
\usepackage{cuted} 
\usepackage[mathscr]{euscript}

\newcommand{\correction}[1]{{\color{red} #1}} 
\renewcommand{\correction}[1]{{#1}}

\newcommand{\correctionII}[1]{{\color{red} #1}} 
\renewcommand{\correctionII}[1]{{#1}}

\newcommand{\correctionIII}[1]{{\color{red} #1}} 
\renewcommand{\correctionIII}[1]{{#1}}

\usepackage{comment}

\usepackage{soul}

\newcolumntype{C}{>{\centering\arraybackslash}p{0.2\textwidth}}
\usepackage{enumitem}
\usepackage{booktabs}
\bibpunct{(}{)}{;}{a}{}{,} 
\begin{document} 
\title{Modeling Local Bubble analogs II: Synthetic Faraday rotation maps}
\author{E.~Maconi \inst{1}, 
S.~Reissl\inst{2},
J.~D.~Soler\inst{1,3},
P.~Girichidis\inst{2},
R.~S.~Klessen\inst{2,4,5,6},
A.~Bracco\inst{7,8},
S.~Hutschenreuter\inst{1}
}

\institute{
1. University of Vienna, Department of Astrophysics, Türkenschanzstrasse 17, 1180 Vienna, Austria.\\
2. Universit\"{a}t Heidelberg, Zentrum f\"{u}r Astronomie, Institut f\"{u}r Theoretische Astrophysik, Albert-Ueberle-Str. 2, 69120, Heidelberg, Germany.\\ 
3. Istituto di Astrofisica e Planetologia Spaziali (IAPS). INAF. Via Fosso del Cavaliere 100, 00133 Roma, Italy.\\
4. Universit\"{a}t Heidelberg, Interdisziplin\"{a}res Zentrum f\"{u}r Wissenschaftliches Rechnen, Im Neuenheimer Feld 205, D-69120 Heidelberg, Germany.\\
5. Harvard-Smithsonian Center for Astrophysics, 60 Garden Street, Cambridge, MA 02138, USA \\ 
6. Elizabeth S. and Richard M. Cashin Fellow at the Radcliffe Institute for Advanced Studies at Harvard University, 10 Garden Street, Cambridge, MA 02138, USA \\ 
7. INAF – Osservatorio Astrofisico di Arcetri, Largo E. Fermi 5, 50125 Firenze, Italy \\
\correction{8. Laboratoire de Physique de l'Ecole Normale Sup\'erieure, ENS, Universit\'e PSL, CNRS, Sorbonne Universit\'e, Universit\'e de Paris, F-75005 Paris, France}
}
\titlerunning{Local Bubble analogs II: Synthetic Faraday rotation maps}
\authorrunning{E.~Maconi and S.~Reissl}

\abstract
    {Faraday rotation describes the change of the linear polarization angle of radiation passing through a magnetized plasma. The Faraday rotation is quantified by the rotation measure (RM), which is related to the line-of-sight (LOS) magnetic field component and the thermal electron density traversed by light along its path toward the observer. 
    However, it is challenging to disentangle the signal from different LOS portions and separate the contribution from the local interstellar medium (ISM).
    \correctionIII{As the Solar System is located within the Local Bubble, a low-density, hot cavity formed by past supernova events, it essential to investigate how this environment may impact the observed RM values.}}
    {The present study investigates the imprint of the local environment on the synthetic RM signal, as measured by an observer within a Local Bubble-like cavity. The RM derived from diffuse polarized synchrotron radiation produced by cosmic ray (CR) electrons at decimeter wavelengths is also analyzed.}
    {We produce synthetic Faraday rotation maps for an observer placed inside a Local Bubble candidate, selected from a magnetohydrodynamic (MHD) simulation that resembles the properties of the ISM in the Solar vicinity.
    Using the capabilities of the radiative transfer code {\tt POLARIS}, we study the imprint of the cavity walls on the RM signal. As the MHD simulation does not account for CR diffusion, we develop a CR toy-model to study the Faraday rotation of the diffuse polarized synchrotron radiation.}
    {We find that 
    (i) the imprint of local structures, such as the walls of the Local Bubble candidate and the edges of other supernovae blown cavities, is of fundamental importance for interpreting the global Faraday sky; 
    (ii) the Local Bubble has a non-negligible contribution to the sinusoidal patterns of RM as a function of Galactic longitude seen in observations;
    and (iii) the RM signal from diffuse synchrotron emission shows a strong correspondence with the RM signal generated by the Local Bubble candidate walls.}
    {}
  \maketitle

\section{Introduction}\label{sec:introduction}

Polarized radiation propagating through magnetized and ionized media experiences the Faraday rotation effect, which consists in the rotation of the polarization angle as the radiation travels through the medium \citep[see, e.g.,][]{cooperANDprice1962,morrisANDberge1964,gardnerANDdavies1966}.
The amplitude of this rotation depends on the radiation frequency and carries information on the line-of-sight (LOS) magnetic field strength weighted by the thermal electron density \citep[see chapter~6 in][for a detailed discussion]{Rybicki1986}. These are key quantities for constraining and characterizing the Milky Way’s magnetic field topology and strength, which plays an important role in regulating the star formation processes and the dynamical evolution of the interstellar medium (ISM; see, e.g. \citealt{McKee2007}, \citealt{Heiles2012}, \citealt{KlessenGlover2016}, \citealt{han2017}, \citealt{Ferriere2020}, \citealt{Reissl2023}).

Interpreting observations of Faraday RM signal is not trivial.
For instance, the distribution of Galactic free electrons is not well constrained and can be described by a variety of \correctionIII{models \citep[e.g.,][]{Taylor1993,Cordes2002,Cordes2004,Gaensler2008,Schnitzeler2012}}. 
The intrinsic \correctionII{RM} of pulsars and extragalactic background sources, used to construct all-sky RM maps of the Milky Way, is also often poorly known \citep{angel1980, impey1990, kim2016} and the diffuse polarized synchrotron radiation emitted by Galactic cosmic ray (CR) electrons may introduce foreground contamination to the Faraday rotation signals \cite[e.g.,][]{Sokoloff1998, beck2003}. 
Another complication in Faraday rotation observations is the cumulative effect of all environments the radiation traverses before reaching the observer, which can be modeled but is not straightforward to separate \citep[see, e.g.,][]{Burn1966}.
\correction{ISM bubbles generated by supernova (SN) explosions and stellar winds also appear to be of particular importance for understanding Faraday rotation data \citep[see, e.g.,][]{Stil2009,Costa2018,Jung2024,Pelgrims2025,Korochkin2025}, as magnetic field lines are wrapped around the expanding cavities and thermal electrons accumulate at the cavities' edges \correctionIII{and in their interface regions}. This is especially relevant as the Solar System is currently located within the Local Bubble \citep[see, e.g.,][]{coxANDreynolds1987,Linsky2021,ONeill2024}, a SN-generated hot void that has already been shown to be an important foreground \citep[e.g.,][]{Alves2018,Skalidis2019,Krause2021,Maconi2023,ONeill2024b}.}
Moreover, the reconstruction of the Faraday spectrum using various techniques, as well as the reconstruction of all-sky RM maps from the \correctionII{available RM catalogs \citep[see, e.g.,][]{VanEck2023}}, also using advanced inference models \citep[e.g.,][]{Oppermann2012,Oppermann2015,hutschenreuter2022}, can introduce artifacts \citep[e.g.,][]{Farnsworth2011}. All of these effects pose significant challenges when modeling the structure of the Milky Way's large-scale magnetic field from Faraday rotation data, as the local environment also appears to play a critical role in RM observations \citep[see, e.g.,][]{Reissl2023}.

\correctionII{In this paper, we use a magnetohydrodynamic (MHD) simulation in which we identified a Local Bubble-like cavity to investigate the imprint of the local environment on the synthetic Faraday rotation measure (RM) signal, as seen by an observer placed at the center of the bubble. The MHD simulation replicates the physical conditions in the Solar neighborhood and was previously utilized in the study by \cite{Maconi2023}. The synthetic observations are performed using the radiative transfer (RT) code {\tt POLARIS} \citep{Reissl2016,Reissl2019}.}
We assume an idealized scenario with ideal background sources uniformly distributed across the sky, one per LOS, without considering any specific instrumental configuration. The RM is directly computed by {\tt POLARIS}. This approach allows us to focus on the imprint of the local environment without additional complications arising from source distribution or instrumental effects. \correctionIII{Observationally, this is equivalent to an idealistic scenario wherein the RM map is determined from background sources whose intrinsic RMs are known and have been subtracted.}
We compare our results with the Milky Way's Faraday sky as reconstructed by \cite{hutschenreuter2022}, while also addressing the relative caveats and limitations of this comparison.
\correctionII{We construct a CR electrons toy-model for our MHD simulation in order to model diffuse Galactic synchrotron emission at various frequencies between 1 and 5 GHz and produce a RM map using RM synthesis.}
We acknowledge that synthetic Faraday rotation has been explored in recent works \citep[e.g.,][]{Pakmor2018,Basu2019,Reissl2020A,Bracco2022,Reissl2023,Erceg2024}. However, to our knowledge, no study has yet placed an observer within a simulation of a Local Bubble-like cavity.

This paper is organized as follows.
In Section~\ref{sec:sims}, the properties of the MHD simulation and of the selected bubble are reviewed. The method used to identify the cavity walls \correctionII{and the CR electron model are also described.}
Section~\ref{sect:SyntheticObservations} describes the post-processing setups used to produce the synthetic Faraday rotation observations and the RM synthesis technique. 
The results are discussed in Section~\ref{sec:results_discussion} and the conclusions are presented in Section~\ref{sec:conclusions}.

\section{Numerical setup}\label{sec:sims}

\subsection{Simulated cavity}\label{sec:simulated_cavity}

\begin{figure*}
    \centering
    \includegraphics[width=0.95\textwidth]{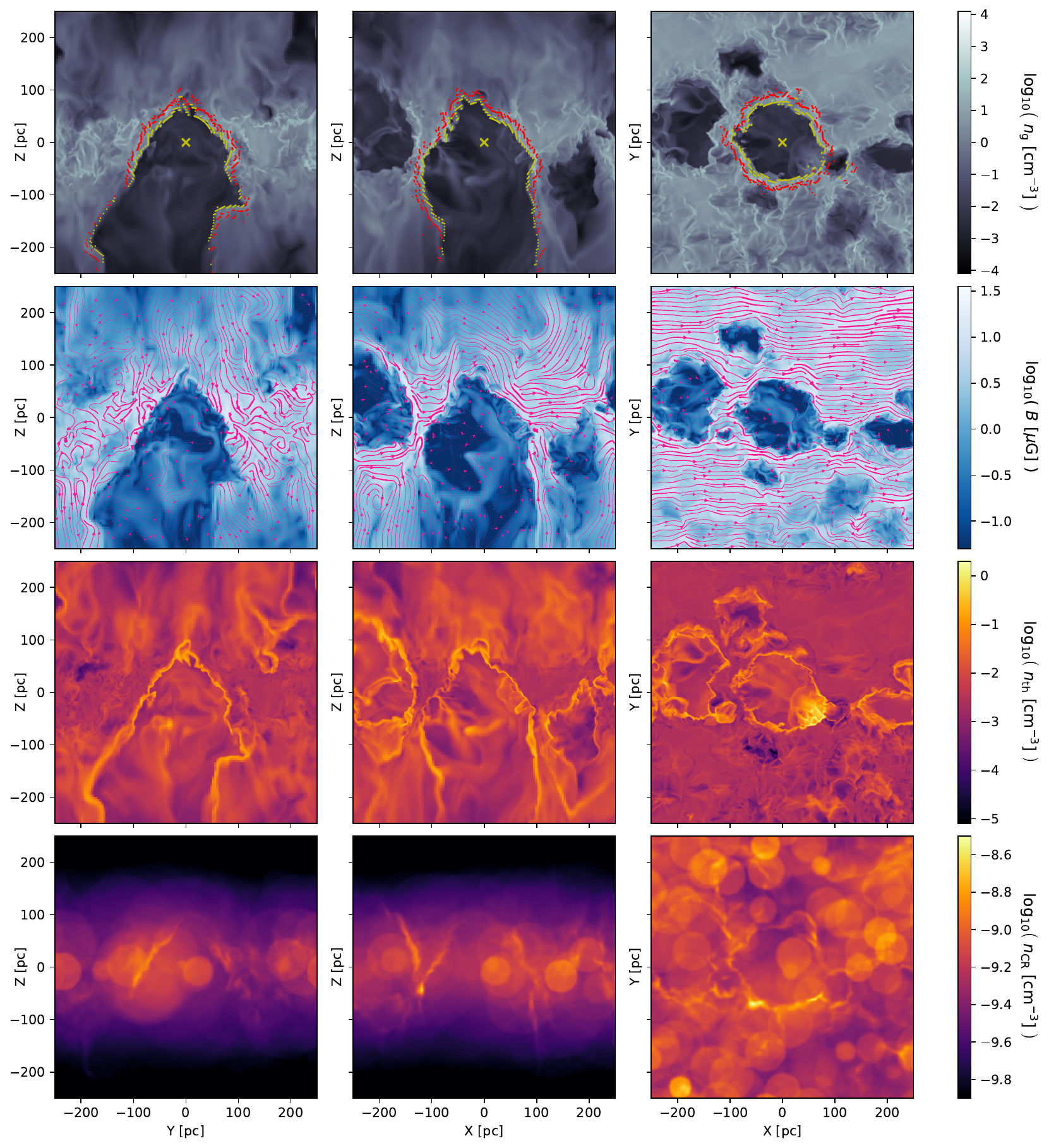}
    \caption{From \textit{top} to \textit{bottom}: gas number density ($n_{\mathrm{g}}$), total magnetic field strength ($B$) with streamlines whose width is proportional to the field strength, thermal electron density ($n_{\mathrm{th}}$), and CR electrons density ($n_{\mathrm{CR}}$) in 1-pc-thick cuts of the simulation domain centered on the Local Bubble candidate.
    In the upper panels, the yellow and red line segments mark the contours of the inner and outer edges of the Local Bubble candidate as identified in \cite{Maconi2023}, respectively. The yellow crosses indicate the position of the observer in the synthetic observations.}
    \label{fig:input_data_overview}
\end{figure*}

\begin{figure*}[]
	\centering
    \includegraphics[width=1.0\textwidth]{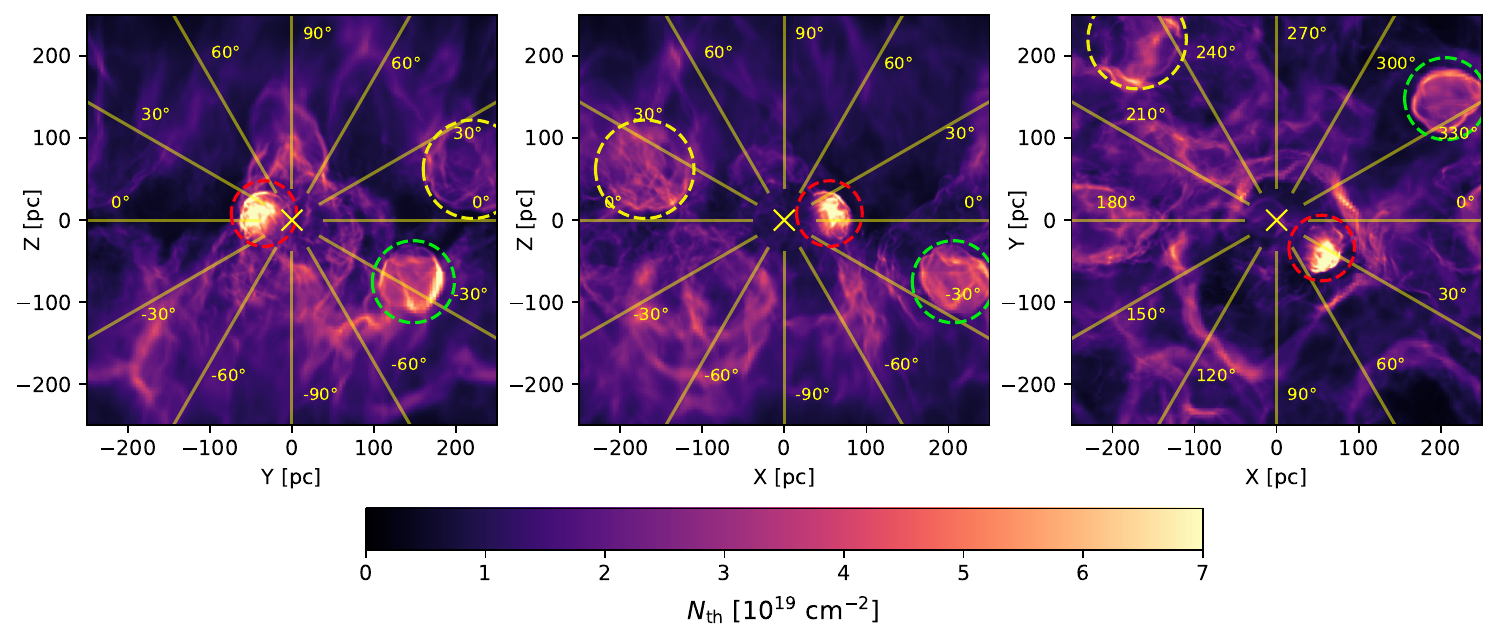}
	\caption{The same as Fig.~\ref{fig:input_data_overview}, but for the column density of thermal electrons. Yellow lines and angles represent the galactic coordinate system. Yellow, red, and green dashed circles indicate highly ionized bubble regions that are most prominent in the RM full-sky maps. This figure is to be compared with Fig.~\ref{fig:FR_fullSkyMaps_reference}.}
	\label{fig:nth_column}
\end{figure*}

\begin{figure}
    \centering
    \includegraphics[width=0.48\textwidth]{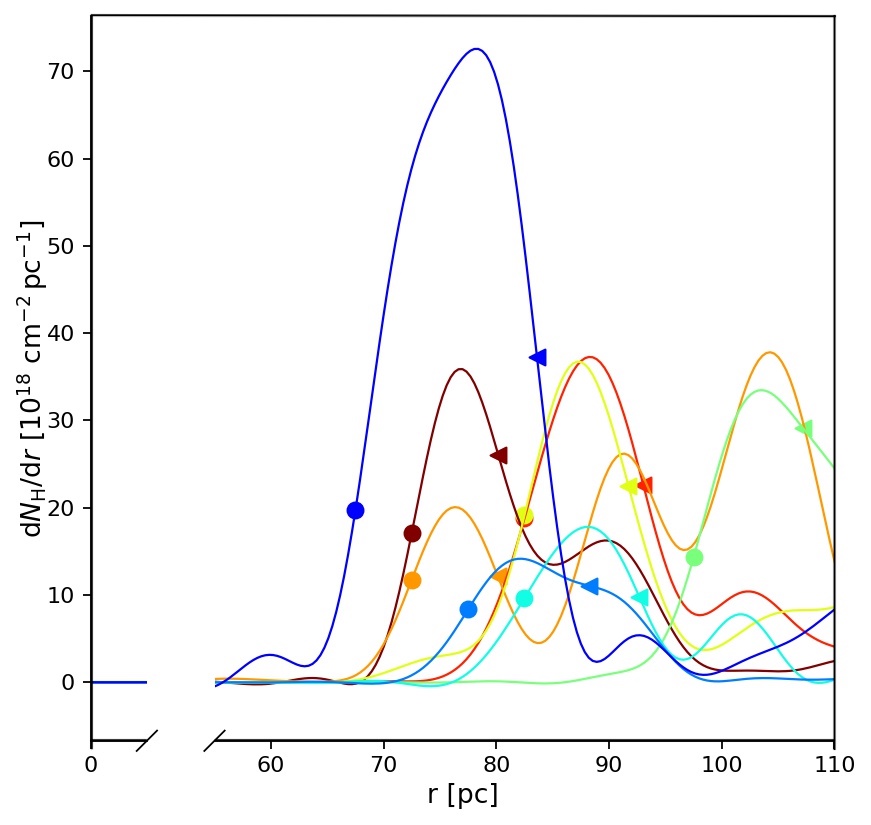}
    \caption{\correction{Radial profiles of the differential column density, $\mathrm{d}N_{\mathrm{H}}\mathrm{(}r\mathrm{)}/\mathrm{d}r$, for eight LOSs selected in the $x-y$ plane. Circles and triangles represent the inner and outer boundaries of the cavity along each specific LOS, determined using the method described in Sect.~\ref{sec:simulated_cavity}. This plot is Fig.~2 in \cite{Maconi2023}.}}
    \label{fig:differential_NH}
\end{figure}

For our study, we use an MHD simulation in which we identified a Local Bubble-like cavity. This simulation was previously used in the study by \cite{Maconi2023} to investigate the imprints of the bubble walls on polarized dust emission at  $353\,$GHz. The candidate bubble was selected from the set of numerical simulations presented in \citet{GirichidisEtAl2018b} and \cite{girichidis2021}, which are part of the Simulating the Life-Cycle of Molecular Clouds (SILCC) project \citep{WalchEtAl2015,GirichidisEtAl2016b}.
\correction{These simulations} cover a patch of the Galactic ISM with dimensions $500 \times 500 \times 500\,\mathrm{pc}^3$, employing outflowing boundary conditions in the $z$ direction and periodic conditions in $x$ and $y$. The total gas mass in the box is $2.5 \times 10^6 \, \mathrm{M}_\odot$, corresponding to a gas surface density of $\Sigma = 10\,\mathrm{M}_\odot\ \mathrm{pc}^{-2}$ similar to the conditions near the solar circle \citep[see, e.g.,][]{Ferriere2015}.
Initially at rest, the gas is permeated by a magnetic field aligned along the $x$ axis. The magnetic field strength is scaled with the gas density in the vertical stratification, that is, $\mathbf{B}(z) = B_0\,(\rho(z)/\rho(z=0))^{1/2} \mathbf{\hat{x}}$. 
For our specific scenario, the central field strength is set to $B_0=3\,\mu\mathrm{G}$. This supercritical setup allows the disk to collapse as it is not supported by magnetic pressure.
The gas heating mechanisms include spatially clustered SNe, CR, and X-ray heating. The rates of CR ionization and heating are denoted as $\zeta_{\mathrm{CR}}=3\times 10^{-17}\,\mathrm{s}^{-1}$ and $\Gamma_{\mathrm{CR}}=3.2\times 10^{-11}\mathrm{erg\,s}^{-1}\,\mathrm{cm}^{-3}$, respectively. We note that the CR ionization rate chosen here is dynamically connected to the MHD equations and the temporal evolution of the simulation. Contrary, the CR electron model applied below is added in post-processing to generate synthetic observables, so they are not self-consistent. We also note that the CR ionization rate can locally vary, which might affect the amount of free electrons across ISM phases \citep[see, e.g.,][and references therein]{Bracco2022,PadovaniEtAl2024}. Heating and radiative cooling are computed through a chemical network monitoring the non-equilibrium concentrations of various species, including ionized hydrogen (H$^+$), atomic hydrogen (H), molecular hydrogen (H$_2$), singly ionized carbon (C$^+$), and carbon monoxide (CO) \citep[for more details see, e.g.,][]{WalchEtAl2015,GirichidisEtAl2018b,girichidis2021}.
The star formation rate used in the simulations follows the Kennicutt-Schmidt relation \citep{KennicuttSchmidt1998}, which is then converted into a SNe rate using the \citet{Chabrier2003} initial mass function (IMF). This results in $\sim15$ SNe per Myr for the simulated volume.  Type Ia SNe (20 percent) and type II SNe (80 percent) are positioned randomly in $x$ and $y$, with the $z$ position sampled from a Gaussian distribution with a vertical scale height of $300\,\mathrm{pc}$ \citep{BahcallSoneira1980, Heiles1987}. Each SN injects $10^{51}\,\mathrm{erg}$ of thermal energy, generating a three-phase medium comparable to the observed local ISM. 
Individual and clustered SNe give rise to dense filaments, clouds of cold gas, and hot voids (i.e., bubbles).

\correction{Our Local Bubble candidate has been selected from a set of cavities identified across multiple simulation snapshots. The chosen bubble is the one that best matches the average density, total ionized gas mass, as well as the mass- and volume-weighted magnetic field and temperature of the Local Bubble.
The candidate cavity is the result of 17 clustered SNe, aligning with the $14$ to $20$ SNe events believed to be responsible for creating the Local Bubble, as suggested by \citet{Fuchs2006}.
For the simulated bubble, the mean value of the magnetic field volume-weighted is below 1\,$\mu$G, while the mass-weighted one, which reflects the field in the shell of the bubble, reaches the value of 2\,$\mu$G. These values are in agreement with the Local Bubble mean field strength, estimated to be $0.5 \leq \left<\left|B\right|\right> \leq 2$ \citep{XuHan2019}.
The total mass of the shell of our simulated cavity is $M_{\rm{shell}} \, \sim \, 2 \times 10^5\,\mathrm{M}_{\odot}$, which is an order of magnitude lower than the mass computed for the Local Bubble by \citet{ZuckerEtAl2022}, \correctionII{namely} $M_{\rm{shell,LB}} \, \sim \, 1.4 \times 10^6\,\mathrm{M}_{\odot}$. This discrepancy can be explained by the fact that the estimation of the mass by \citet{ZuckerEtAl2022} includes high density nearby clouds, like Taurus and Ophiuchus (private communication), which are not present in our simulation. We note that these clouds occupy a very small solid angle and do not significantly affect the analysis of the full sky maps. However, our shell mass estimate is of the same order of magnitude as the one computed by \citet{ONeill2024}, $M_{\rm{shell,LB}} \, \sim \, 6 \times 10^5\,\mathrm{M}_{\odot}$.
For our candidate, the mean shell thickness is estimated to be $\sim$\,$14\,$pc, whereas for the Local Bubble, \citet{ONeill2024} estimate it to be around $35\,$pc. As already noted in \citet{Maconi2023}, the thickness of the bubble walls do not significantly affect our results.
Additionally, we emphasize that our goal is to select a cavity that closely matches the properties of the Local Bubble, while acknowledging that finding a perfect match is not possible.}
For a more detailed overview of the simulations setup and the selection of the candidate cavity, we refer the reader to \cite{GirichidisEtAl2018b}.

In Fig.~\ref{fig:input_data_overview}, we present the gas density, the total magnetic field strength, and the thermal electron density for the candidate cavity in cuts through the center. These quantities are the outcome of the MHD simulations. The bottom panel of the figure additionally displays the density of CR electrons as obtained with a CR toy model, which is further discussed in Sect.~\ref{Sect:CR_electrons_model}. 
The figure outlines a complex environment shaped by expanding SNe bubbles that interact with each other, generating a characteristic medium similar to what is observed in the Milky Way's local environment \citep[e.g.,][]{Soler2018,Bracco2020,ZuckerEtAl2022}.
\correction{The selected cavity also present an open and a closed cap. This is of observational significance because evidence suggests that the Local Bubble may be a ``Local Chimney'' \citep[][]{Cox1974, Welsh1999, Lallement2003,ZuckerEtAl2022,ONeill2024}.}
In Fig.~\ref{fig:nth_column}, we show the column density of thermal electrons along the three projection planes for our simulation. It is worth noting that the SN cavities shaping the environment are also visible in these maps.

To assess the potential impact of the bubble's edges on Faraday rotation, we identify the inner and outer surfaces of the cavity using the column density maps provided by {\tt POLARIS}, following the approach previously employed in \cite{Maconi2023}. In summary, by sampling the celestial sphere on a {\tt HEALPix} pixelization with $N_{\rm side}$\,$=$\,$512$ \citep[corresponding to $3\,145\,728$ pixels with an angular size of 6.87\arcmin\ each;][]{Gorski05} and by defining 70 concentric spheres centered on the cavity with an increasing radius step of $5\,$pc, we construct a radial profile of the column density $N_{\rm H}(r)$, where $r$ represents the distance from the center of the cavity (i.e. the observer).
Following the method outlined in \cite{PelgrimsEtAl2020} and \cite{Maconi2023}, we then compute the differential of the column density $\Delta N_{\rm H}(r)$ and its first and second derivatives relative to $r$ for each LOS. The inner and outer boundaries of the Local Bubble candidate are determined as the distance where $N_{\rm H}$ exhibits its first relative increase greater than 0.9\,\% and where the profiles present their first inflection point beyond the inner wall, respectively. 
\correction{In Fig.~\ref{fig:differential_NH}, taken from \cite{Maconi2023}, we present, as an example, the differential column density curves for eight LOSs selected in the $x-y$ plane.} 
In the top panel of Fig.~\ref{fig:input_data_overview}, the inner and outer walls of the bubble are highlighted as yellow and red segments, respectively.
The computed distance to the Local Bubble walls along each LOS is subsequently used to select \correctionII{and isolate} the volume enclosed by the outer wall, thus allowing the study of its \correctionII{imprint on the Faraday RM signal.} 

\subsection{Cosmic ray electron model}\label{Sect:CR_electrons_model}

The numerical simulation to which our candidate cavity belongs does not include a diffusion model for CR electrons, although it does consider the CR ionization and heating effects, as pointed out in Sect.~\ref{sec:simulated_cavity}.
\correctionII{In order to derive the RM signal from diffuse polarized synchrotron radiation emitted by CR electrons,} we construct a basic CR diffusion model, which aims to provide a reasonable approximation for our simulation.

We begin by locating the SNe events within the simulation before the selected time snapshot. 
We inject an amount of CR proton energy equal to $10^{51}\,\mathrm{erg}$ for each SN \citep[e.g.,][]{RuszkowskiPfrommer2023}. This energy is uniformly distributed across a sphere with a radius equivalent to four times the original hydro-injection radius computed in the MHD simulation ($r_{\mathrm{SN},e} = 4\,r_{\mathrm{SN,th}}$)\footnote{\correction{This is an ad-hoc assumption which we chose based on the resulting effective CR distribution. Further details are included in Sect.~\ref{sect:caveats}.}}. In case the Sedov-Taylor radius is resolved, we inject thermal energy into a radius ($r_{\mathrm{SN,th}}$) that encompasses $800\,\mathrm{M}_\odot$ \citep{WalchEtAl2015, GirichidisEtAl2016b}. Otherwise, we inject momentum as described in \citet{GattoEtAl2015}. The injection and transport of CR electrons is fairly complicated and depends on the injection efficiency at the shock, the magnetic field obliquity, the escape fraction from the shock front, the transport speed in the upstream region, and the effective losses. Many of these parameters are unknown by orders of magnitude. We therefore chose a factor of 4 \correction{in radius for the effective injection of the CRs compared to the thermal injection radius} to approximately match the effects. As live cooling mechanisms are not accounted for, we normalize the total energy to a value comparable to that observed in our Solar neighborhood. To do so, we compute the total energy within the midplane of our simulation, approximately $\pm100\,\mathrm{pc}$, and subsequently re-scaled it to match with observed values of $\langle \varepsilon_{\mathrm{cr,p}}\rangle = 1\, \mathrm{eV}\,\mathrm{cm^{-3}}$ for the protons \citep[e.g.,][]{BoularesCox1990}. Following this normalization, the electron energy is set to 0.01 times the proton energy, $\langle \varepsilon_{\mathrm{cr,e}}\rangle = 0.01 \langle \varepsilon_{\mathrm{cr,p}}\rangle$ \citep{CummingsEtAl2016}. The number and energy density of CRs are computed as integrals over the particle distribution function, $f(p)$,
\begin{align}
    n_{\mathrm{cr}} &= \int_{p_0}^{p_1} 4\pi p^2 f(p) \mathrm{d}p\\
    e_{\mathrm{cr}} &= \int_{p_0}^{p_1} 4\pi p^2 f(p) E_{\mathrm{kin}}(p) \mathrm{d}p \, ,
\end{align}
\correction{where $E_{\mathrm{kin}}$ is the kinetic energy of the particle.}
We derive the CR electron number density from the estimated electron energy by assuming a power-law $f(p)\propto p^{-\alpha}$ with an index of $\alpha$\,$=$\,$3.2$, $p_0$\,$=$\,$3\,m_{\mathrm{e}}c$, and $p_1$\,$=$\,$100\,m_{\mathrm{e}}c$ \citep{WerhahnEtAl2021c}. The CR electron density obtained with this simple model is shown in the bottom panel of Fig.~\ref{fig:input_data_overview}.

\section{Synthetic observations}\label{sect:SyntheticObservations}

\correctionII{We use the RT code {\tt POLARIS} \citep{Reissl2016,Reissl2019}, which is capable of fully solving the RT problem, including Faraday rotation and Faraday conversion, to perform a set of synthetic observations for our MHD simulations. In this section, after a brief overview of Faraday rotation, we describe the post-processing routine, the various setups, and the RM synthesis technique used in this work.}

\subsection{Faraday rotation}

Faraday rotation refers to the rotation of the linear polarization angle, $\chi$, experienced by polarized radiation as it propagates through a magnetized plasma \citep[][]{Rybicki1979}.
The observed polarization angle, $\chi_{\mathrm{obs}}$, can be expressed as
\begin{equation}
    \chi_{\mathrm{obs}} = \chi_{\mathrm{source}} + \lambda^2 \times \mathrm{RM} \,,
    \label{eq:obs_angle_FR}
\end{equation}
where $\chi_{\mathrm{source}}$ is the polarization angle of the radiation at the source, $\lambda$ is the wavelength of the radiation, and RM is the Faraday rotation measure experienced \correctionII{along} the path toward the observer.
The RM is a wavelength-independent quantity equal to 
\begin{equation}
    \mathrm{RM} = \frac{1}{2\pi} \frac{e^3}{m^2_{\mathrm{e}}c^4}\int^{\ell_{\mathrm{source}}}_{\ell_{\mathrm{obs}}} \, n_{\mathrm{th}}(\ell) \times B_{\parallel}(\ell) \, \mathrm{d}\ell \,,
    \label{eqn:FaradayDepth}
\end{equation}
with $e$ and $m_{\mathrm{e}}$ representing the charge and mass of the electron, respectively, and $c$ is the speed of light. The source and observer are positioned at $l_{\mathrm{obs}}$ and $l_{\mathrm{source}}$, respectively, $n_{\mathrm{th}}$ is the thermal electron density, and $B_{\parallel}$ denotes the LOS magnetic field strength. An average parallel component of the magnetic field pointing toward the observer would result in a positive RM, while a component pointing away would result in a negative value. 
\correction{Equation~\ref{eq:obs_angle_FR} is properly defined only for a background source. It holds in the case of a medium causing solely Faraday rotation but breaks down if the medium also emits synchrotron radiation or if significant Faraday dispersion occurs within the volume probed by the telescope beam.}
For this reason, the quantity RM is typically replaced with the generalized quantity $\phi(\ell)$, called Faraday depth, where $\ell$ represents the distance from the observer. $\phi(\ell)$ has the same formal expression as the RM in Eq.~\ref{eqn:FaradayDepth}, with the difference that it can be defined at any point in the ISM, independent of any background source. For more details, we refer to the works by \citet{Burn1966}, \citet{Rybicki1979}, \citet{Huang2011}, and \citet{Ferriere2021}.

\subsection{Radiative post-processing}\label{sect:PostProcessing}

\begin{figure}[]
	\centering
    \includegraphics[width=0.49\textwidth]{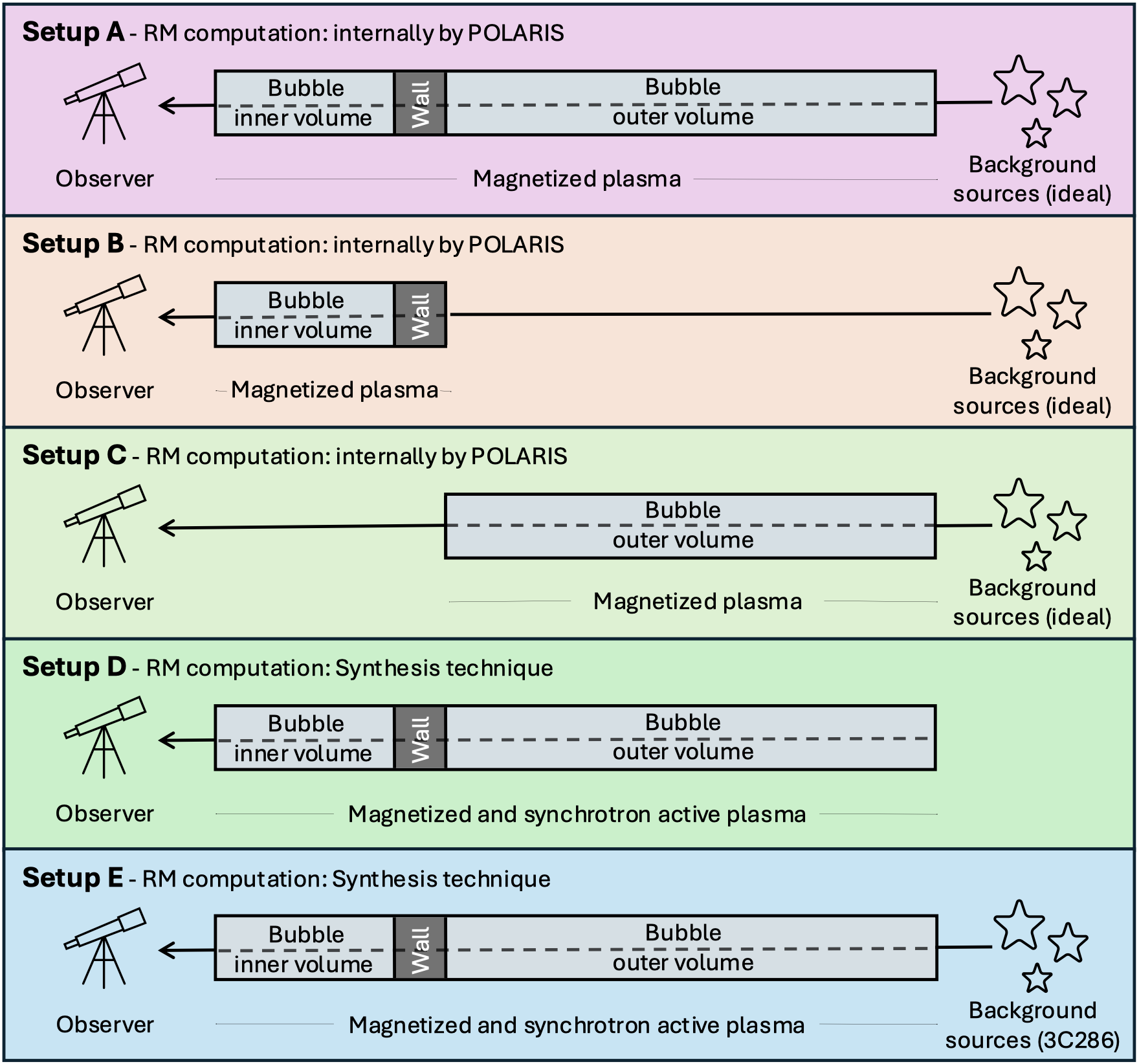}
	\caption{\correctionII{Setups for the RT simulations. This schematic representation depicts the situation along a single LOS. The method used to compute the RM is reported. The observer is positioned at the center of the cavity, marked by the yellow cross in the top-panel of Fig.~\ref{fig:input_data_overview}. The \textit{bubble inner volume} refers to the region enclosed by the inner edge of the cavity walls, highlighted by the yellow line segments in the same figure. The \textit{wall} corresponds to the gas overdensity at the bubble boundaries and it is the region enclosed by the yellow and red segments in the top panel of Fig.~\ref{fig:input_data_overview}. The bubble outer volume refers to the region outside the outer edge of the walls, marked by red segments. This figure is not to scale, as the specific configuration varies depending on the LOS.
    For a discussion of the individual schemes, we refer the reader to the text and Appendix~\ref{app:FR_synthesis_appendix}. Additional details about the cavity boundaries can be found in Sect.~\ref{sec:simulated_cavity}.}}
	\label{fig:set_ups_RT_simulations}
\end{figure}

In our RT simulations the observer is positioned at the center of the selected \correctionII{Local Bubble-like} cavity (see the yellow cross in the top panel of Fig.~\ref{fig:input_data_overview}), receiving radiation from the background sources and, in cases where synchrotron emission is considered, from the intervening medium. 
We consider an idealized scenario where background sources are \correctionII{ideal and}  evenly distributed across the sky, with one source per LOS, and no specific instrumental configuration is taken into account.
In Fig.~\ref{fig:set_ups_RT_simulations}, we present a schematic representation of the situation along a given LOS for different {\tt POLARIS} setups used in this work. \correction{More details on these configurations are provided later in the text. We note that these configurations are made feasible by the use of simulations, highlighting one of the key advantages of this approach. This allows us, \correctionII{for example,} to include or exclude diffuse synchrotron emission \correctionII{from CR electrons} (setups A and D) \correctionII{or to} simulate a scenario where only the bubble is considered \correctionII{or removed} (setup B \correctionII{and C}).}

For each LOS, the output from {\tt POLARIS} includes by default the RM and the Stokes vector $\textbf{S}=(I,Q,U,V)^{T}$, where $I$ is the total intensity, $Q$ and $U$ represent the linear polarization, $V$ the circular polarization. Given the Stokes vector, it is therefore possible to determine the linear polarized intensity, $P_{\mathrm{l}}$, the degree of linear polarization, $p_{\mathrm{l}}$, and the polarization angle, $\chi$, as
\begin{equation}
    \label{eqn:PolAngle}
        P_{\mathrm{l}} = \sqrt{Q^2+U^2}, \quad  p_{\mathrm{l}} = \frac{P_{\mathrm{l}}}{I}, \quad \chi = \frac{1}{2}\tan^{-1} \left(\frac{U}{Q}\right) \,.
\end{equation}

\begin{figure*}[]
	\centering
    \includegraphics[width=0.49\textwidth]{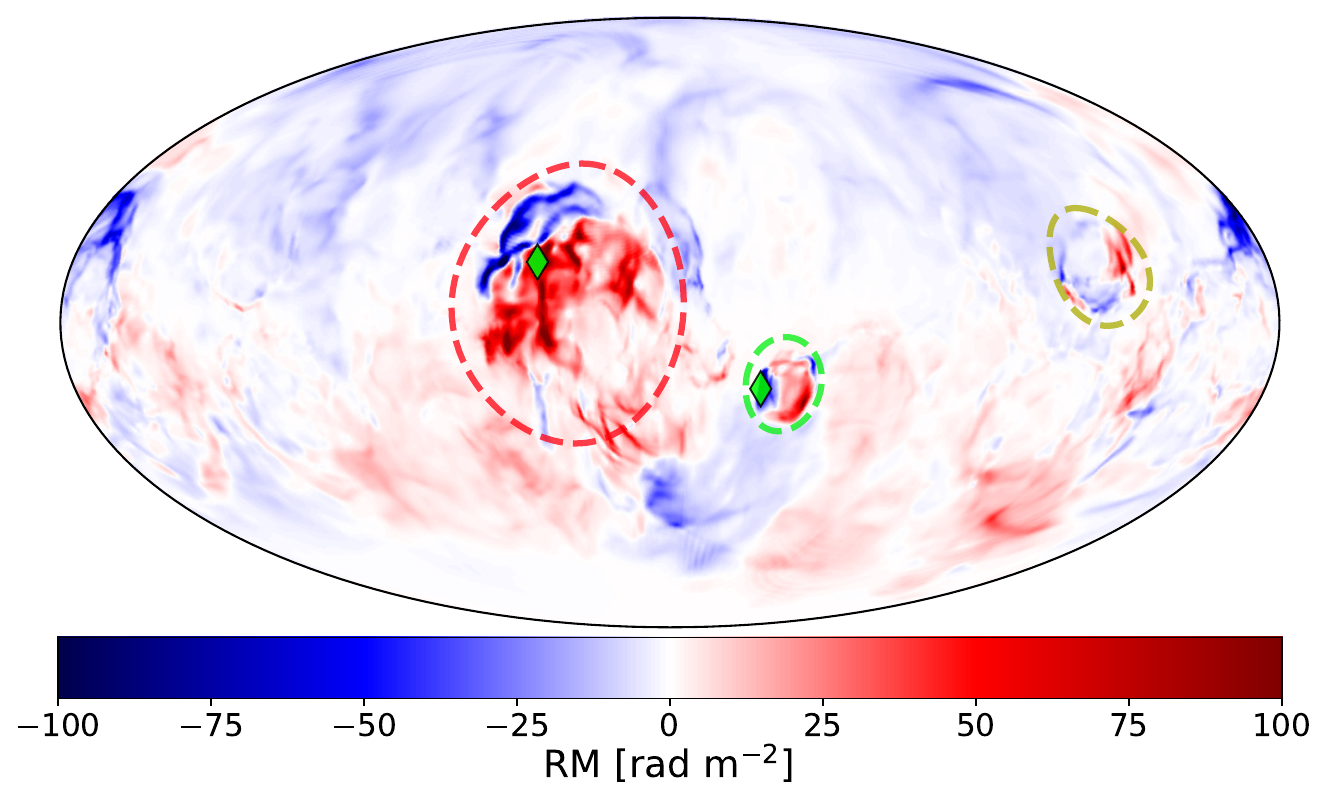}
    \includegraphics[width=0.49\textwidth]{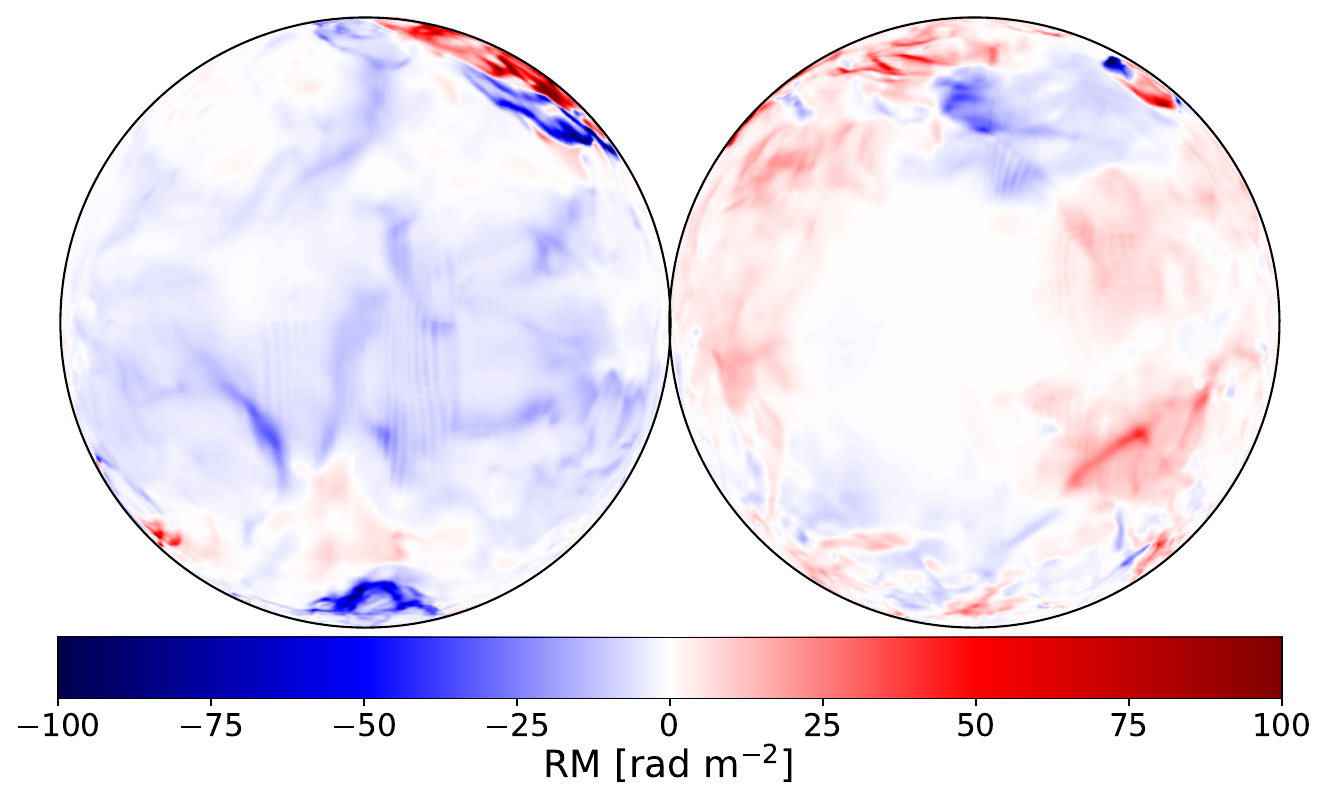}
	\caption{Full-sky RM maps in Mollweide (\textit{left}) and orthographic projections centered at the Galactic poles (\textit{right}) for our reference case (setup A). The \correctionII{RM signal} is internally computed by {\tt{POLARIS}} and is obtained using the full-data cube. These maps are addressed in the text as reference map. Dashed circles indicate the most prominent features corresponding to the regions highlighted in Fig.~\ref{fig:nth_column}. Marked positions (green diamonds) correspond to minimal RM and maximal RM, respectively, of the total map.}
	\label{fig:FR_fullSkyMaps_reference}
\end{figure*}

\correctionII{To study the imprint of the local environment on the RM as measured by an observer within the cavity, we rely on the RM internally computed by {\tt POLARIS} (see Sect.~\ref{Sect:LOS-analysis}, for more details). We refer to the RM map obtained for the case in which the full data cube is considered as our ``reference map''. The reference map is shown in Fig.~\ref{fig:FR_fullSkyMaps_reference}, in both its Mollweide and orthographic projections and will be described in more detail later.}
We also derive a RM map by using the diffuse polarized emission from Galactic CR electrons. We follow the typical observational approach: we perform a series of observations (i.e., RT simulations) of the Stokes parameters between 1 and 5 GHz. Since we want to focus on the diffuse synchrotron radiation, no background source is used (see Sect.~\ref{Sect.set_up_synchrotron}, for more details). \correctionIII{We note that thermal free–free emission is not considered here, although it can contribute at high radio frequencies due to its relatively flat continuum spectrum ($S_\nu \propto \nu^{-1}$).} The Faraday rotation is then computed using a RM synthesis technique, whose details are discussed in Sect.~\ref{Sect:FR_synthesis} and Appendix~\ref{app:FR_synthesis_appendix}.

\begin{figure*}[]
	\centering
    \includegraphics[width=0.49\textwidth]{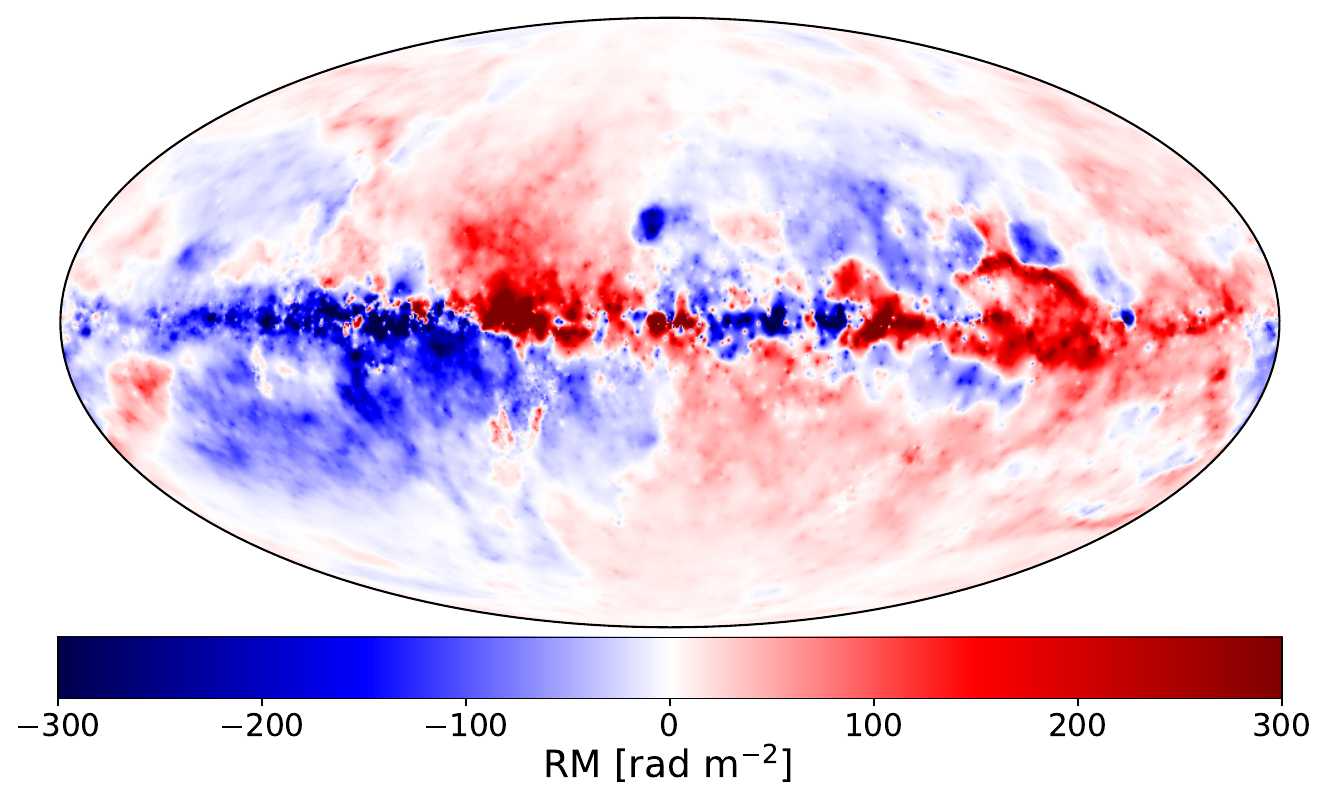}
    \includegraphics[width=0.49\textwidth]{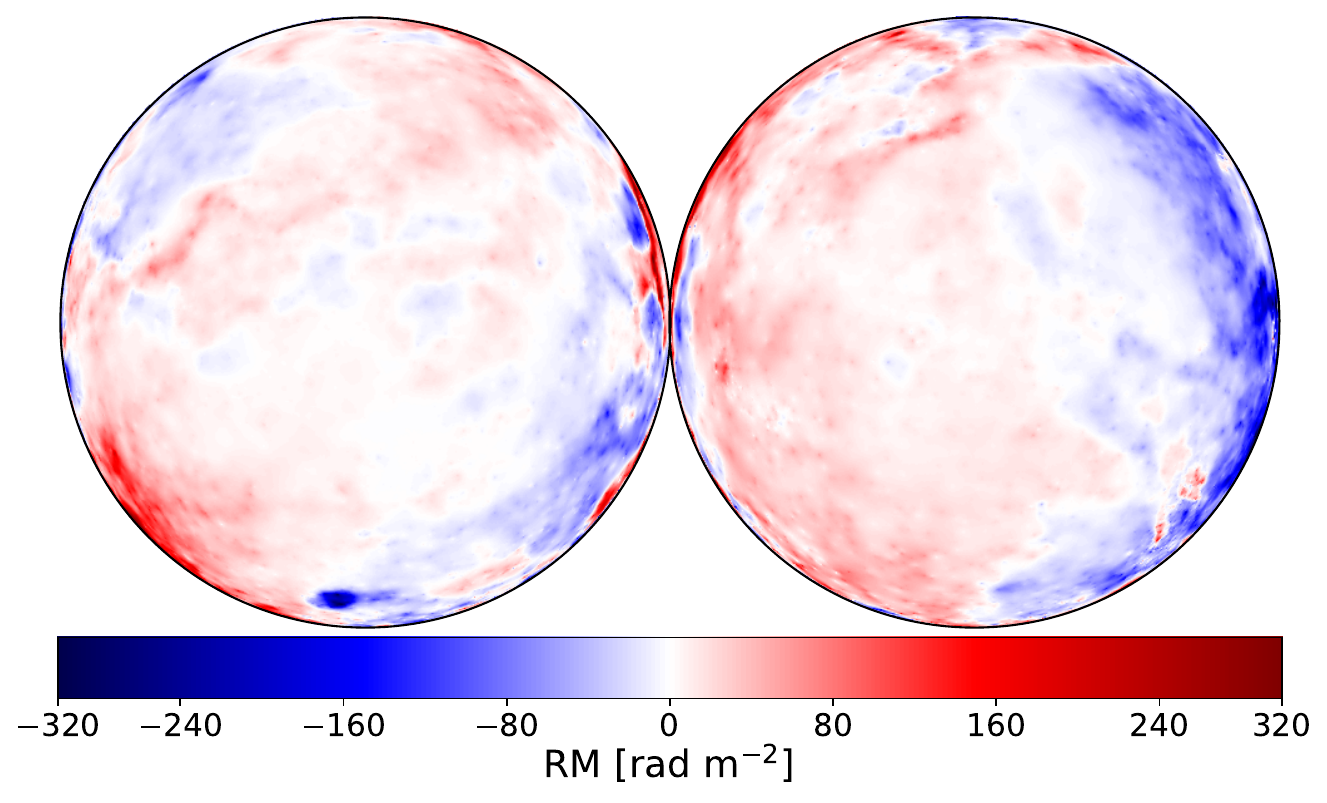}
	\caption{The same as in Fig.~\ref{fig:FR_fullSkyMaps_reference} but for actual observational RM data, as presented in \cite{hutschenreuter2022}. Note the different color-bar scaling with respect to Fig.~\ref{fig:FR_fullSkyMaps_reference}.}
	\label{fig:FR_fullSkyMaps_hutschenreuter}
\end{figure*}

\subsection{Faraday rotation measure synthesis}\label{Sect:FR_synthesis}

To compute the RM, the simplest method would be to perform a fit of Eq.~\ref{eq:obs_angle_FR}. However, this method may suffer from $n\pi$ ambiguities \citep[e.g.,][]{Ma2019}, overlap of signals from different sources with distinct RM values, and the inability to detect faint sources with high RM values. \correctionII{Therefore, the implementation of either a Faraday RM synthesis technique \citep[see, e.g.,][]{Burn1966,Brentjens2005} or the Stokes QU-fitting technique \citep[see, e.g.,][]{Sokoloff1998, Pasetto2021} is required. In this paper, we decided to use a Faraday RM synthesis routine, as this approach does not require to assume any specific analytical model for the polarization signal.}

The idea is to write the observed polarized emission in its complex form, \correctionII{$\mathcal{P}=Q+iU$, which can be rewritten as a function of the fractional polarization $p_{\mathrm{l}}$ as}
\begin{equation}
    \mathcal{P}( \lambda^2 ) = p_{\mathrm{l}}I ( \lambda^2 )e^{2i\chi ( \lambda^2 )} \,,
    \label{eq:polIntensity_complex}
\end{equation}
By introducing Eq.~\ref{eq:obs_angle_FR}, this becomes
\begin{equation}
    \mathcal{P}( \lambda^2 ) = \int_{-\infty}^{\ +\infty} F(\phi) e^{2i\phi\lambda^2} \mathrm{d}\phi \,.
    \label{eq:polIntensity_complex_2}
\end{equation}
The quantity $F(\phi)$ in Eq.~\ref{eq:polIntensity_complex_2} is the Faraday dispersion function, which represents the intrinsic polarized flux as a function of Faraday depth.
Equation~\ref{eq:polIntensity_complex_2} can be rewritten in a form that allows for inversion to obtain the Faraday dispersion function in terms of observable quantities \correctionII{and the RM can then be derived}. 

\correctionII{We analyze and process the synthetic maps using the python code {\tt RM-tools} \citep[see][]{Purcell2020}, which we modified to handle the {\tt POLARIS} output data.}
\correctionII{For more details, we refer to Appendix~\ref{app:FR_synthesis_appendix}, where we also discuss the shortcomings and limitations of the RM synthesis technique, as well as a test case for our routine.}

\subsection{Setup: Bubble edges and local environment}\label{Sect:LOS-analysis}

To study the influence of the bubble on the observed Faraday rotation, we produce a synthetic map where only the volume enclosed by the outer surface of the cavity is considered (for more details, see Sect.~\ref{sec:simulated_cavity} and top panel of Fig.~\ref{fig:input_data_overview}) and compare the RM directly computed by {\tt POLARIS} with the reference map. A schematic representation of this setup is depicted in panel B of Fig.~\ref{fig:set_ups_RT_simulations}.
We also examine the case in which the bubble has been removed from the simulation data, as shown for a given LOS in panel C of Fig.~\ref{fig:set_ups_RT_simulations}. This allows us to study the sinusoidal patterns of Faraday rotation as a function of Galactic longitude, as seen in observations \correctionIII{\citep[see e.g.,][]{Taylor2009,Dickey2022}.}

Additionally, to identify the main contributors to the RM signal, we use the LOS analysis technique, which is currently not feasible in real observations. Indeed, current observational data do not allow the construction of a 3D spatial differential map of the Faraday rotation; rather, they provide 2D sky maps posing challenges in understanding the local environment's contribution to the maps. 
The advantage of synthetic observations made with a RT post-processing technique is that the 3D information remains accessible along each LOS. As a result, it becomes possible to explore in detail the contribution of different structures to the Faraday rotation signal.

To keep track of the RM experienced by radiation as it travels toward the observer, we define a series of 70 concentric spheres spaced by $5\,$pc apart from each other. Each sphere is organized according to a {\tt HEALPix} scheme, \correction{with the same parameters as the ones} described in Sect.~\ref{sec:simulated_cavity}. 
In this way, we are able to evaluate the first derivative of the Faraday depth along each path element $\mathrm{d}\ell$ of the LOS as
\begin{equation}
    \frac{\mathrm{d}\phi_{{i}}(\ell)}{\mathrm{d}\ell} = \frac{\phi_{{i}}(\ell+\mathrm{d}\ell)-\phi_{{i}}(\ell)}{\mathrm{d}\ell} \,,
\end{equation}
where $\ell$ is the distance from the observer, \correction{$\mathrm{d}\ell$ is the distance between two adjacent spheres (i.e.\ 5$\,$pc)} and the index $i$ stands for a distinct LOS, that is a pixel of the {\tt HEALPix} map. This technique \correction{has already been} successfully used in \cite{Reissl2020A}, \cite{Reissl2021MNRAS}, and \cite{Maconi2023} to \correctionII{explore in detail the actual origin of any polarization or RM signal.}
Using the quantity we just defined, it is possible to identify the small and/or highly heterogeneous structures within the ISM that contribute to the rotation of the radiation's polarization angle.

\begin{figure*}[]
	\centering
    \includegraphics[width=1\textwidth]{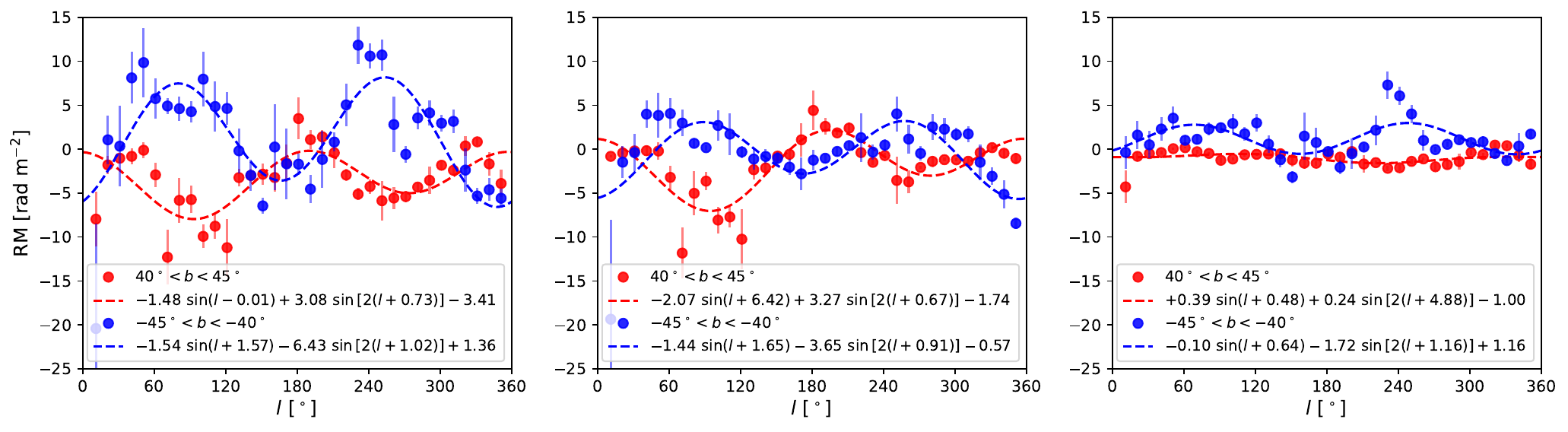}
	\caption{\correction{Sinusoidal patterns of RM as a function of Galactic longitude ($l$) in bins of $10^\circ$, are presented for three different configurations: the entire data cube is considered (\textit{left}; \correctionII{setup A}), only the volume enclosed by the outer edges of the bubble is used (\textit{center}; \correctionII{setup B}), and the bubble is carved out from the data cube (\textit{right}; \correctionII{setup C}).}
    The profiles are obtained from averaging the latitude range ${40^\circ<b<45^\circ}$ (red) and ${-45^\circ<b<-40^\circ}$ (blue). The error bars on the points show the standard deviation of the values in each bin. The least-squares fit parameters of the first three terms of a Fourier series in $l$, ${\,\mathrm{RM}(l)=C_0+C_1\sin{(l+\phi_1)}+C_2\sin{2(l+\phi_2)}\,}$, are reported.}
	\label{fig:RMlon}
\end{figure*}

\subsection{Setup: Synchrotron radiation from Galactic CR electrons}
\label{Sect.set_up_synchrotron}

Relativistic CR electrons moving through a magnetized environment emit synchrotron radiation. For the energy spectrum of the CR electrons we assume a power-law distribution ${ N_{\mathrm{CR}}(\gamma)=n_{\mathrm{CR}}\gamma^{-\beta} }$, where $\gamma$ is the Lorentz factor and $\beta$ is the power-law index \citep[][]{Rybicki1979,Webber1998}. This power-law assumption is a valid approximation for synchrotron observations above $1\ \mathrm{GHz}$ \correction{in the ISM} \citep{Padovani2021}.

Here, by implementing the CR electrons toy model described in Sect.~\ref{Sect:CR_electrons_model}, we use the capability of {\tt POLARIS} to simulate synchrotron radiation \citep[see, e.g.,][for further details]{Reissl2019}. \correctionII{We run a set of RT simulations where the medium is synchrotron active (see panel D of Fig.~\ref{fig:set_ups_RT_simulations}) and apply the RM synthesis technique described in Sect.~\ref{Sect:FR_synthesis}.} 
\correctionII{We observe the Stokes vector over the observational frequency interval $1-5$\,GHz, with a bin size of 200\,MHz and we then derive the RM. Computational constraints limit us to use this channel width, which is broader than what will be achieved in future surveys \citep[see, e.g.,][for a review]{Heald2020}. However, this resolution is sufficient for our purposes.} 
The RM map obtained is then compared to the reference one. 

\section{Results and discussion}\label{sec:results_discussion}

\subsection{Our Faraday sky}

The full-sky RM map for an observer placed at the center of \correctionII{our Local Bubble-like} cavity is presented in Fig.~\ref{fig:FR_fullSkyMaps_reference} in both its Mollweide and orthographic projections. 
We refer to this map as our reference map, as it is computed directly by {\tt POLARIS} and \correctionII{pertains to the full extent of the MHD simulation.}
\correction{To help in the interpretation of our results, we show in Fig.~\ref{fig:FR_fullSkyMaps_hutschenreuter} the most recent full-sky RM map of the Milky Way \citep[][]{hutschenreuter2022}. While we acknowledge that this comparison comes with some caveats, it provides a meaningful connection to observational data and is intended to contextualize our synthetic observations, highlighting potential similarities and differences. We discuss the caveats of our work in Sect.~\ref{sect:caveats}.}

By visually comparing the two maps, \correction{we note that our synthetic simulation does not exhibit} strong RM values at low latitudes ($|b|$\,$<$\,$15^{\circ}$), in contrast to the map by \cite{hutschenreuter2022}. This can be explained by the fact that our MHD simulation only samples the Galactic midplane out to a distance of $250\,$pc, lowering the amount of cumulative rotation of the polarization angle.
\correctionII{Also, the size of the simulation domain limits the enhancement of magnetic fields through large-scale processes via dynamo action \citep[see, e.g.,][]{Gent2024}.}
However, despite our simulation covers only a cube with a side length of 500\,pc, we obtain RM values exceeding 100 $\mathrm{rad}\,\mathrm{m}^{-2}$. This supports the idea that a non-negligible portion of the signal observed in the map by \cite{hutschenreuter2022} is of local origin, in agreement with the findings by \cite{Reissl2023}. In Fig.~\ref{fig:FR_fullSkyMaps_reference}, we highlight with red, green, and yellow dashed circles some of these high RM regions, \correction{which correspond to the highly ionized bubbles} identified in Fig.~\ref{fig:nth_column}. We further discuss \correction{the imprint of the SN-blown cavities on RM maps} in Sect.~\ref{sect:local_enviroment}. \correctionIII{In Appendix~\ref{app:polarizationMaps}, we present the maps of the polarized intensity and linear polarization fraction to illustrate how these structures impact these quantities.} A more quantitative comparison between the two RM maps can be made using the power spectrum, which we report in Appendix~\ref{app:power_spectra} due to the uncertainties in its interpretation.

Our MHD simulations do not incorporate a global toroidal magnetic field component, which is typically present in Milky Way-like galaxies \citep[see, e.g.,][]{beck2013,BorlaffEtAl2021,unger2024}. This absence is likely reflected by the lack of changes between positive and negative values in RM between the quadrants of the same celestial hemisphere, as is observed in the Milky Way map of Fig.~\ref{fig:FR_fullSkyMaps_hutschenreuter}.
\correction{Despite this limitation, we successfully reproduce the sinusoidal patterns of RM as a function of Galactic longitude, although not their amplitude. These patterns} are observed both in the RM signals toward extragalactic \correctionIII{sources \citep[see e.g.,][]{Taylor2009,hutschenreuter2022}} and in RM synthesis of radio data from the Galactic Ionic Medium \correctionIII{Survey \citep[see e.g.,][]{Wolleben2021,Dickey2022}.}
In Fig.~\ref{fig:RMlon}, we show these trends with sinusoidal fits to the synthetic data in the intermediate latitude range ($40^\circ<|b|<45^\circ$) for three different cases: considering the entire data cube, considering only the bubble, and considering the data with the bubble carved out. The sinusoidal pattern is clearly visible when the full data cube is used and also when only the volume enclosed by the outer edges of the bubble is considered. However, the pattern disappears for positive latitudes and is mostly absent for negative latitudes when the bubble is removed from the data. \correctionII{The amplitude of the pattern does not vanish entirely at negative latitudes likely because the cavity selected from our MHD simulations has an open chimney in the southern direction (see Fig.~\ref{fig:input_data_overview}). As a result, the bubble's walls in that region are less dense, with shallower boundaries, possibly allowing a residual signal from the surrounding regions.} This finding suggests that the observed RM sinusoidal pattern originates in the Solar neighborhood under the influence of the Local Bubble.

\begin{figure*}
    \centering
    \includegraphics[width=0.95\textwidth]{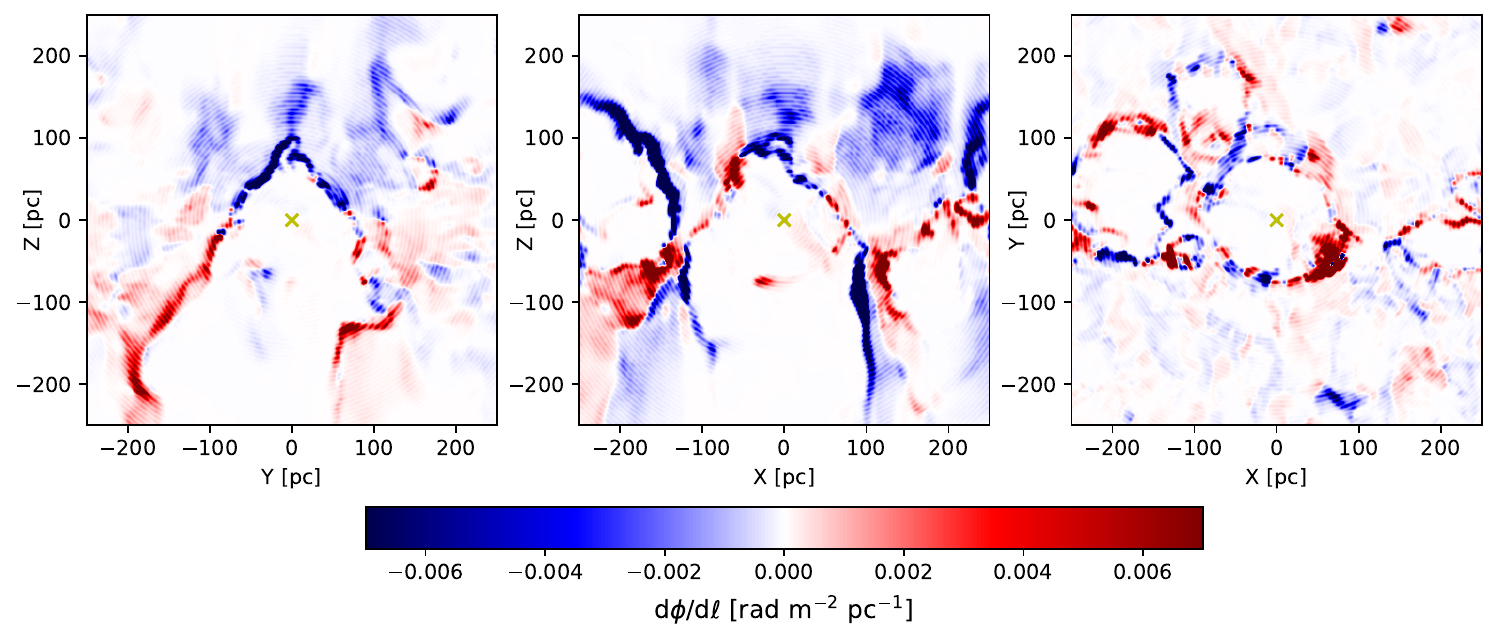}
    \caption{Midplane cuts, as in Fig.~\ref{fig:input_data_overview}, but for the first derivative of the Faraday rotation depth along the LOS, ${ \mathrm{d}\phi/\mathrm{d}\ell=(\phi(\ell+\mathrm{d}\ell)-\phi(\ell))/\mathrm{d}\ell }$, computed from the synthetic observation produced using {\tt POLARIS}. The red (blue) colors correspond to increases (decreases) in RM along each LOS toward an observer placed at the center of the selected bubble. The position of the observer is marked by a yellow cross. We note that the finite number of shells used to sample the rays causes the concentric ring regular pattern visible in the plots.}
    \label{fig:deltaFR}
\end{figure*}

\subsection{The importance of the bubbles edges and local environment}
\label{sect:local_enviroment}

Recent studies highlight how the ISM and its structure are shaped by the expansion and interaction of shells generated by SNe \citep[see, e.g.,][]{Krause2018,Soler2018,Bracco2020,ZuckerEtAl2022}. 
The injection of energy by SNe and stellar feedback are responsible for sweeping away and ionizing the medium. The thermal electrons thus formed are usually accumulated at the edges of the cavities, at their interaction points, and in the SNe outflows. Furthermore, an enhancement in magnetic field strength is usually observed at the cavity walls due to the squeezing of the field lines into a thin shell \citep[see, e.g.,][]{FerriereEtAl1991}.
We therefore expect local structures to have a strong imprint on the Faraday rotation maps \citep[see, e.g.,][]{Reissl2023}.

Using the advantage that the RT technique offers, we are able to compute the first derivative of the Faraday depth along each LOS, as described in Sect.~\ref{Sect:LOS-analysis}.
We present the results for the midplane cuts in Fig.~\ref{fig:deltaFR}.
This figure illustrates that the edges of cavities and the outflow of gas at high latitudes significantly contribute to the full RM signal, \correction{as these are the regions where the majority of free electrons are accumulated.} This correspondence becomes more apparent when comparing Fig.~\ref{fig:deltaFR} with the thermal electron density, $n_{\mathrm{th}}$, shown in the third panel of Fig.~\ref{fig:input_data_overview}. \correction{The Faraday rotation within the cavities is instead significantly lower due to the low-density environment and the more chaotic, weaker magnetic field \citep[see also][]{Stil2009}.}
By also examining the thermal electron column density in  Fig.~\ref{fig:nth_column}, \correction{it is possible to see} how the overdensity of thermal electrons at the edges of the cavities (highlighted by the red, yellow, and green dashed circles) generate some of the most intense Faraday rotation features in the maps presented in Fig.~\ref{fig:FR_fullSkyMaps_reference}.
We acknowledge that large uniform features could also contribute significantly to the RM signal but would not be discernible in the derivative map.

\begin{figure*}[]
	\centering
	\includegraphics[width=0.49\textwidth]{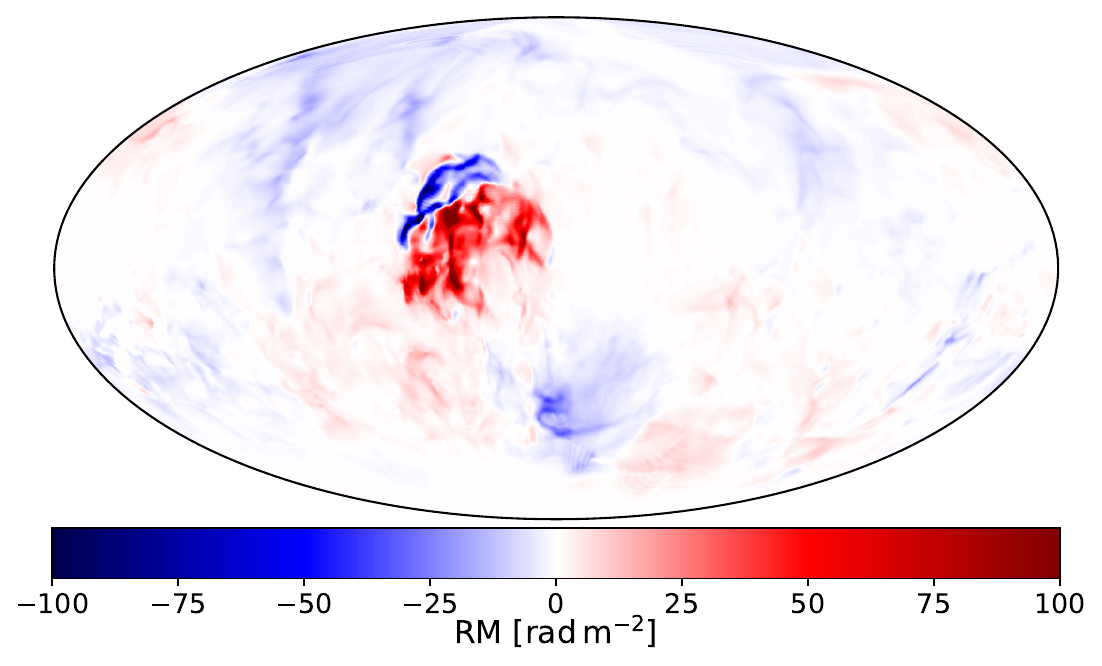}
 	\includegraphics[width=0.49\textwidth]{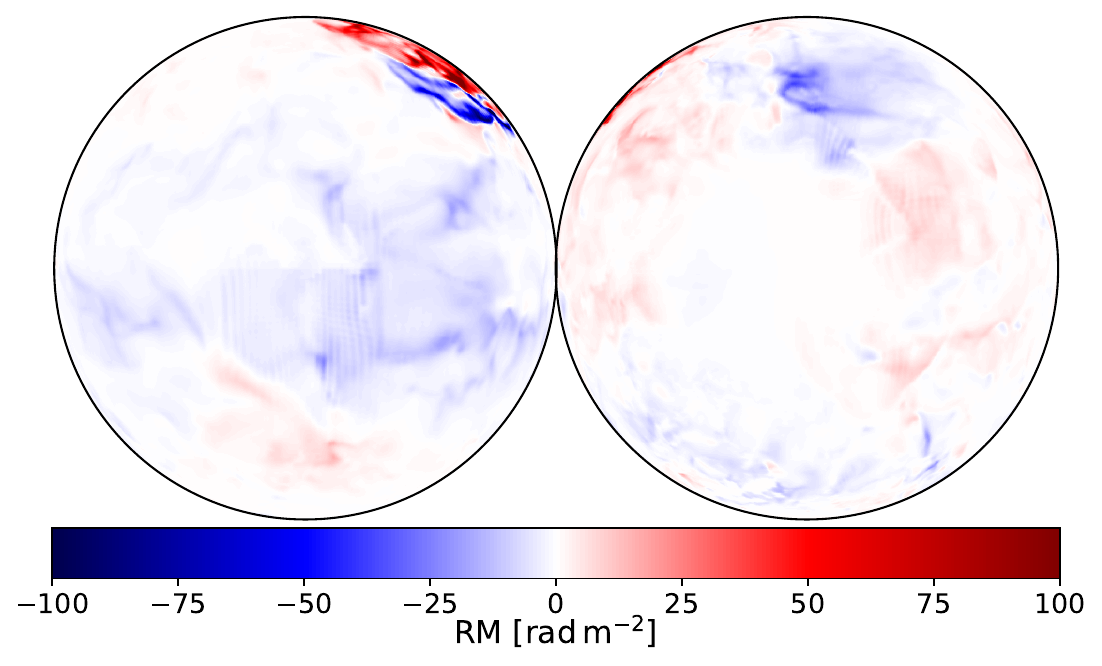}\\
    \includegraphics[width=0.49\textwidth]{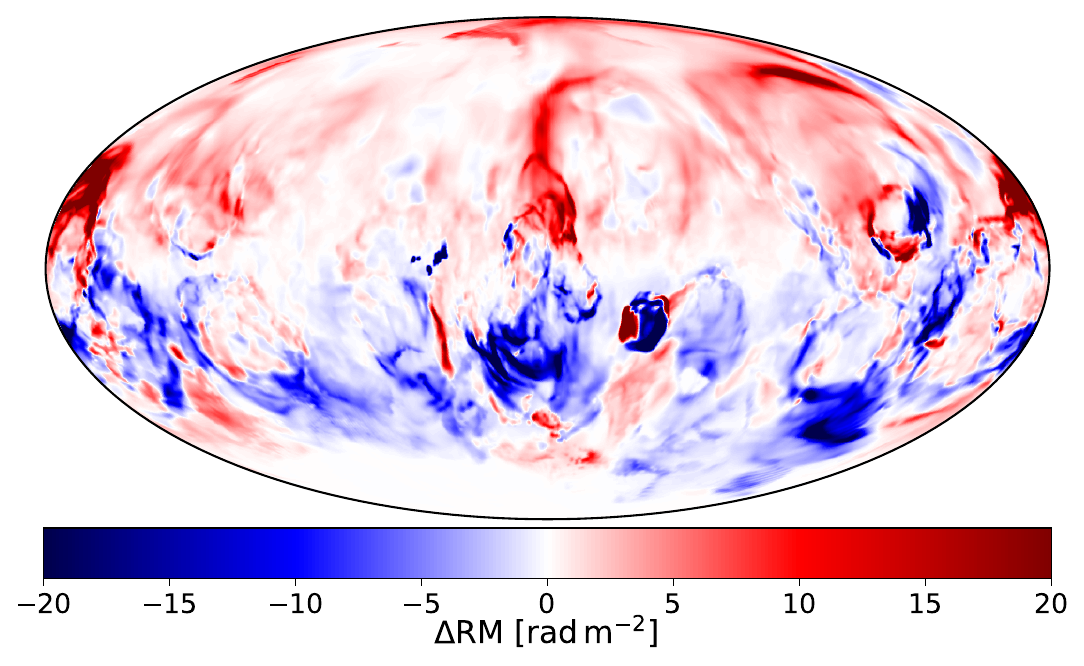}
    \includegraphics[width=0.49\textwidth]{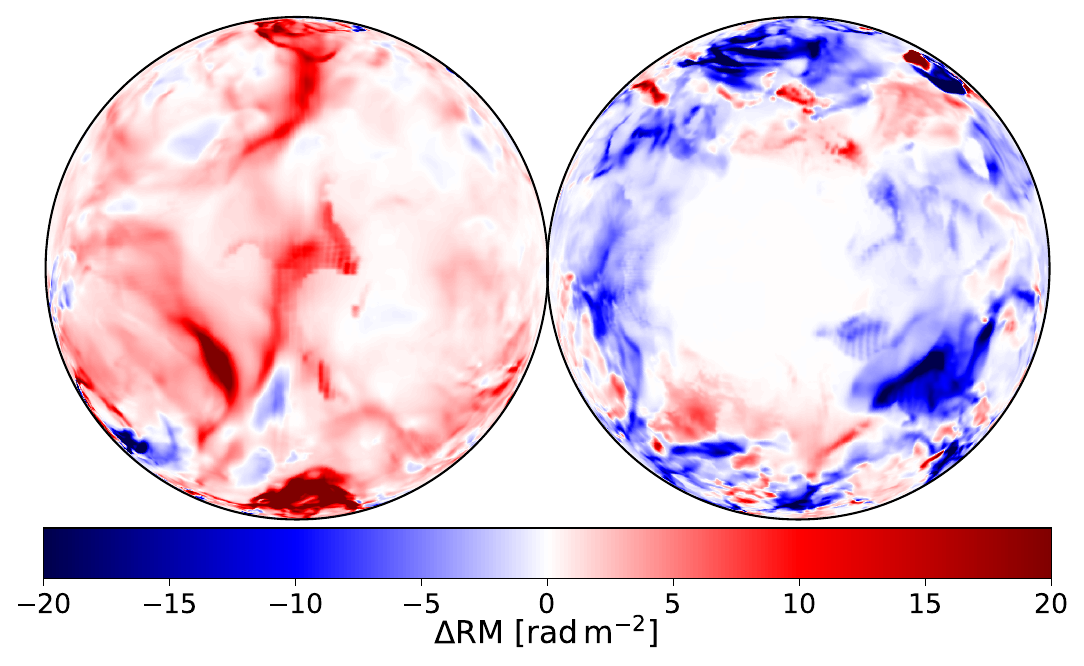}\\
	\caption{\textit{Top panel}: \correctionII{The same as in} Fig.~\ref{fig:FR_fullSkyMaps_reference}, but \correctionII{for the RM signal} obtained by considering only the volume enclosed by the outer edges of the selected bubble (setup B).
    \textit{Bottom panel}: the difference between the RM reference map (see Fig. \ref{fig:FR_fullSkyMaps_reference}) and the one presented in the top panel of this figure.}
	\label{fig:FR_fullSkyMaps_reference_onlyEdges}
\end{figure*}

To evaluate the impact of the candidate bubble on the RM maps, we conduct a RT simulation considering only the \correctionII{walls and the} inner volume of the cavity within which the observer is located (see also Sects.~\ref{sec:simulated_cavity}, \ref{Sect:LOS-analysis}, and \citet{Maconi2023}).
In the top panel of Fig.~\ref{fig:FR_fullSkyMaps_reference_onlyEdges}, we present the Faraday rotation map obtained for this setup {(corresponding to panel B in Fig.~\ref{fig:set_ups_RT_simulations}). In the bottom panel of Fig.~\ref{fig:FR_fullSkyMaps_reference_onlyEdges}, we show the difference between this map and the reference one.}
From this comparison it becomes clear that the very close environment is responsible for the strongest and largest angular feature of our reference map, supporting the idea that large angular correlations in Faraday data likely result from local structures \citep[see, e.g.,][]{Hutschenreuter2020}.

Our results corroborate the idea that the local environment and the Local Bubble strongly influence the \correctionII{observed RM for an observer} situated within such a cavity \citep[see e.g.,][]{Reissl2023,Pelgrims2025}. The importance of the impact of local structures therefore poses limitations on the extent to which Faraday rotation maps can be used to model the structure of the Milky Way's global magnetic fields and might be underestimated in some studies \citep[see, e.g.,][]{unger2024}.

\begin{figure*}[]
	\centering
	\includegraphics[width=0.49\textwidth]{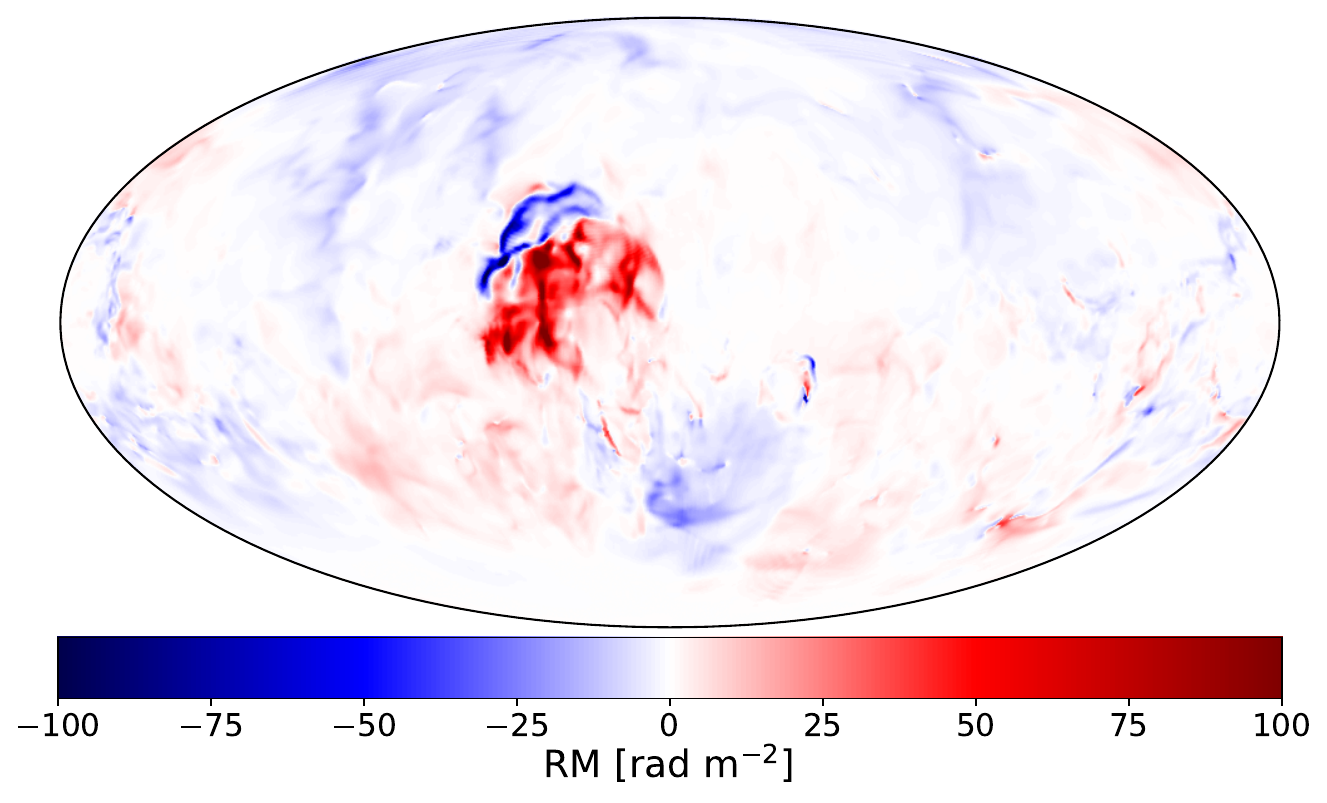}
 	\includegraphics[width=0.49\textwidth]{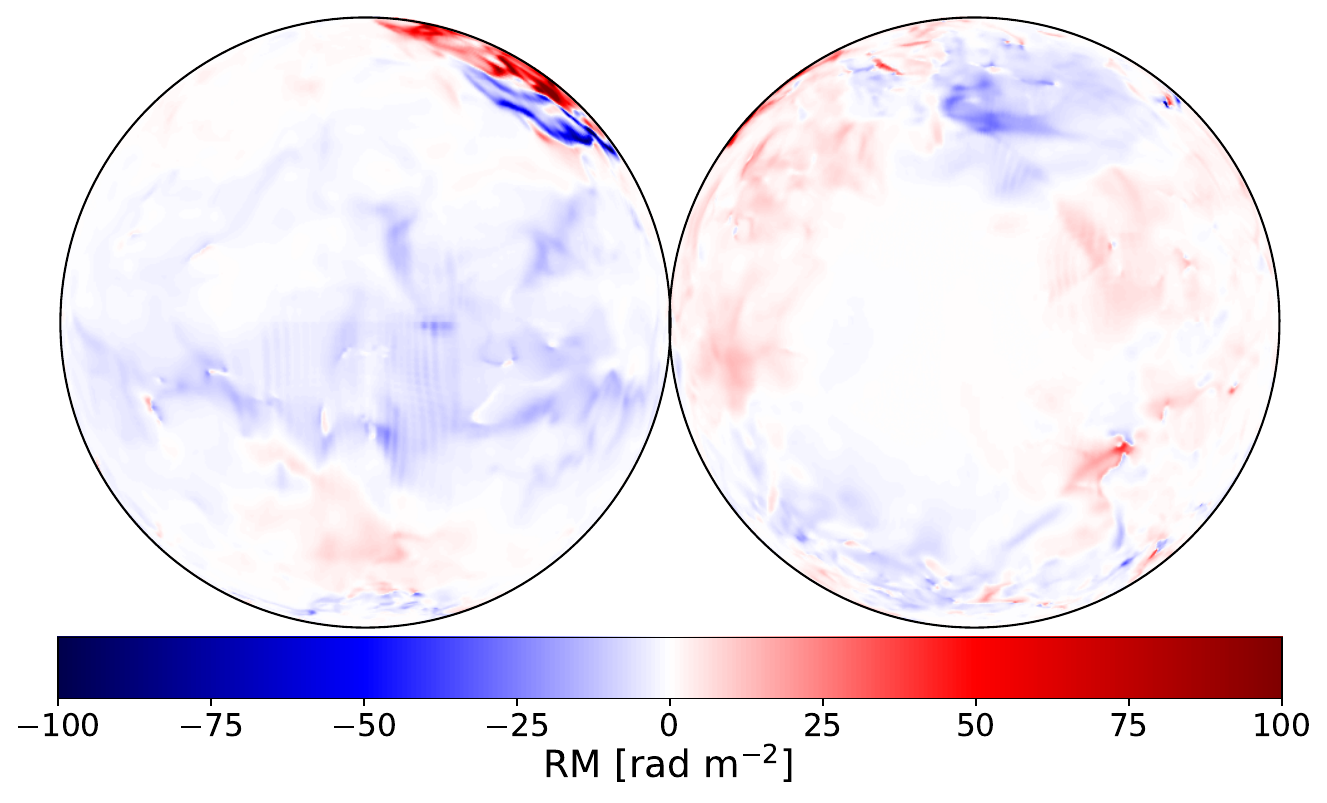}\\
    \includegraphics[width=0.49\textwidth]{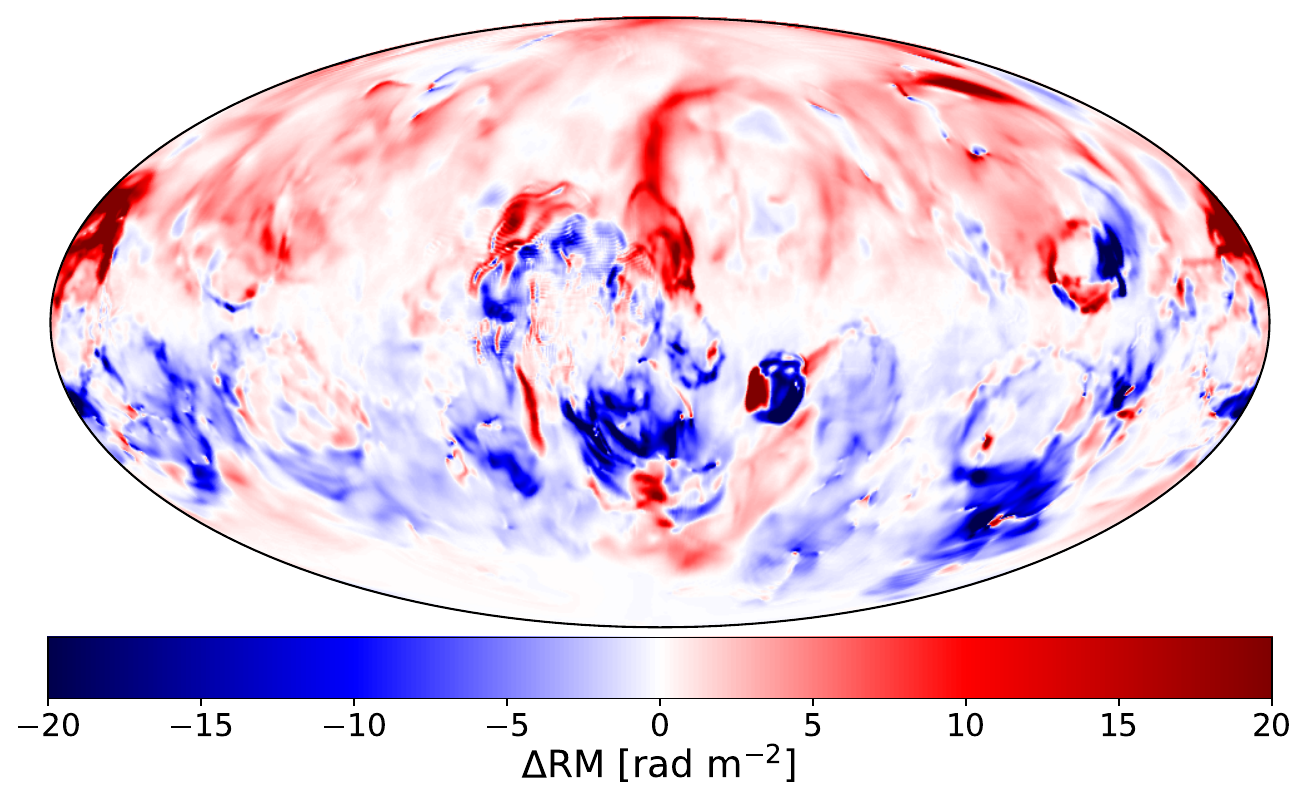}
    \includegraphics[width=0.49\textwidth]{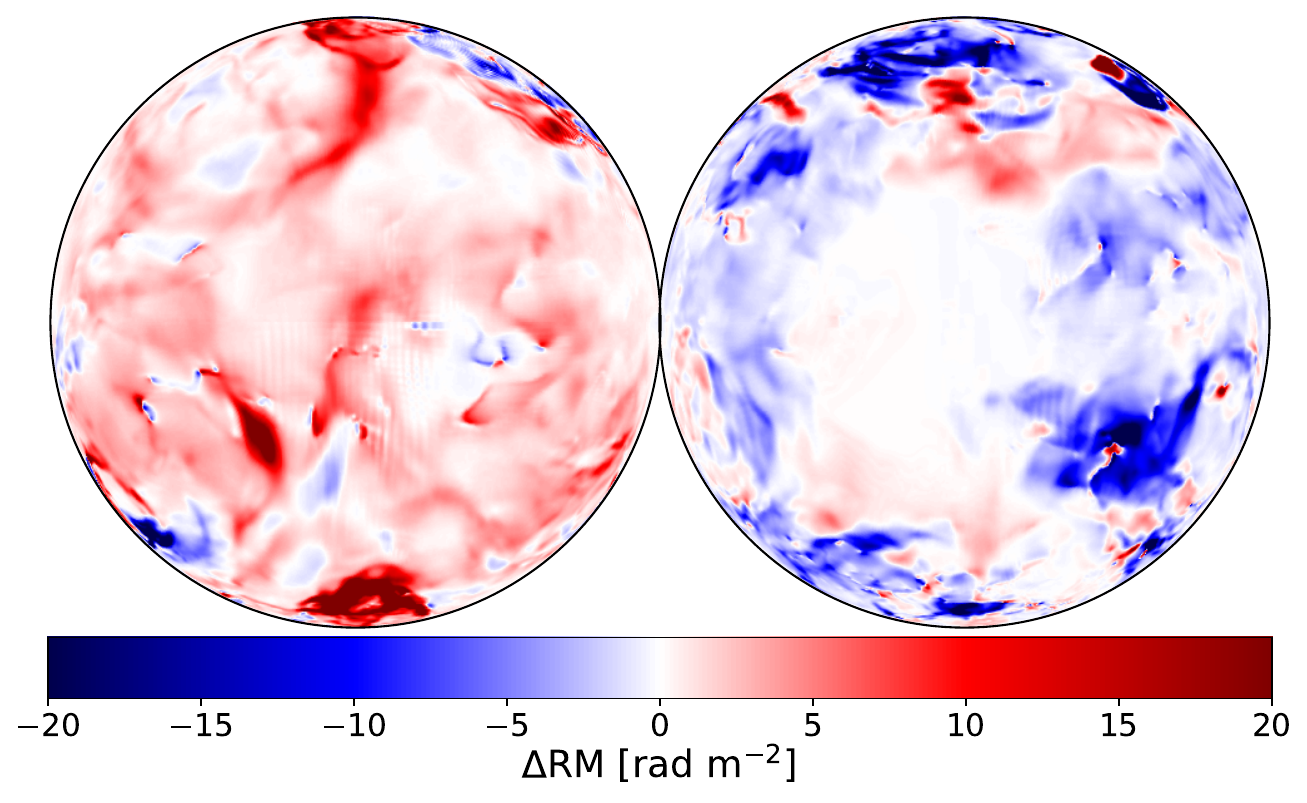}\\
    \includegraphics[width=0.49\textwidth]{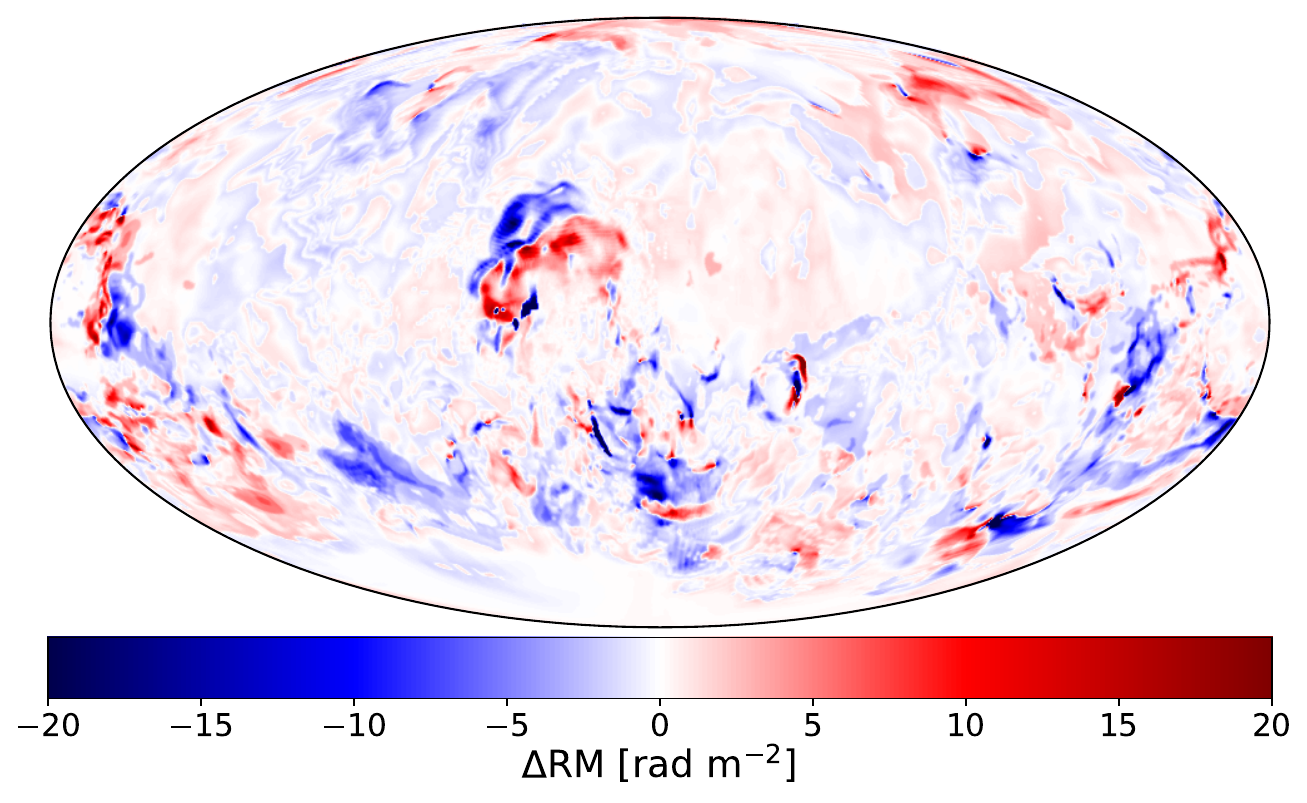}
    \includegraphics[width=0.49\textwidth]{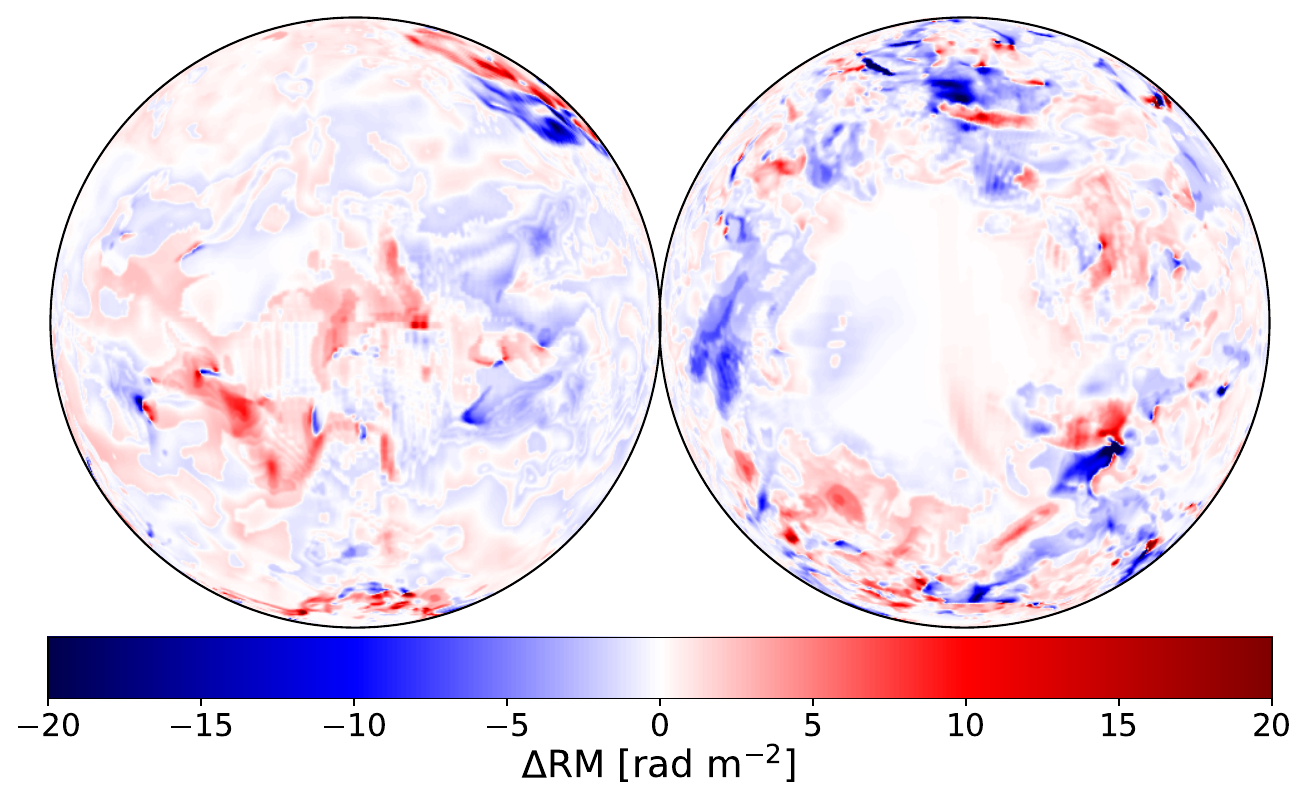}\\
	\caption{\correctionII{
    \textit{Top panel}: The same as in Fig.~\ref{fig:FR_fullSkyMaps_reference}, but for the RM signal obtained from synchrotron radiation without any background source  (setup D).
    \textit{Middle panel}: the difference between the RM reference map (see Fig.~\ref{fig:FR_fullSkyMaps_reference}) and the one presented in the top panel of this figure. 
    \textit{Bottom panel}: the difference between the RM map obtained by considering only the volume enclosed by the bubble walls (see Fig.~\ref{fig:FR_fullSkyMaps_reference_onlyEdges}) and the one presented in the top panel of this figure.}
     }	\label{fig:FR_fullSkyMaps_synchrotron_only}
\end{figure*}

\correctionII{
\subsection{The Faraday rotation without background sources}\label{Sect:FR_maps_and_synch_emission}
}

Relativistic CR electrons, as they spiral along magnetic field lines, emit synchrotron radiation across a broad spectrum of wavelengths. 
This leads to complex configurations along the LOS, where a region can simultaneously emit polarized radiation and induce rotation in the polarization angle of the radiation passing through it. \correctionII{Separating these two effects can be challenging.}
\correctionII{In this Section, we limit our study to the case in which no background sources are considered at all, meaning that the RM signal is derived by applying the RM synthesis technique (see Sect.~\ref{Sect:FR_synthesis}) to the diffuse polarized emission from the synchrotron active medium.} We refer to Sect.~\ref{Sect.set_up_synchrotron} for further details on the RT setup and to Sect.~\ref{Sect:CR_electrons_model} for the CR electron model used.

We present the Faraday rotation map obtained for the analyzed case and the difference with the reference one in the top and middle panels of Fig.~\ref{fig:FR_fullSkyMaps_synchrotron_only}, respectively. Remarkably, we find that the resulting RM map still closely resembles the one obtained when considering only the walls and the inner volume of the selected cavity (see Fig.~\ref{fig:FR_fullSkyMaps_reference_onlyEdges}). \correctionII{The bottom panel of Fig.~\ref{fig:FR_fullSkyMaps_synchrotron_only} presents the corresponding difference map.}
This suggests that RM values generated by local structures, such as the Local Bubble, can be recovered by analyzing diffuse synchrotron radiation alone \citep[see, e.g.,][]{Jelic2015,Jelic2018,Erceg2022,Erceg2024,Boulanger2024}. This is likely because polarization originating far from the observer is depolarized along the LOS and only nearby structures survive. \correctionII{While this effect is generally more prominent at low frequencies, such as those of LOFAR \citep[see, e.g.,][]{Shimwell2022}, we still observe it at higher frequencies in our simulations, despite the conditions for significant Faraday depolarization not being fully met in our simulated volume.}

\subsection{Caveats}\label{sect:caveats}
The synthetic Faraday rotation observations presented here regard an observer placed inside a Local Bubble-like cavity. This cavity has been selected within a set of simulations that aim to reproduce the dynamical evolution of a complex multi-phase ISM with physical properties comparable to the ones of the solar neighborhood. However, due to limitations in resolution required to detail the local environment, our investigation is confined to a specific region of the ISM rather than encompassing the entire galaxy. Consequently, we lack a comprehensive view of the full galactic disk.
A global toroidal magnetic field component, that may contribute to the RM signal, is also missing from our simulations.

To study the RM from diffuse synchrotron radiation, we develop a simplified CR diffusion model. We note that a detailed model \correction{accounting for all aspects of CR electron physics} is very complex and beyond the scope of this paper. 
\correction{We highlight some aspects that justify the simplified model used here. Firstly, our understanding of CR electron acceleration is incomplete. In particular the acceleration efficiency as a function of magnetic obliquity differs between particle-in-cell and plasma simulations and the effective models that match observations \citep{WinnerEtAl2020}. Secondly, the transition of the effective transport speed from the shock front into the ISM is completely unknown. The diffusion speed just ahead of the shock should follow the Bohm's diffusion, which is of the order of $D\sim10^{21}\,\mathrm{cm}^2\,\mathrm{s}^{-1}$. In the ISM, the diffusion is likely to follow the galactic diffusion value of the order $D\sim10^{28}\,\mathrm{cm}^2\,\mathrm{s}^{-1}$. The transition from one to the other transport mode is highly uncertain and depends on the details of the small-scale magnetic field structure, which is not resolved in our simulations. Moreover, the hypothesis that electrons predominantly diffuse has recently been questioned by CR streaming simulations compared to observations \citep{ThomasEtAl2020}.
If CR stream rather than diffuse, their effective transport speed is likely to have similarly large dynamical range as the protons, which can be as large as eight orders of magnitude \citep{ThomasPfrommerPakmor2023}.
Thirdly, the effective losses depend on the energy of the spectrum as described in detail in \citet{WinnerEtAl2019,WinnerEtAl2020}. 
It is thus evident that the effective transport and resulting distribution of CR electrons constitute a highly complex topic that cannot be addressed within this paper. 
For these reasons, we adopt an ad-hoc model that incorporates acceleration, transport, and cooling processes.} This model aims to provide a reasonable estimate of the potential effects of diffuse synchrotron radiation on Faraday rotation maps.

In conclusion, the optimal numerical configuration for this investigation would include the large-scale field arising from Galactic dynamics, achieve the necessary local environment resolution, and incorporate CR diffusion models. However, current computational capabilities pose challenges to conducting such a comprehensive numerical experiment.

\section{Conclusions}\label{sec:conclusions}

We conducted an analysis on synthetic Faraday rotation observations for an observer placed at the center of a Local Bubble-like cavity, chosen within a simulation with properties similar to the local Milky Way environment.

We found that the local environment is fundamental in determining the Faraday rotation signal. Specifically, the edges of the SN-blown cavities contribute the most to the Faraday signal \citep[see also][]{Stil2009,Pakmor2018}.
This suggests the importance of characterizing the solar neighborhood and the Local Bubble in detail, as well as quantifying their impact on observations, \correctionIII{as also pointed out by other recent works \citep[see e.g.,][]{Pelgrims2025,Korochkin2025}.}This is further corroborated by the fact that, even though the numerical simulation used to produce the synthetic maps traces the Galactic midplane out to a distance of only $250\,$pc \citep{GirichidisEtAl2018b, girichidis2021}, it is possible to obtain high values of RM, suggesting that a portion of the signal observed in the map by \cite{hutschenreuter2022} comes from the local environment \citep[see, e.g.,][]{Hutschenreuter2020,Reissl2023}. 
We are also able to reproduce the sinusoidal patterns of RM as a function of Galactic longitude, \correction{demonstrating their disappearance when the bubble is excluded, thereby suggesting a local origin} \citep[see, e.g.,][]{Taylor2009,hutschenreuter2022,Dickey2022}.
However, our synthetic maps do not exhibit a shift from positive to negative values in the Faraday signal between the quadrants within the same celestial hemisphere, as seen in the map by \cite{hutschenreuter2022}. This discrepancy may be attributed to the absence of a global toroidal magnetic field component in our MHD simulation \citep[see, e.g.,][]{beck2013,BorlaffEtAl2021}.

We show that by using the diffuse polarized synchrotron radiation alone, it is possible to recover the Faraday rotation signal generated by the SN-blown cavity within which the observer is placed. This corroborate the observational results obtained using radio surveys data \citep[see, e.g.,][]{Jelic2018,Thomson2021,Erceg2024}.

\correction{The simulations and analysis presented in this paper highlight the importance and the need of upcoming Milky Way multi-physics simulations and their post-processing for a comprehensive understanding and accurate interpretation of future high-resolution Faraday rotation maps.}



\section*{Data Availability}
The Local Bubble candidate simulations data are part of the SILCC project and are available at \url{http://silcc.mpa-garching.mpg.de}. The radiative transfer code {\tt POLARIS} is publicly available at \url{https://portia.astrophysik.uni-kiel.de/polaris/}. The {\tt RM-Tools} are hosted at \url{https://github.com/CIRADA-Tools/RM-Tools}. The synthetic Faraday rotation observations maps and the analysis scripts will be shared upon request.


\begin{acknowledgements}
\correction{We thank the anonymous referee for the insightful comments, which have significantly improved the quality and readability of the paper.} This work has received financial support from the European Research Council via the ERC Synergy Grant “ECOGAL” (project ID 855130), from the German Excellence Strategy via the Heidelberg Cluster of Excellence (EXC 2181 - 390900948) “STRUCTURES”, and from the German Ministry for Economic Affairs and Climate Action in the project “MAINN” (funding ID 50OO2206). The team in Heidelberg also thanks for computing resources provided by the Ministry of Science, Research and the Arts (MWK) of {\em The L\"{a}nd} through bwHPC and the German Science Foundation (DFG) through grant INST 35/1134-1 FUGG and 35/1597-1 FUGG, and also for data storage at SDS@hd funded through grants INST 35/1314-1 FUGG and INST 35/1503-1 FUGG. EM and SH were funded by the European Union (ERC, ISM-FLOW, 101055318). Part of the crucial discussions that led to this work took part under the program Milky-Way-Gaia of the PSI2 project funded by the IDEX Paris-Saclay, ANR-11-IDEX-0003-02. AB acknowledges financial support from the INAF initiative ``IAF Astronomy Fellowships in Italy'' (grant name MEGASKAT). \correction{RSK also thanks the 2024/25 Class of Radcliffe Fellows for their company and for highly interesting and stimulating discussions.}
\end{acknowledgements}

\bibliographystyle{aa}
\bibliography{./bibtex,astro,girichidis}

\appendix

\section{Radiative Transfer simulations with {\tt POLARIS}}\label{app:polaris}

{\tt POLARIS} is a three-dimensional Monte-Carlo (MC) continuum and line radiative transfer (RT) code employed to post-process magnetohydrodynamic (MHD) simulations \citep{Reissl2016}. 
The code, making use of the physical quantities provided by MHD simulations (e.g., gas density, gas temperature, magnetic fields, electron densities, etc.), along with an arbitrary number of radiation sources, can compute, among other parameters, the dust temperature, grain alignment efficiency, synchrotron emission, and generate synthetic multi-wavelength intensity, polarization, and Faraday rotation maps \citep{Reissl2016,Reissl2019}.

\correction{From a physical point of view,} the propagation of radiation through a medium can be described by the RT equation \citep[see, e.g.,][]{Rybicki1979},
\begin{equation}
    \frac{d}{\mathrm{d}\ell}\textbf{S} = - \hat{K} \textbf{S} + \textbf{J} \,,
    \label{eq:RT}
\end{equation}
where $\textbf{J}$ is the emissivity and $\hat{K}$ is a 4$\times$4 Müller matrix describing the extinction, absorption, as well as Faraday rotation. The Müller matrix coefficients for modeling the Faraday effect are 
\begin{equation}
    K_{\mathrm{23}} = -K_{\mathrm{32}} =  \frac{1}{2\pi}\frac{n_{\rm th}(\ell)e^2B_{||}(\ell)}{m_{\rm e}^2c^4}\lambda^2 \,,
    \label{eq:kappaV}
\end{equation}
where $B_{||} = B \cos(\vartheta)$ is the line-of-sight (LOS) magnetic field strength \correction{and $\ell$ indicates the distance between the source and the observer.}
The rotation of the polarization vector over the path element d$\ell$ is then $\mathrm{d}\phi = \lambda^{-2}K_{\mathrm{23}}\mathrm{d}\ell$.
{\tt POLARIS} solves Eq.~\ref{eq:RT} for each path element $\mathrm{d}\ell$ along the LOS. We refer the reader to \cite{Reissl2019,Reissl2020} for more technical details about the solving technique.

\section{Rotation measure synthesis}\label{app:FR_synthesis_appendix}

We construct a Faraday rotation measure (RM) synthesis routine following the work of \cite{Burn1966}, \cite{Brentjens2005}, and \cite{Heald2009}, and \correctionII{by adapting the Python code {\tt RM-tools} \citep{Purcell2020} to handle the {\tt POLARIS} output.}

\correctionII{As pointed out in Sect.~\ref{Sect:FR_synthesis},
the ideal approach to obtain the RM} would be to fit the data with the relation \mbox{$\chi_{\rm obs}=\chi_{\mathrm{source}}+\lambda^2 \times \mathrm{RM}$} but this may suffer from the shortcomings outlined in the same section. \correctionII{A Faraday RM synthesis technique \citep[see, e.g.,][]{Brentjens2005} or the Stokes QU-fitting technique \citep[see, e.g.,][]{Sokoloff1998} is therefore required. In this work, we focus on the Faraday RM synthesis approach.} 

\correctionII{In this context, the idea is to rewrite} the observed polarized emission $\mathcal{P}$ in Eq.~\ref{eq:polIntensity_complex_2} as 
\begin{equation}
    \Tilde{\mathcal{P}}( \lambda^2 ) = \mathcal{W}(\lambda^2) \mathcal{P}(\lambda^2) = \mathcal{W}(\lambda^2)\int_{-\infty}^{\ +\infty} F(\phi) e^{2i\phi\lambda^2} \mathrm{d}\phi \,,
    \label{eq:polIntensity_complex_3_appendix}
\end{equation}
where $F(\phi)$ is the Faraday dispersion function and $\mathcal{W}(\lambda^2)$ a window function, which is different from zero only at the sampled wavelengths.
The Faraday dispersion function is defined in terms of the observable quantities as
\begin{equation}
    F(\phi) = \int_{-\infty}^{\ +\infty}  \mathcal{P}( \lambda^2 ) e^{-2i\phi\lambda^2} \mathrm{d}\lambda^2 \,.
    \label{eq:faraday_dispersion_appendix}
\end{equation}
The computation of the quantity $F(\phi)$ is the main objective of RM synthesis, as it contains information about the Faraday depth $\phi$, which is ultimately related to the magnetic fields and thermal electrons. However, it is necessary to note that observations are not performed where $\lambda^2<0$ nor at all the values where $\lambda^2>0$ in Eq.~\ref{eq:faraday_dispersion_appendix}. To mathematically deal with this problem, Eq.~\ref{eq:polIntensity_complex_3_appendix} can be manipulated and inverted to obtain 
\begin{equation}
    \tilde{F}(\phi) = K \int_{-\infty}^{\ +\infty}  \tilde{\mathcal{P}}( \lambda^2 ) e^{-2i\phi(\lambda^2-\lambda_0^2)} \mathrm{d}\lambda^2 = F(\phi)\ast R(\phi) \,,\label{eq:observed_faraday_dispersion_appendix}
\end{equation}
where $K$ is the inverse of the integral over $\mathcal{W}(\lambda^2)$, $R(\phi)$ is the RM spread function, $\lambda_0$ is a factor introduced for the behavior of $R(\phi)$, and $\ast$ is the convolution operator. The rotation measure spread function (RMSF) is defined as
\begin{equation}
    R(\phi) \equiv K \int_{-\infty}^{\ +\infty}  \mathcal{W}( \lambda^2 ) e^{-2i\phi(\lambda^2-\lambda_0^2)} \mathrm{d}\lambda^2 \,.
    \label{eq:RM_spread_function}
\end{equation}
It has been shown \citep{Brentjens2005} that the quantities $\tilde{F}(\phi)$ and $R(\phi)$ can be written as sums, which are a suitable form for practical cases. The RM synthesis technique proceed by deconvolving Eq.~\ref{eq:observed_faraday_dispersion_appendix} and by applying a cleaning routine. 

The RM synthesis technique also presents different issues and limitations that should be considered when doing a RM synthesis experiment \citep[see, e.g.,][]{Brentjens2005,Heald2009}. 
For example, the precision of determining the RM at the peak of the Faraday dispersion function is influenced by the full width at half maximum (FWHM) of the RMSF. The FWHM is inversely proportional to the width of the observed $\lambda^2$ space, $\Delta \lambda ^2$, and can be computed as $\mathrm{FWHM}=2\sqrt{3}/\Delta \lambda ^2$ \citep{Brentjens2005} or, by introducing an empirical correction, $\mathrm{FWHM}=3.8/\Delta\lambda ^2$ \citep{Schnitzeler2009}.
Moreover, extended Faraday structures along the LOS can cause high depolarization at large $\lambda^2$ values, making sensitivity to these structures inversely proportional to the minimum $\lambda^2$ sampled.
An advantage of the RM synthesis technique is that bandwidth depolarization effects can be reduced by using narrow channels. However, the depolarization is not entirely eliminated, especially at low frequencies where it can be significant. Consequently, the technique of RM synthesis itself affects the results depending on the frequency coverage \citep[see, e.g.,][]{Basu2019}.
We refer the reader to the previously mentioned references for further details.

\correctionII{In our study, we apply the RM synthesis described above to setup D and E of Fig.~\ref{fig:set_ups_RT_simulations}. For these setups}
we consider the frequency range of $1-5\ \mathrm{GHz}$ with a channel width of 200\,MHz. This corresponds to a range in RM of $\pm 873\ \mathrm{rad\ m}^{-2}$ resulting in a RMSF with a FWHM of $44.0\ \mathrm{rad\ m}^{-2}$. The number of Faraday depth channels used is 161 and the cleaning is performed one time. 
The chosen frequency interval partially overlaps with the range covered by the forthcoming surveys generating RM grids \citep[see, e.g.,][for a review]{Heald2020}. We acknowledge that the width of our frequency bin is far from representing a real-case scenario, as forthcoming surveys will provide much denser sampling of the frequency interval. Nevertheless, our choice still represents an improvement compared, for example, to the \cite{Taylor2009}’s catalog mainly used by \cite{hutschenreuter2022} in their work, where only two bands were available. \correctionII{Moreover, for the purposes of our work, this channel width is sufficient, as also shown by the test on this routine described in Appendix~\ref{app:test_case}.}

\correctionII{
\subsection{A test case}\label{app:test_case}

In this section, we describe the test conducted for the routine outlined above and in Sect.~\ref{Sect:FR_synthesis}.
We assumed a distribution of background sources, with one source per LOS. As background sources, we adopt one of the standard polarization calibrators analyzed by \citet{Perley2013a,Perley2013b}, as described later. The conditions along a given LOS correspond to those shown in setup E of Fig.~\ref{fig:set_ups_RT_simulations}.
We then applied the RM synthesis and compared the resulting map with the reference one.}

\begin{figure*}
\centering
	\includegraphics[width=0.99\textwidth]{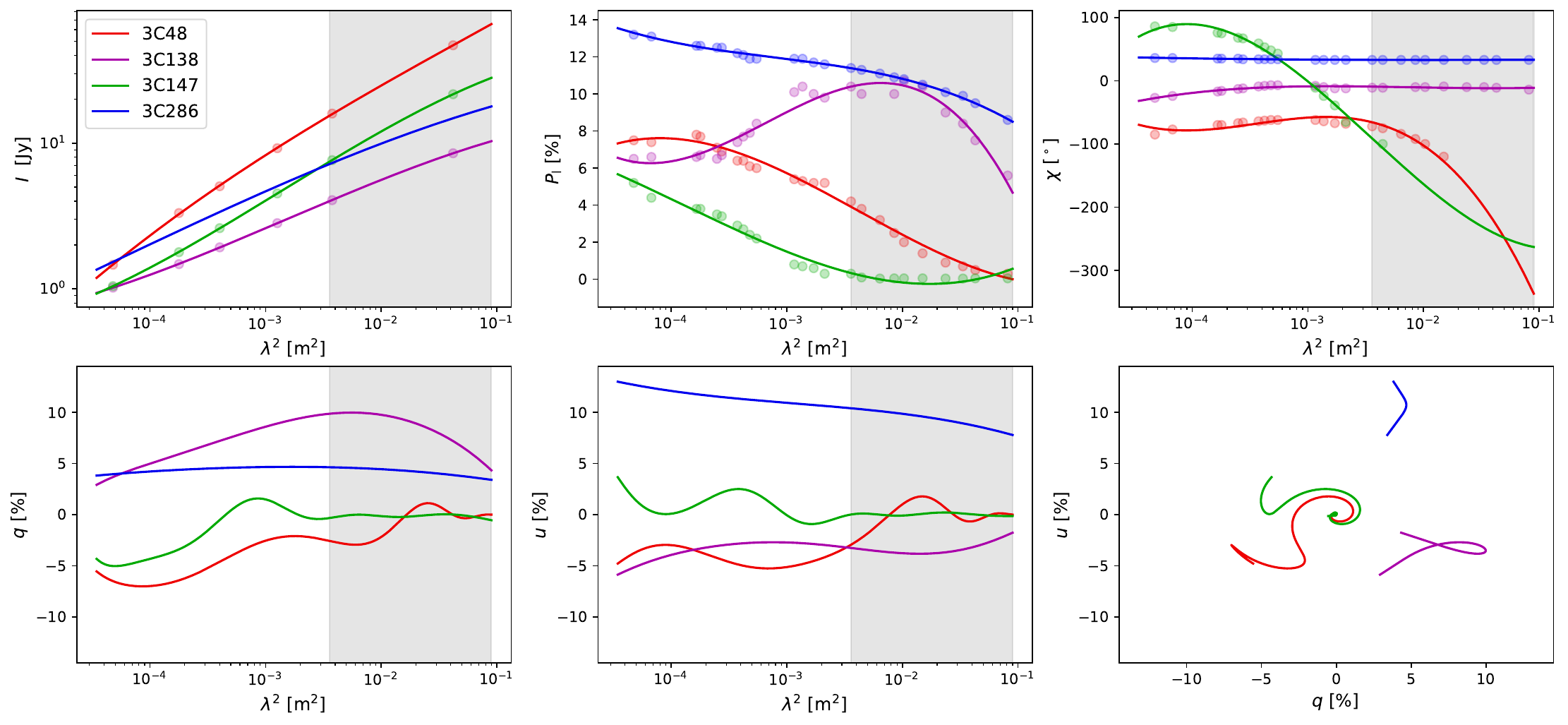}
    \caption{\correctionII{Polarization properties for the standard calibrators from \citet{Perley2013a,Perley2013b}.} \textit{Top row}: observed values of $I$, $p_{\mathrm{l}}$, and $\chi$, along with the best-fit functions. \textit{Bottom row}: Stokes quantities $q=Q/I$, $u=U/I$, and $q-u$ plane, respectively. The shaded gray area in the panels highlights the wavelength range covered by our synthetic Faraday rotation observations. This range corresponds to the frequency interval of 1-5\,GHz.}
    \label{fig:SourceFit}
\end{figure*}

As {\tt POLARIS} allows us to define the properties of the background sources, such as intensity and polarization state, along each LOS, to test our routine we chose to use the 3C286 source, which we select among the four radio sources (3C48, 3C138, 3C147, and 3C286) analyzed by \citet[][]{Perley2013a,Perley2013b}. These sources are standard polarization calibrators, as they are bright and their radiation properties are known at different frequencies. For these reasons, they exhibit significantly better and more desirable polarization properties compared to typical extragalactic sources \citep[see, e.g.,][]{OSullivan2012,Anderson2016,Pasetto2018,Schnitzeler2019,Ma2019,Livingston2022}. We selected the 3C286 source as its intrinsic linear polarization angle is generally constant across different wavelengths. This simplifies the interpretation of our results, as it avoids the complications introduced by an intrinsically complex polarization spectrum.
In order to select this source,  we model, following the approach by \cite{Perley2013a}, the intensity ($I\,\mathrm{[Jy]}$), linear polarization fraction ($p_{\mathrm{l}}\,\mathrm{[\%]}$), and linear polarization angle ($\chi\,\mathrm{[^\circ]}$) of four the sources with a cubic polynomial function of the form
\begin{equation}
    \begin{split}
    \{\ \log_{\mathrm{10}}\left( I \right),\ p_l,\ \chi\ \}=a_{\mathrm{0}}  + a_{\mathrm{1}} \log_{\mathrm{10}}\left(  
     \frac{\nu}{1\ \mathrm{GHz}}  \right) +\qquad\qquad \\ \qquad\qquad a_{\mathrm{2}} \log_{\mathrm{10}}^2\left(  
     \frac{\nu}{1\ \mathrm{GHz}}  \right)  + a_{\mathrm{3}} \log_{\mathrm{10}}^3\left(  
     \frac{\nu}{1\ \mathrm{GHz}}  \right) \,,
     \end{split}
    \label{eq:FitSource}
\end{equation}
where $\nu$ is the frequency in GHz.
The resulting fit values for the coefficients are given in Tables~\ref{tab:FitInt}, \ref{tab:FitPol}, and \ref{tab:FitAng}. In the top panels of Fig.~\ref{fig:SourceFit}, we show the observed values of $I$, $p_{\mathrm{l}}$, and $\chi$ for the selected sources, along with the best-fit functions, as a function of $\lambda^2$. For completeness, we report in the bottom panels of Fig.~\ref{fig:SourceFit} the quantities
\begin{align}
    q &= Q/I  = p \cos\ 2\chi \,, \\
    u &= U/I  = p \sin\ 2\chi \,,
    \label{eq:FitQ0-U0}
\end{align}
as a function of $\lambda^2$, and the $q - u$ plane. From Fig.~\ref{fig:SourceFit} (top row, third column), it is possible to observe that the intrinsic polarization angle $\chi$ \correctionII{varies over wavelength for 3C48 and 3C147, while it remains constant for 3C138 and 3C286, which justifies our choice of the latter.}

\begin{figure*}[]
	\centering
	\includegraphics[width=0.49\textwidth]{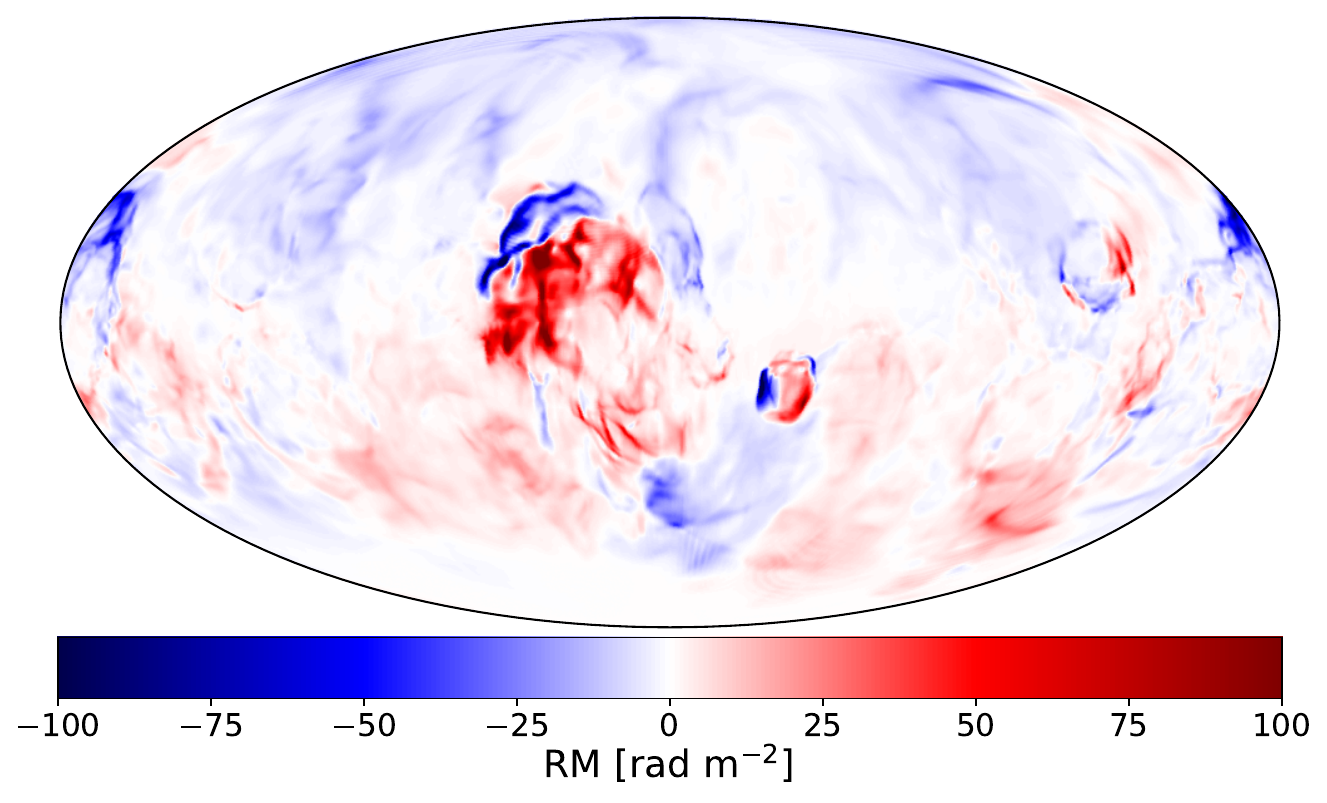}
 	\includegraphics[width=0.49\textwidth]{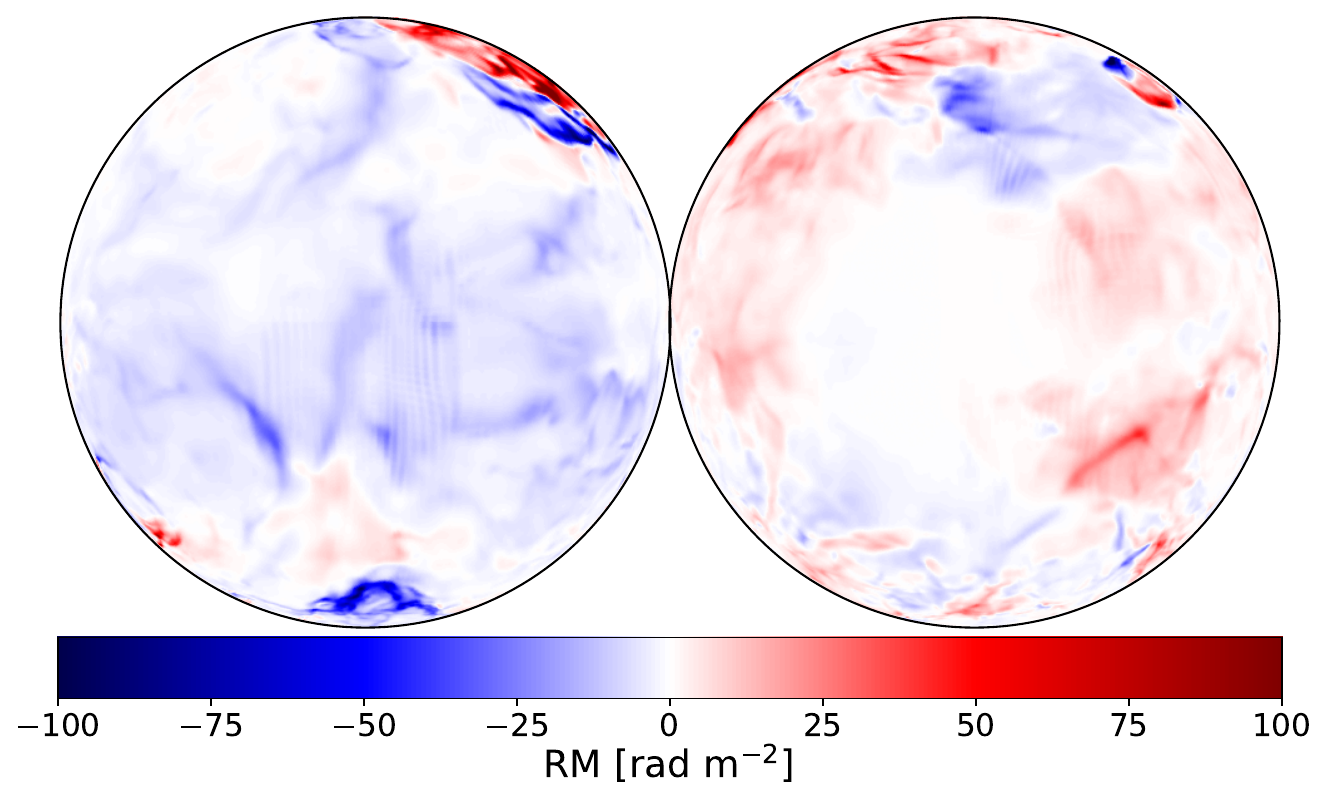}\\
    \includegraphics[width=0.49\textwidth]{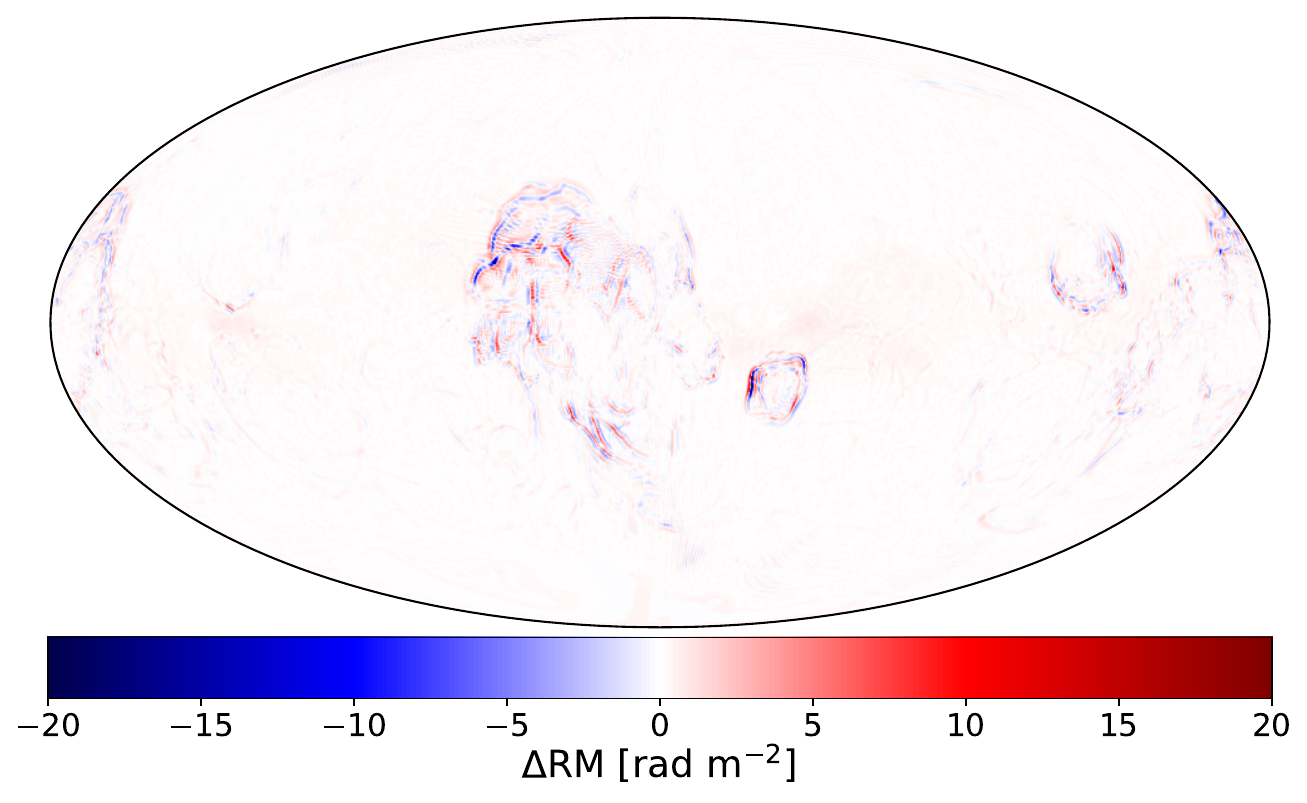}
    \includegraphics[width=0.49\textwidth]{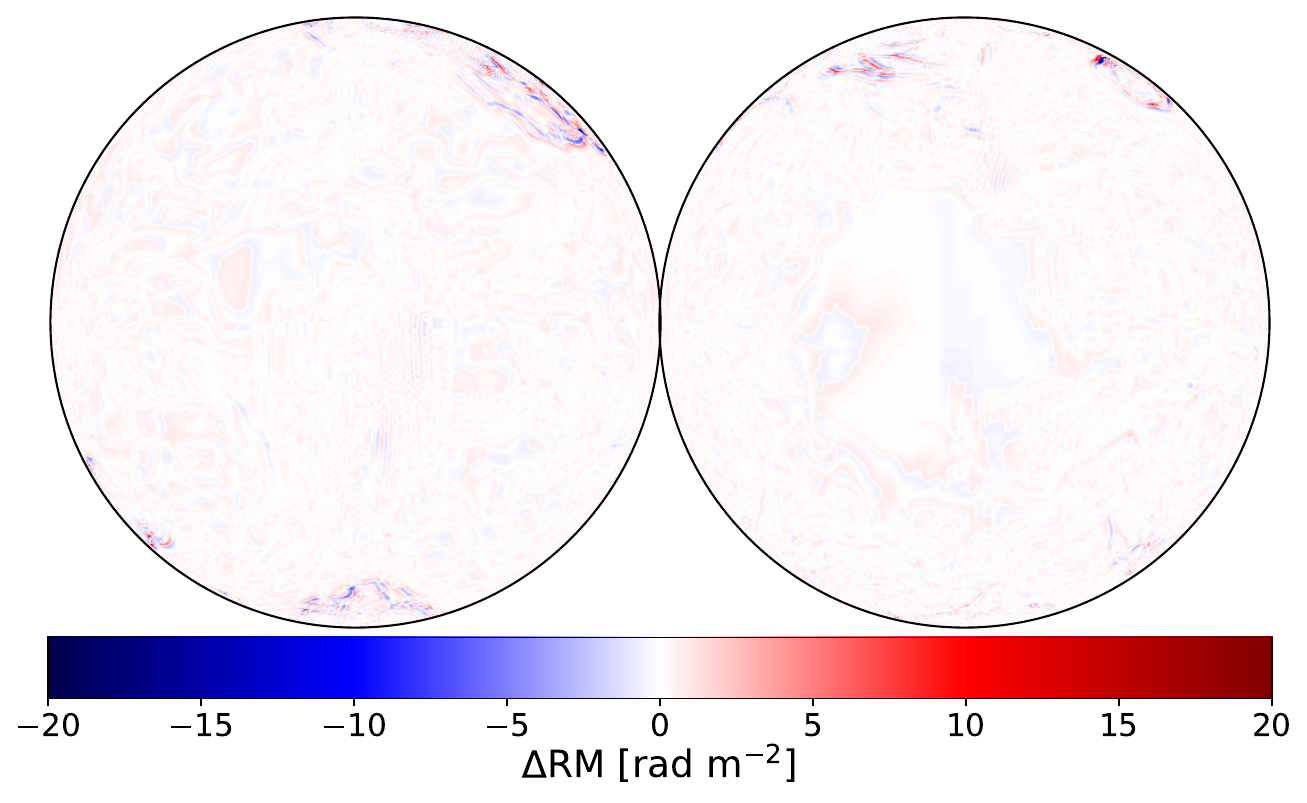}\\
	\caption{\textit{Top panel}: \correctionII{The same as in Fig.~\ref{fig:FR_fullSkyMaps_reference}, but for the RM signal obtained using the RM synthesis technique described in Sect.~\ref{Sect:FR_synthesis} and Appendix~\ref{app:FR_synthesis_appendix}. 
    The polarization properties of the background sources are set to the ones of the 3C286 source (setup E).
    \textit{Bottom panel}: the difference between the RM reference map (see Fig.~\ref{fig:FR_fullSkyMaps_reference}) and the one presented in the top panel of this figure.}   
    }
	\label{fig:FR_fullSkyMaps_indirect_3C286}
\end{figure*}

In Fig.~\ref{fig:FR_fullSkyMaps_indirect_3C286}, we present the Faraday rotation map obtained for our test case. 
The top panel displays the resulting Mollweide and orthographic projections of the map, which closely resemble our reference map (see Fig.\ref{fig:FR_fullSkyMaps_reference}). This similarity is confirmed by the difference maps shown in the bottom panel of Fig.~\ref{fig:FR_fullSkyMaps_indirect_3C286}. The close resemblance of these two maps can be attributed to the well-behaved nature of 3C286, whose intrinsic linear polarization remains constant across frequencies, as illustrated in Fig.~\ref{fig:SourceFit}, \correctionIII{meaning that this source has a RM close to $0\,\mathrm{rad}/\mathrm{m}^2$.} \correctionII{We acknowledge that, in a real-case scenario, the situation is significantly more complex due to the non-uniform distribution of background sources across the celestial sphere, their potentially low brightness, and the unknown intrinsic \correctionII{RM} spectrum. However, despite these caveats, our RM synthesis technique is able to satisfactorily recover the true RM map in this ideal case.}

\begin{table}
    \caption{Resulting coefficients of the cubic polynomial fit (refer to Eq.~\ref{eq:FitSource}) of the intensity of the radio sources 3C48, 3C138, 3C147, and 3C286  \citep{Perley2013a,Perley2013b}.}
    \centering
    \resizebox{0.49\textwidth}{!}{
    \begin{tabular}{l | c c c c | c} 
        \toprule
        Source & $a_{\mathrm{0}} \ \mathrm{[Jy]}$  & $a_{\mathrm{1}}\ \mathrm{[Jy]}$ & $a_{\mathrm{2}}\ \mathrm{[Jy]}$ & $a_{\mathrm{3}}\ \mathrm{[Jy]}$ & Ref.\\
        \midrule
        3C48 & 1.82 & -0.874 & 0.002 & -0.051 & this work \\
        3C138 & 1.01 & -0.462 & -0.242 & 0.091  & this work \\
        3C147 & 1.45 & -0.618 & -0.39 & 0.143  & this work \\
        3C286 & 1.25 & -0.461 & -0.172 & 0.034  & \citet{Perley2013a,Perley2013b} \\
        \bottomrule
    \end{tabular}}
    \label{tab:FitInt}
\end{table}

\begin{table}
    \caption{Same as Table~\ref{tab:FitInt}, but for the linear polarization fraction.}
    \centering
    \resizebox{0.49\textwidth}{!}{
    \begin{tabular}{l | c c c c | c} 
        \toprule
        Source & $a_{\mathrm{0}} \ \mathrm{[\%]}$  & $a_{\mathrm{1}}\ \mathrm{[\%]}$ & $a_{\mathrm{2}}\ \mathrm{[\%]}$ & $a_{\mathrm{3}}\ \mathrm{[\%]}$ & Ref.\\
        \midrule
        3C48 & -3.07 & 6.1 & 2.84 & -0.07  & this work \\
        3C138 & 8.79 & -28.2 & 23.6 & 4.68  & this work \\
        3C147 & -1.79 & 7.6 & -4.77 & 0.57  & this work \\
        3C286 & 1.71 & -5.29 & 7.0 & 8.50  & this work \\
        \bottomrule
    \end{tabular}}
    \label{tab:FitPol}
\end{table}

\begin{table}
    \caption{Same as Table~\ref{tab:FitInt}, but for the linear polarization angle.}
    \centering
    \resizebox{0.49\textwidth}{!}{
    \begin{tabular}{l | c c c c | c} 
        \toprule
        Source & $a_{\mathrm{0}} \ \mathrm{[deg]}$  & $a_{\mathrm{1}}\ \mathrm{[deg]}$ & $a_{\mathrm{2}}\ \mathrm{[deg]}$ & $a_{\mathrm{3}}\ \mathrm{[deg]}$ & Ref.\\
        \midrule
        3C48 & -336 & 777 & -691 & 192 & this work \\
        3C138 & -11.1 & -8.56 & 29.0 & -18.1 & this work \\
        3C147 & -262 & 68.3 & 378 & -178 & this work \\
        3C286 & 33.2 & -1.36 & 1.18 & 0.502 & this work \\
        \bottomrule
    \end{tabular}}
    \label{tab:FitAng}
\end{table}

\correctionIII{
\section{Polarization maps}\label{app:polarizationMaps}

In Fig.~\ref{fig:polarizationMaps}, we present two additional maps for our case study: the polarized intensity (left panel) and the linear polarization fraction (right panel), as seen by an observer located at the center of the Local Bubble candidate cavity. These maps are obtained using the full data cube configuration, which includes both the bubble (i.e., its walls and interior) and the surrounding environment.

It is interesting to note that regions with lower fractional polarization correspond to the highly ionized bubble regions, which are also prominent in the RM full-sky maps (see Fig.~\ref{fig:FR_fullSkyMaps_reference}), and some of which are highlighted in Fig.~\ref{fig:nth_column}.
These regions exhibit strong ionization and turbulence, with chaotic magnetic fields lines (see, for example, the second panel of Fig.~\ref{fig:input_data_overview}), which in turn reduce the degree of polarization. This evidence may offer insight into the polarization properties of HII regions.

\begin{figure*}[]
	\centering
	\includegraphics[width=0.49\textwidth]{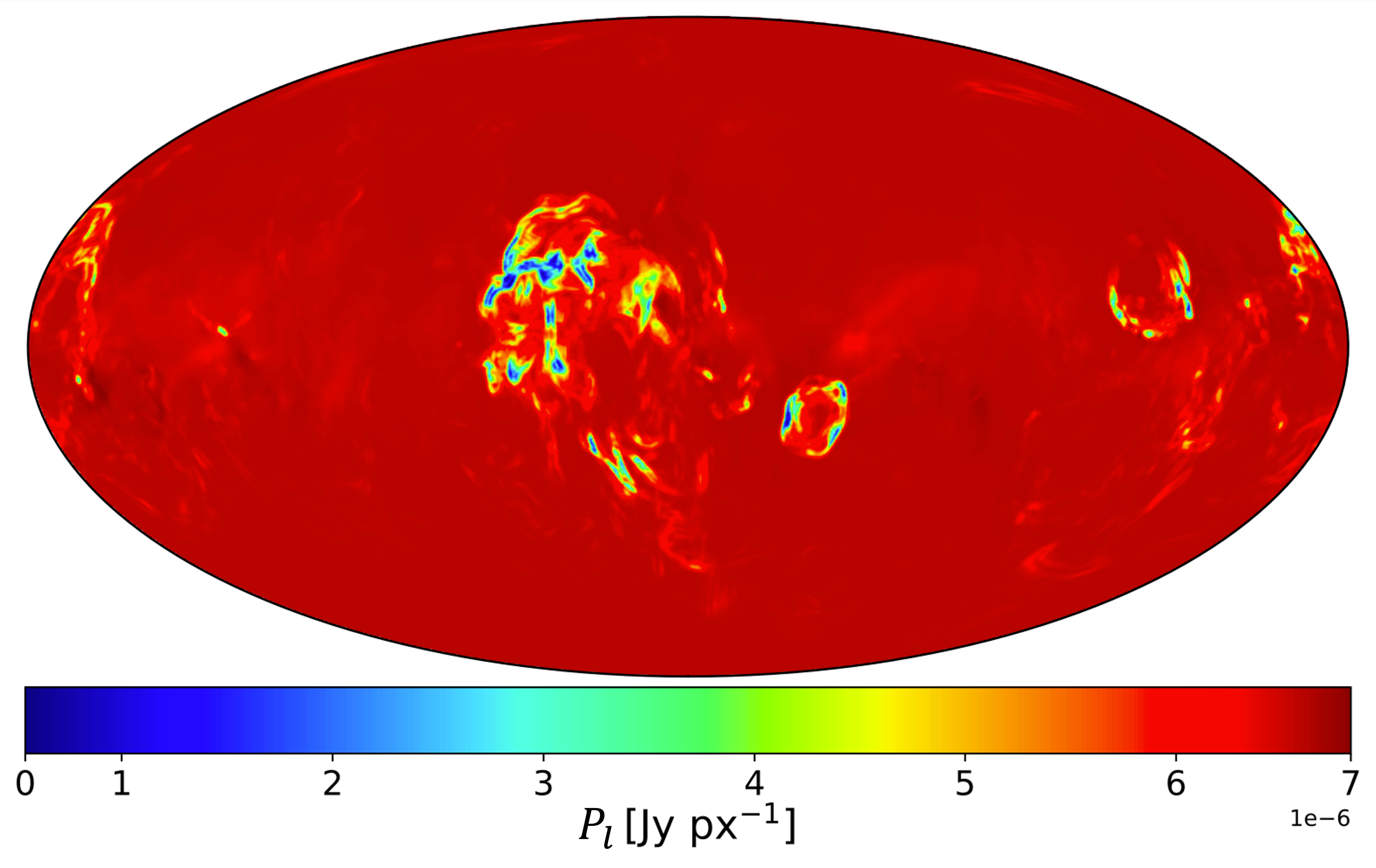}
 	\includegraphics[width=0.49\textwidth]{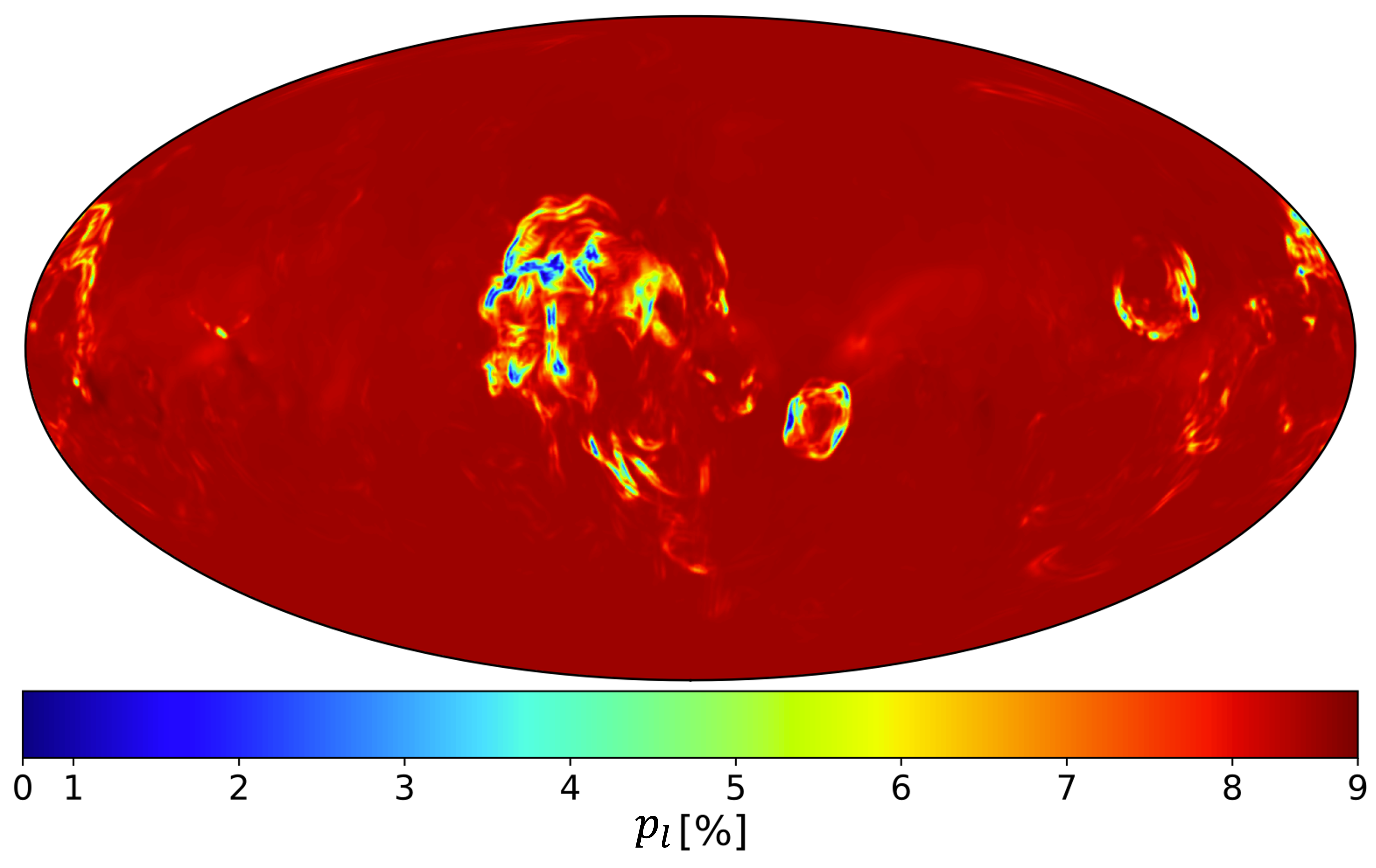}\\
	\caption{\correctionIII{Polarized intensity (\textit{left panel}) and linear polarization fraction (\textit{right panel}) of radiation as measured by the observer within the Local Bubble like cavity.}
    }
	\label{fig:polarizationMaps}
\end{figure*}
}

\section{Power spectra}\label{app:power_spectra}

The observed RM signal contains information about both the Milky Way's global structure and its small-scale features, such as ionization caused by individual SN events or star-forming regions.
One way to separate the structures on multiple scales is by decomposing the RM all-sky maps into spherical harmonics. In this approach, the small multipole moments $\ell$ represent large-scale structures, while the large multipole moments correspond to small-scale features. 

\begin{figure*}[]
	\centering
	\includegraphics[width=0.32\textwidth]{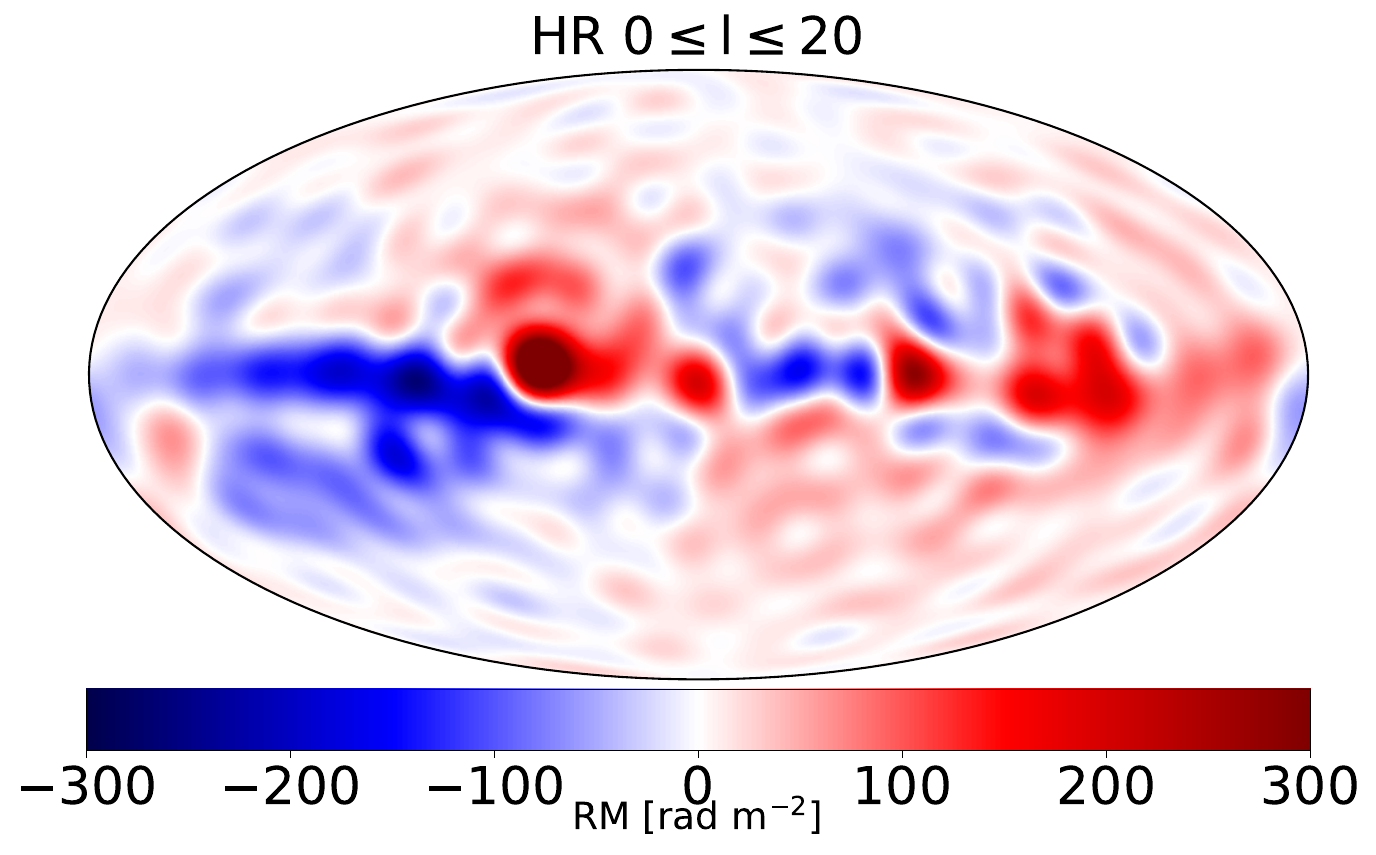}
    \includegraphics[width=0.32\textwidth]{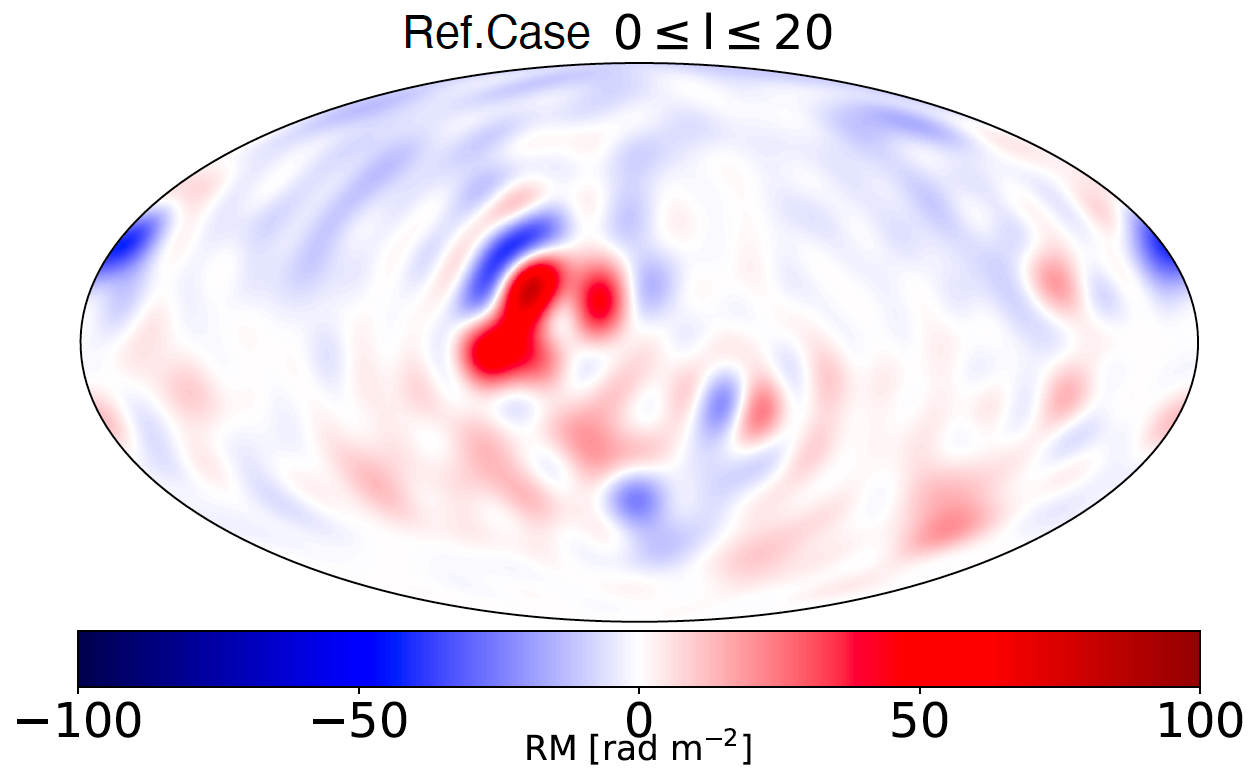}
    \includegraphics[width=0.32\textwidth]{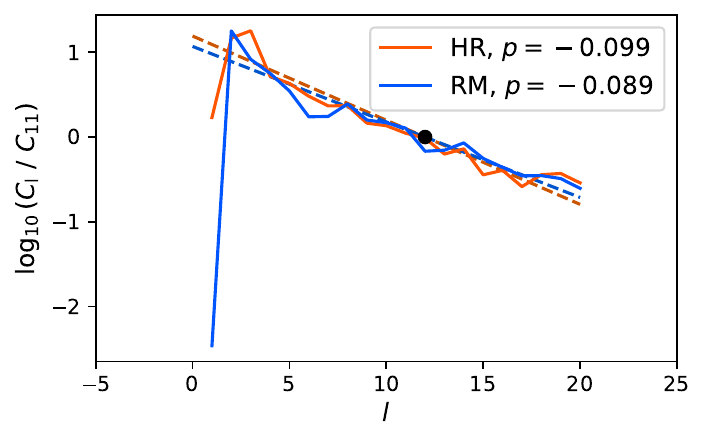}\\

    \includegraphics[width=0.32\textwidth]{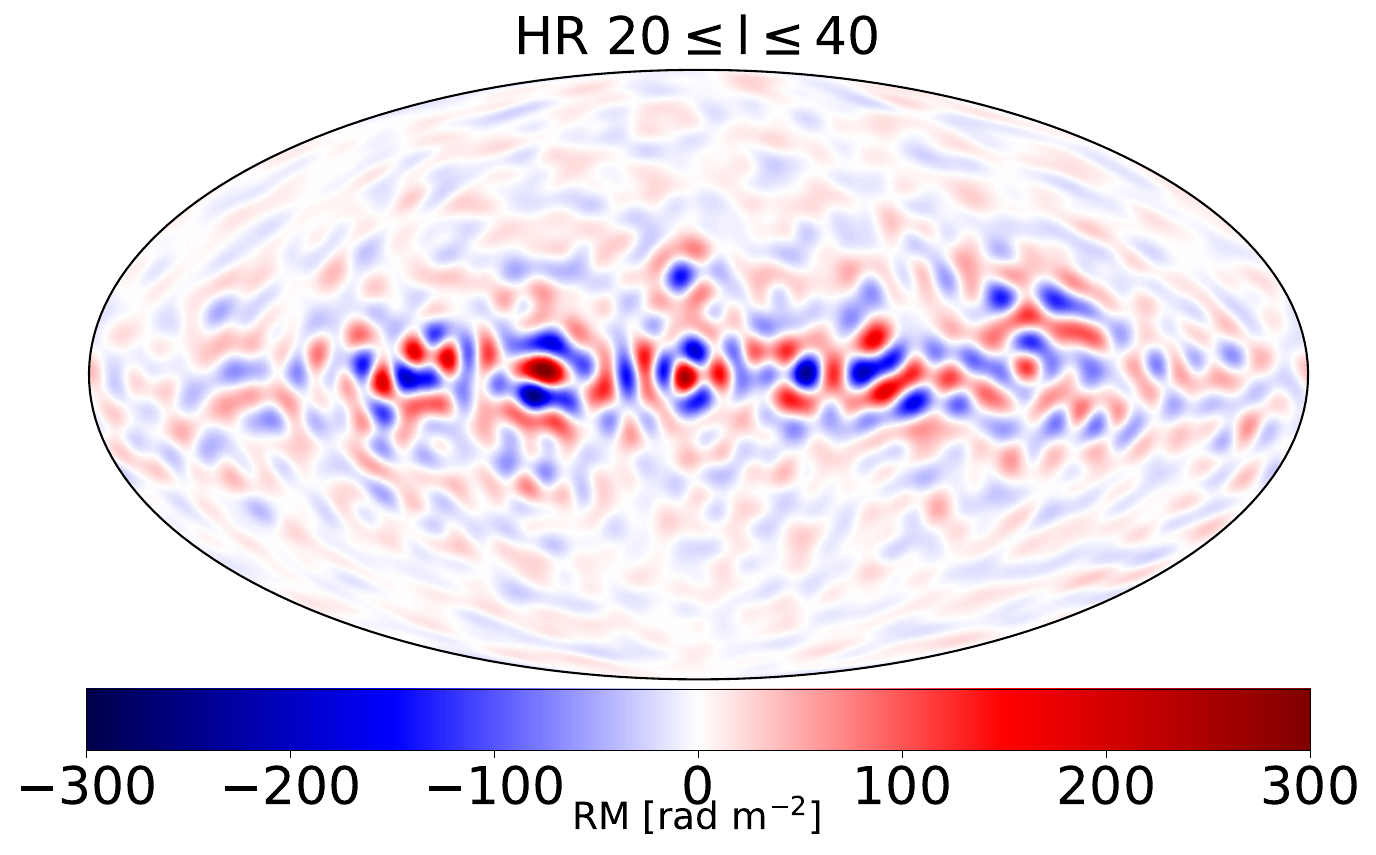}
    \includegraphics[width=0.32\textwidth]{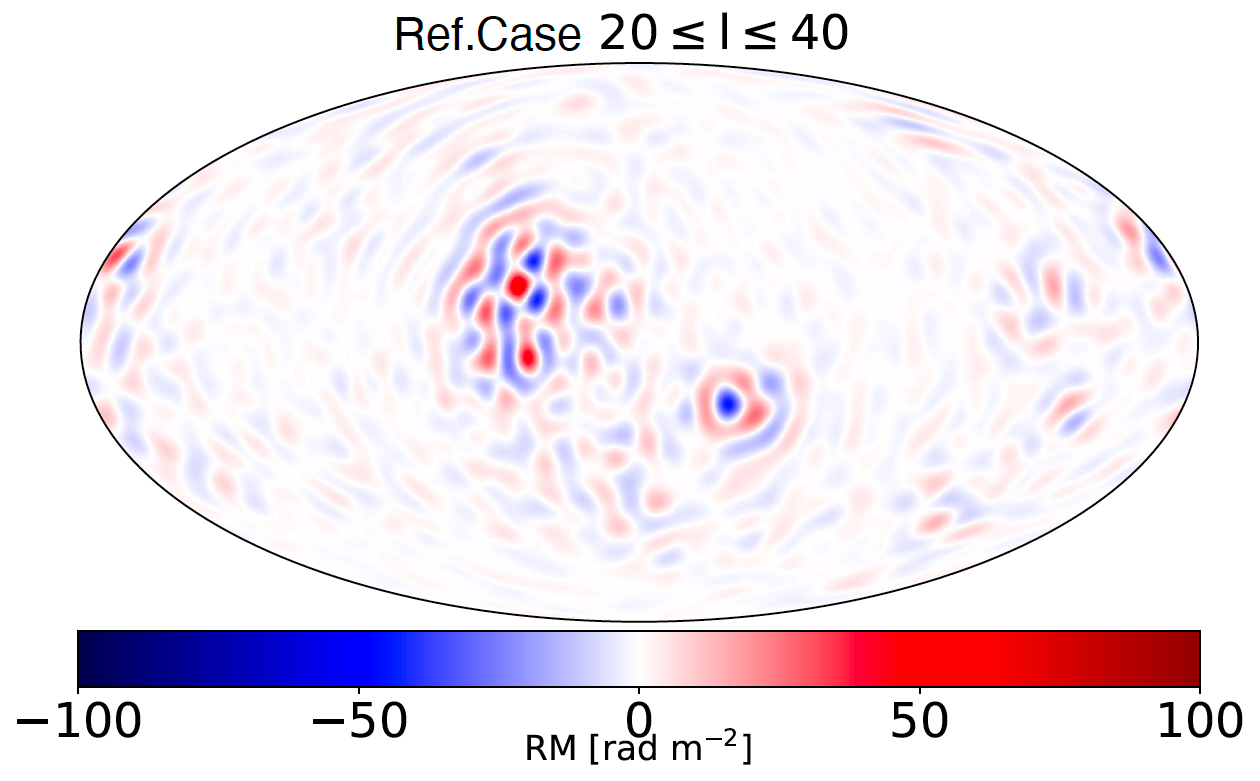}
    \includegraphics[width=0.32\textwidth]{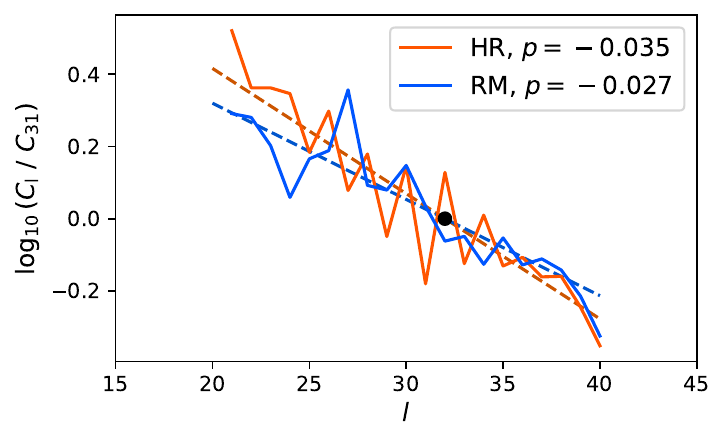}\\

    \includegraphics[width=0.32\textwidth]{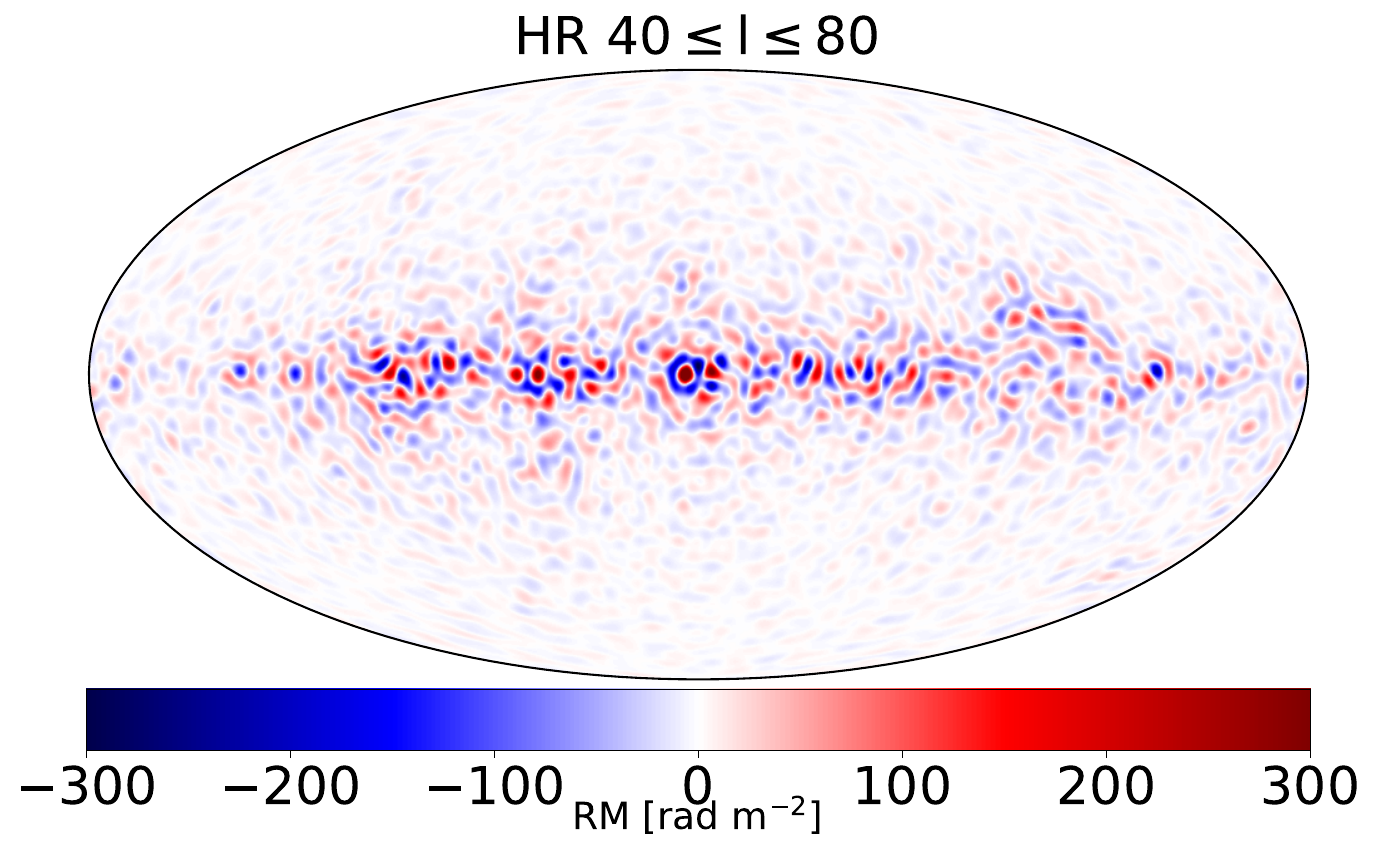}
    \includegraphics[width=0.32\textwidth]{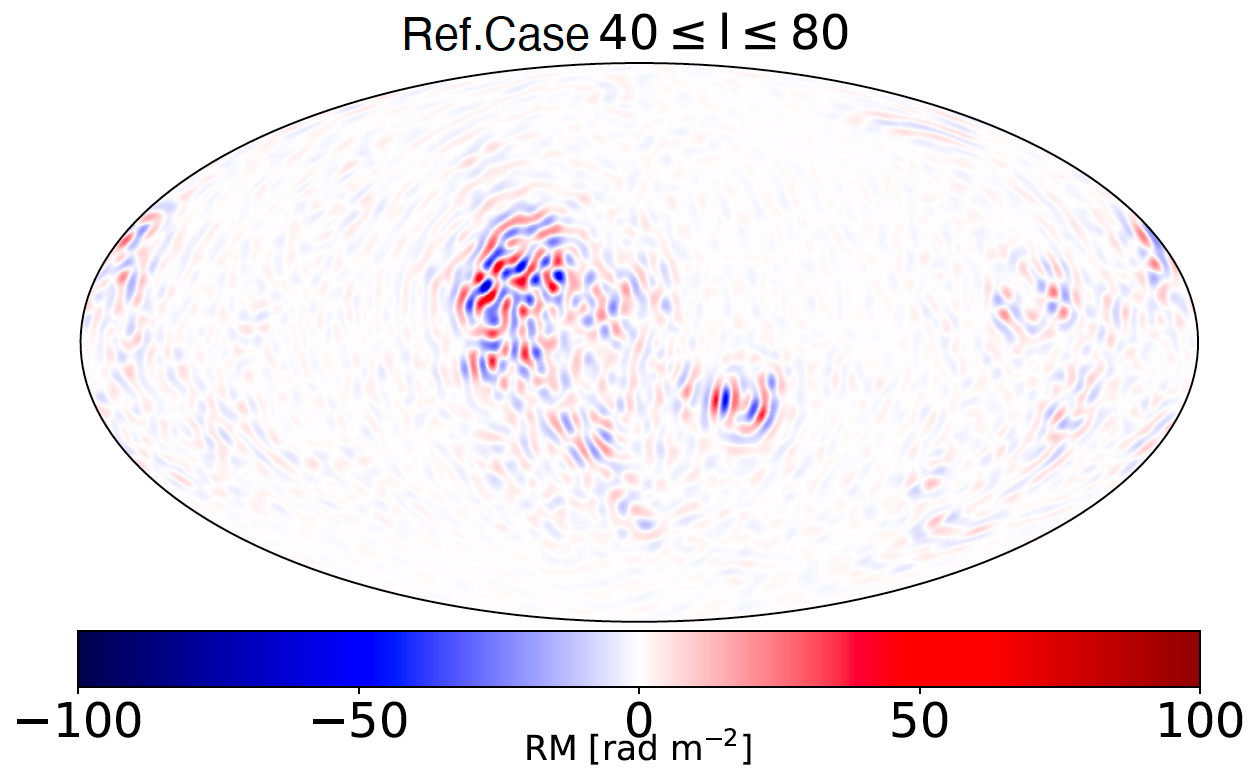}
    \includegraphics[width=0.32\textwidth]{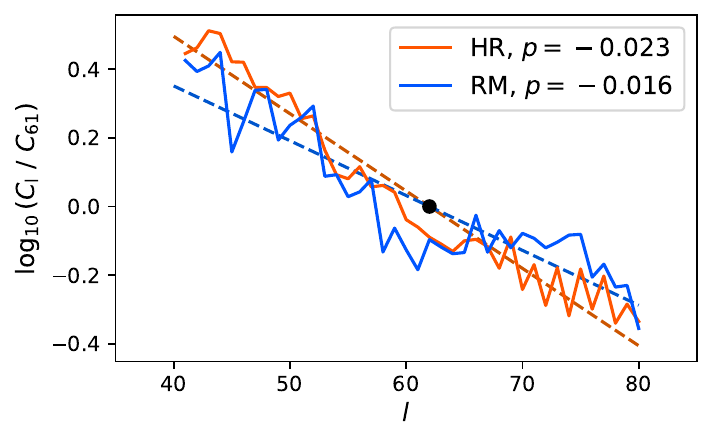}\\

    \includegraphics[width=0.32\textwidth]{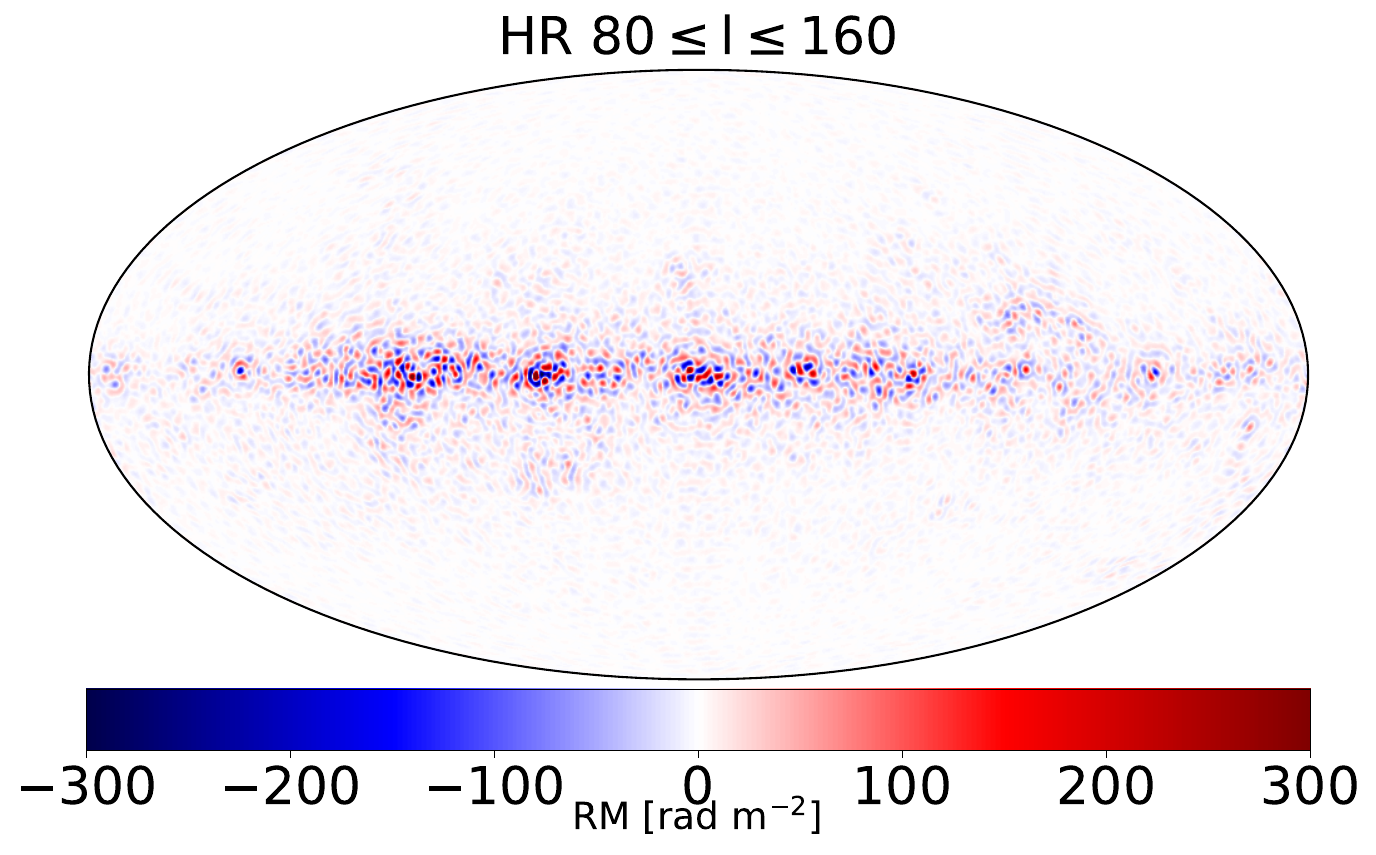}
    \includegraphics[width=0.32\textwidth]{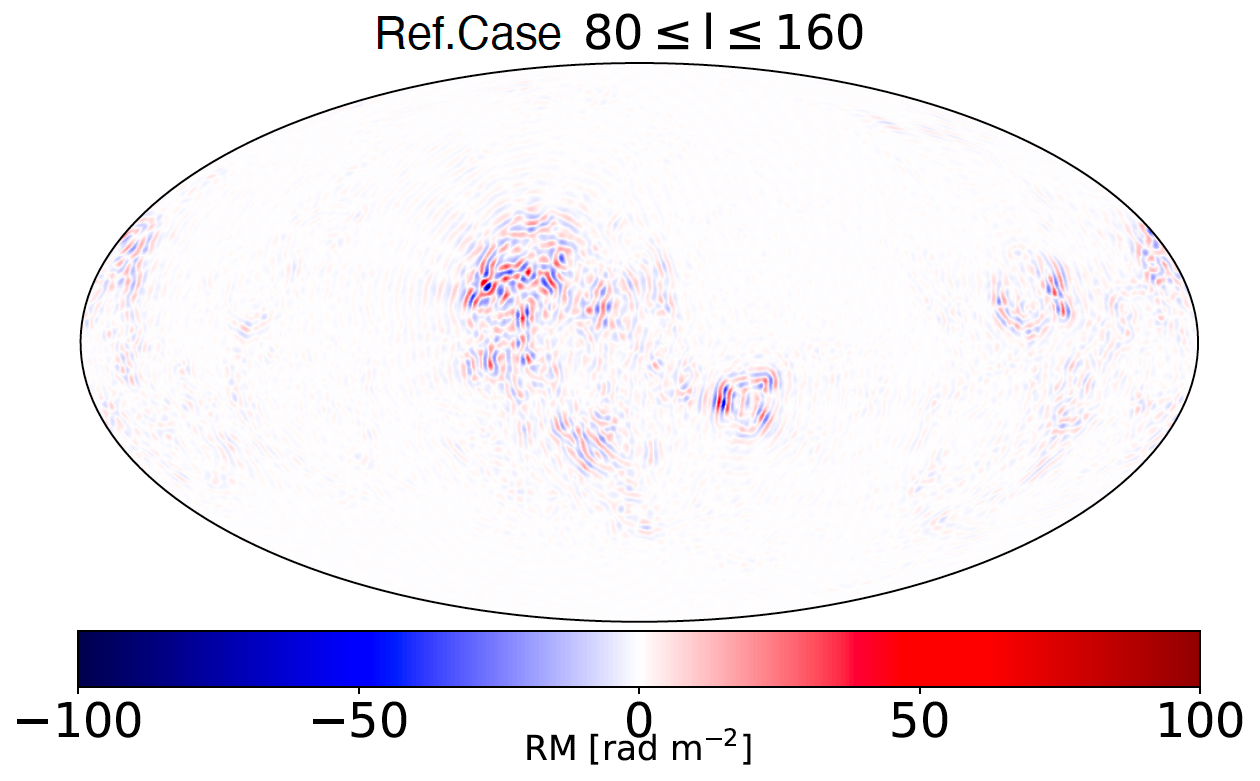}
    \includegraphics[width=0.32\textwidth]{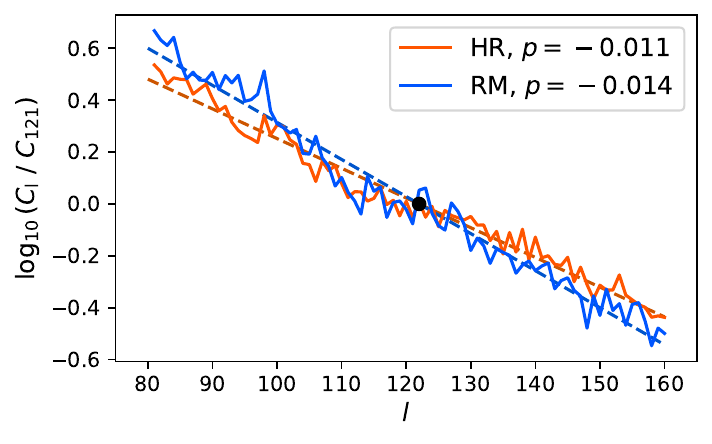}\\

    \includegraphics[width=0.32\textwidth]{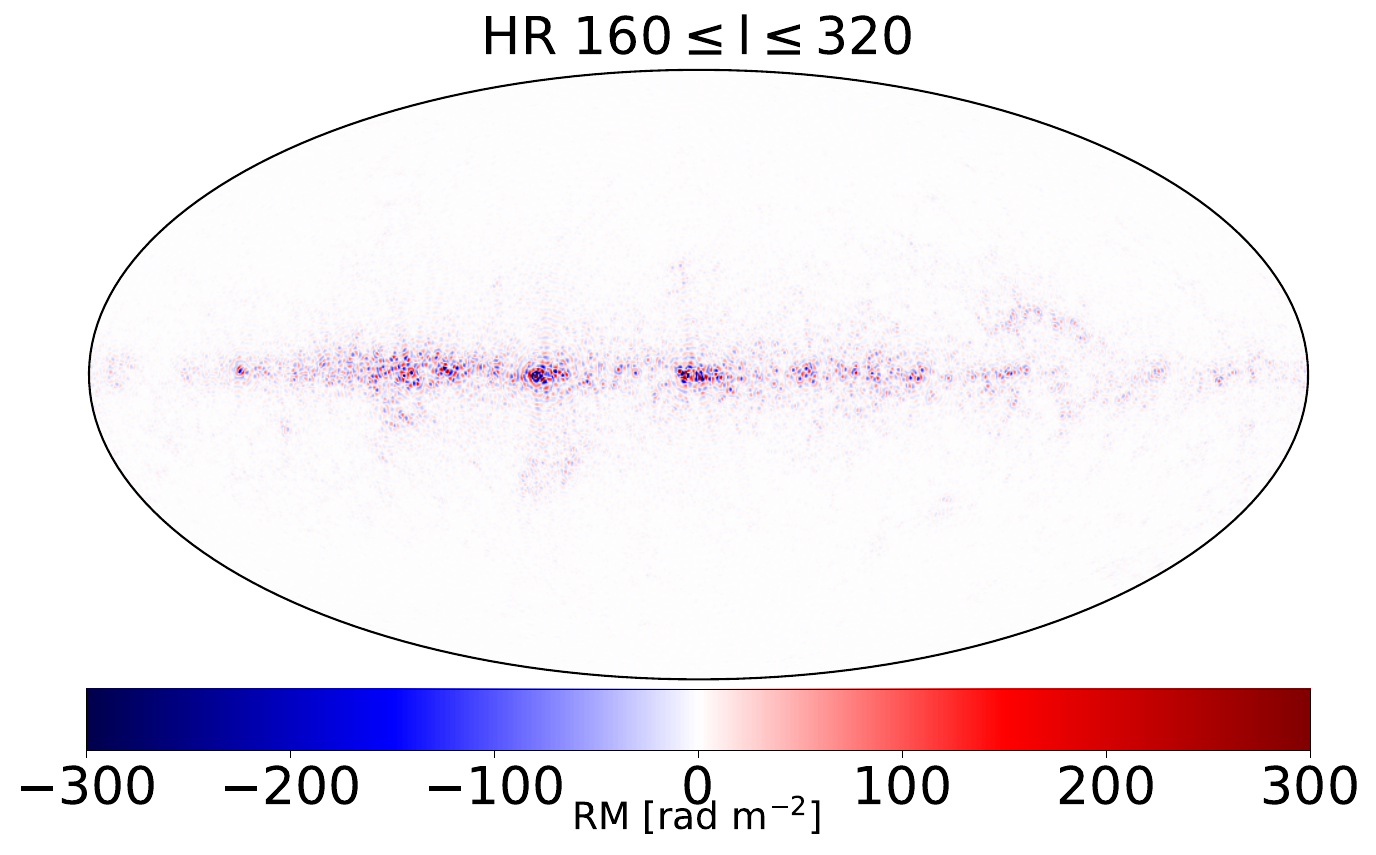}
    \includegraphics[width=0.32\textwidth]{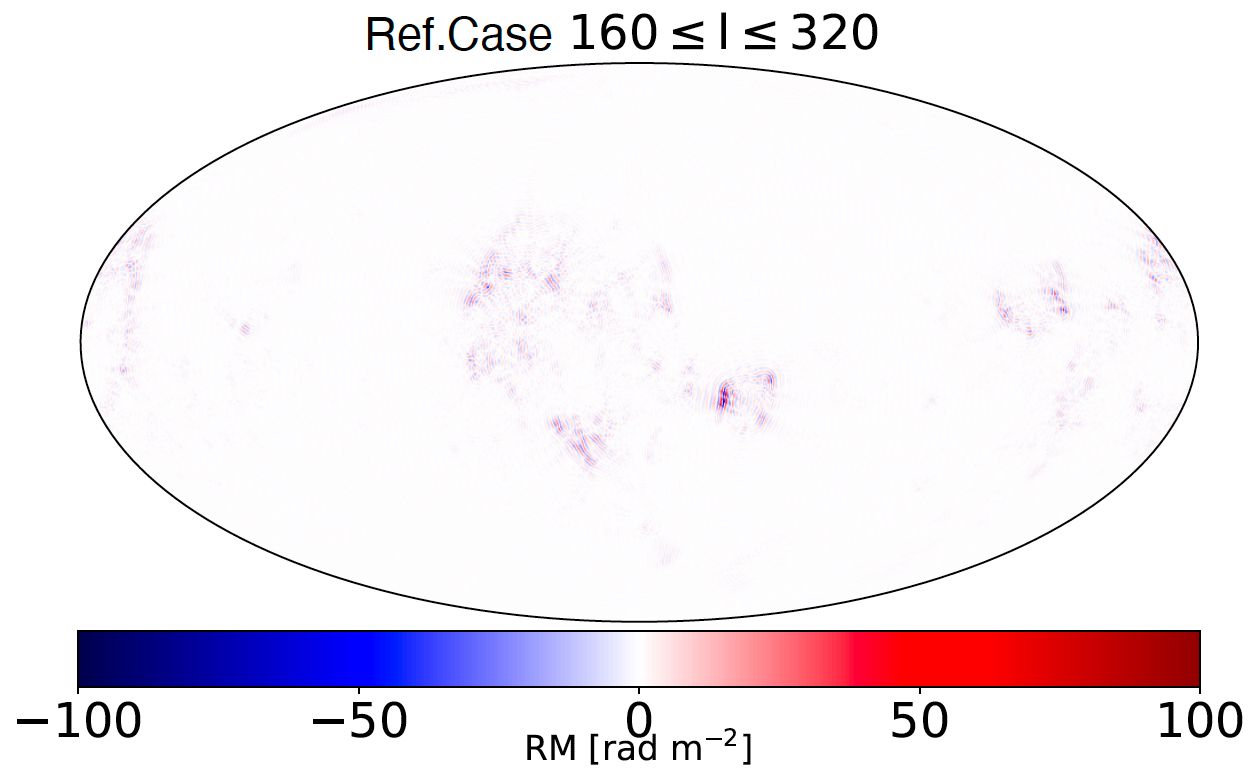}
    \includegraphics[width=0.32\textwidth]{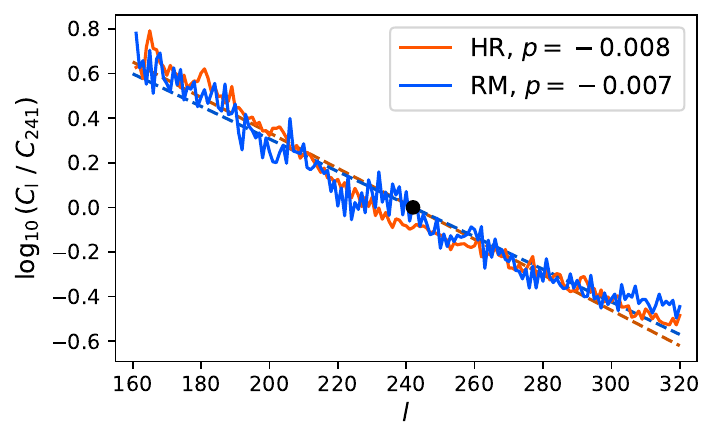}\\
	\caption{Harmonic multipole decomposition of the full-sky RM map presented in \cite{hutschenreuter2022} (\textit{left}), in conjunction with the simulated RM map for the reference map (\textit{center}), and the corresponding power spectra of the harmonic multipole expansion (\textit{right}). Each row represent a specific range of the multipole moment $\ell$ ($\ell_{\mathrm{min}} < \ell < \ell_{\mathrm{max}}$). Note that the maps from \cite{hutschenreuter2022} and the synthetic ones are not on the same scale. Hence, the resulting spectra for the map from \cite{hutschenreuter2022} (solid orange) and the synthetic observations (solid blue) are normalized by the central value of the fitted slope with exponent $p$ (dashed lines), for comparison. The normalization is indicated by a black dot.}
	\label{fig:FR_fullSkySpectrum}
\end{figure*}

\begin{figure*}[]
	\centering
	\includegraphics[width=0.32\textwidth]{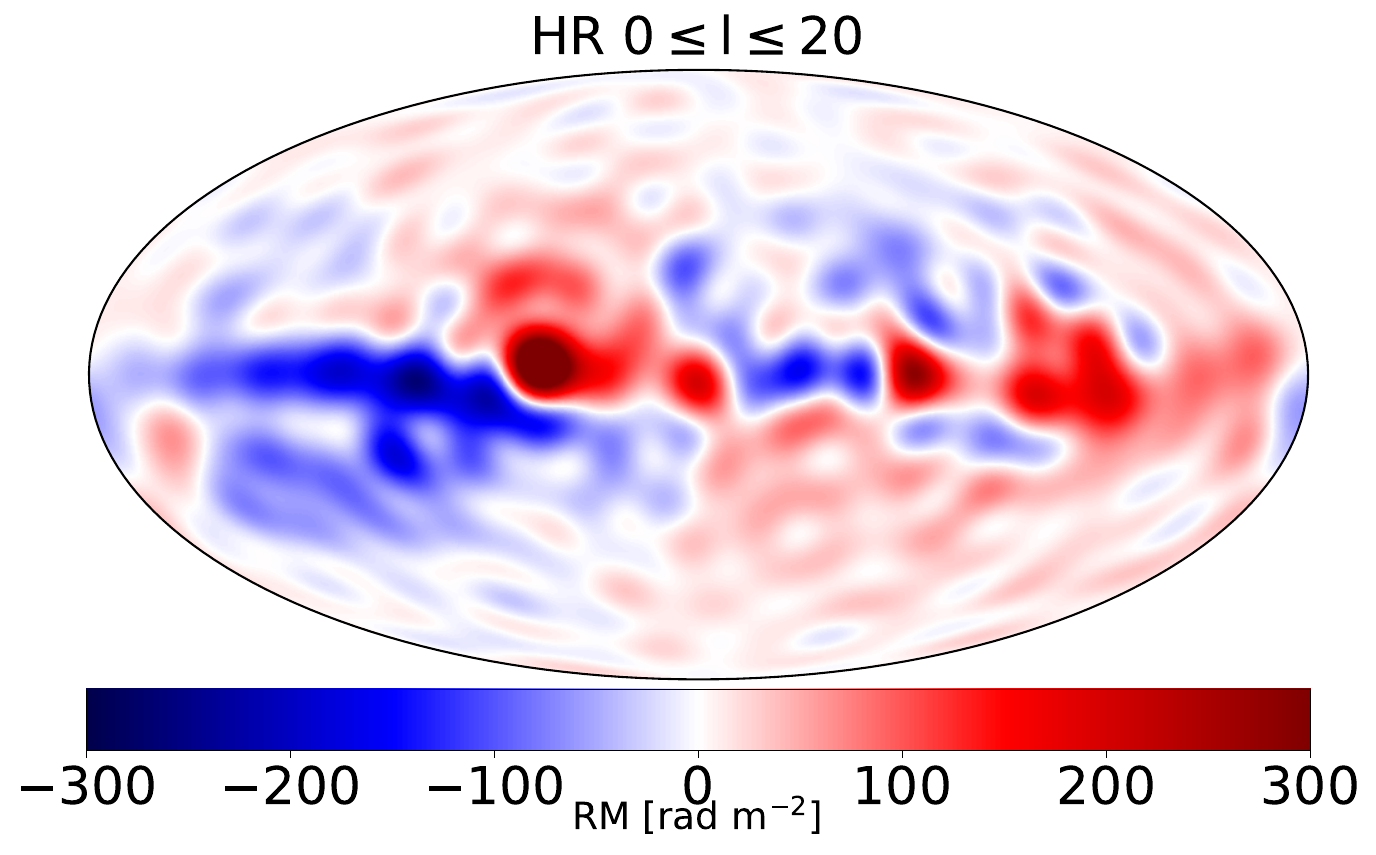}
    \includegraphics[width=0.32\textwidth]{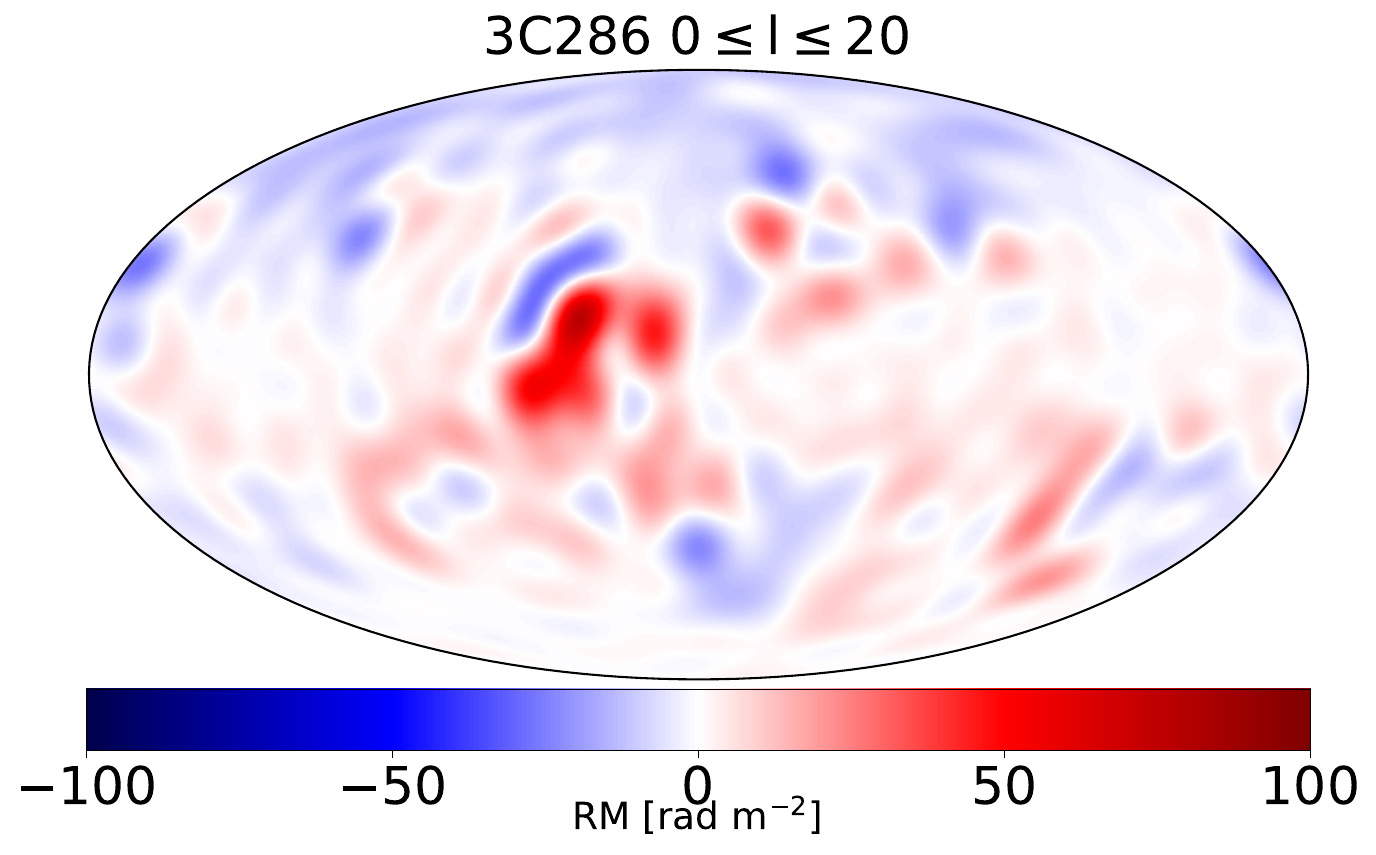}
    \includegraphics[width=0.32\textwidth]{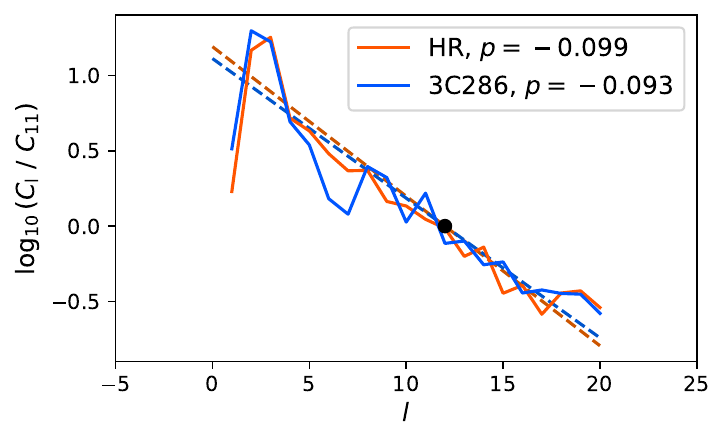}\\

    \includegraphics[width=0.32\textwidth]{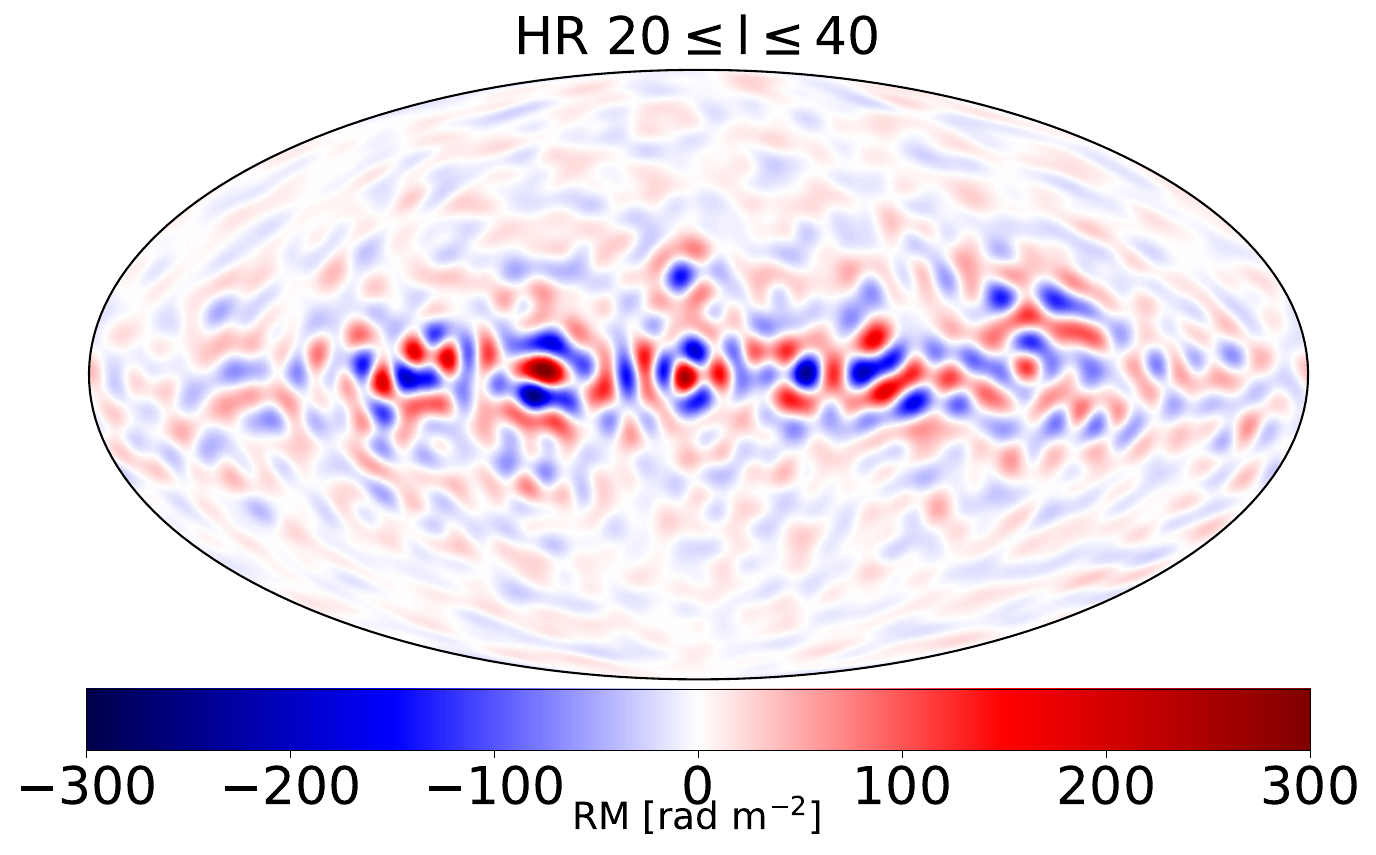}
    \includegraphics[width=0.32\textwidth]{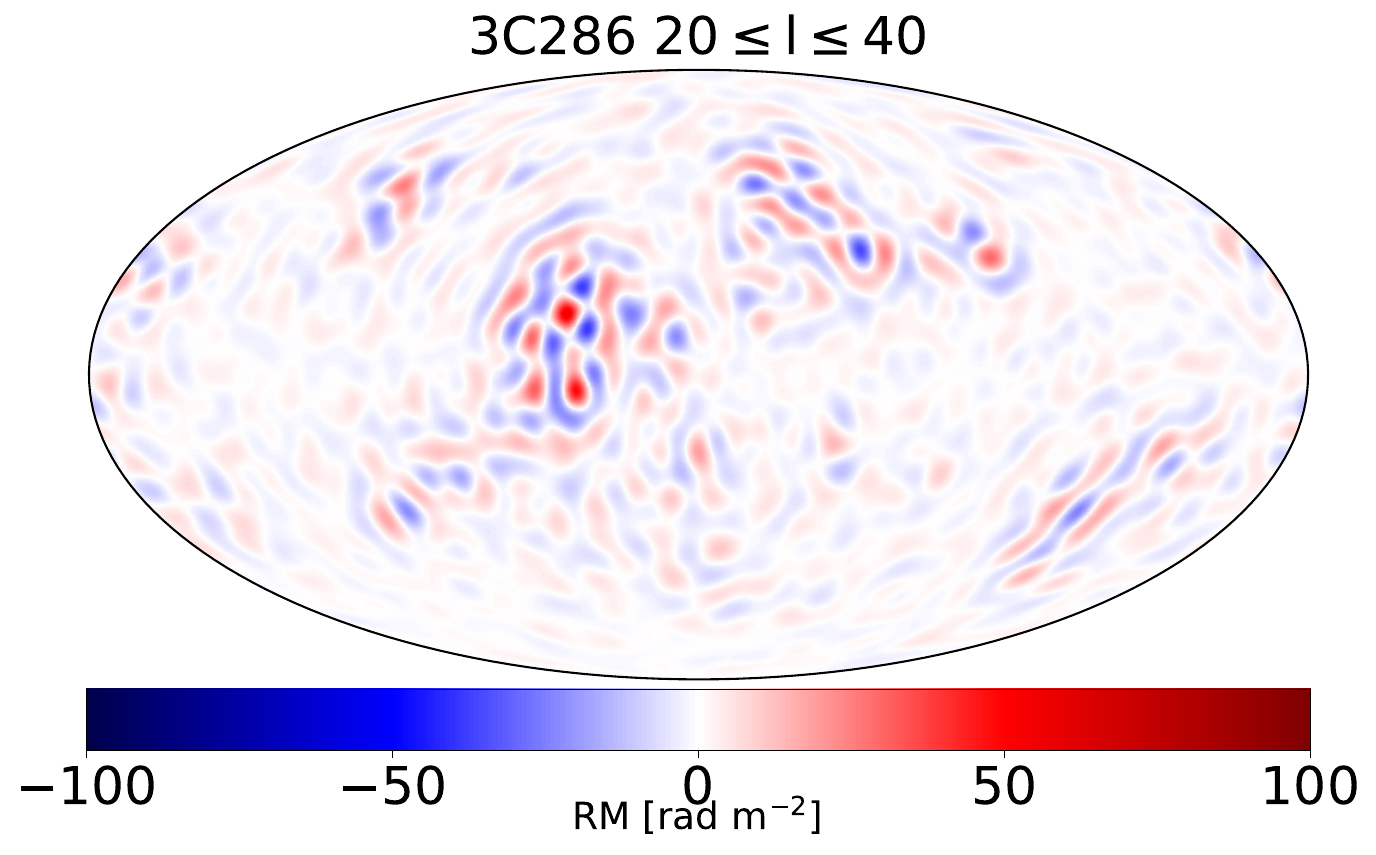}
    \includegraphics[width=0.32\textwidth]{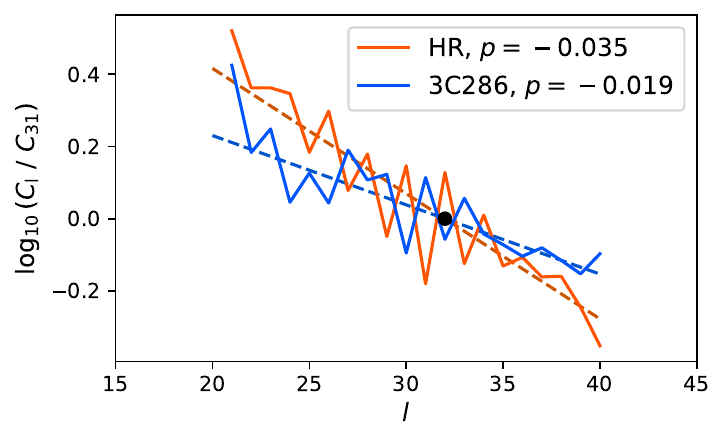}\\

    \includegraphics[width=0.32\textwidth]{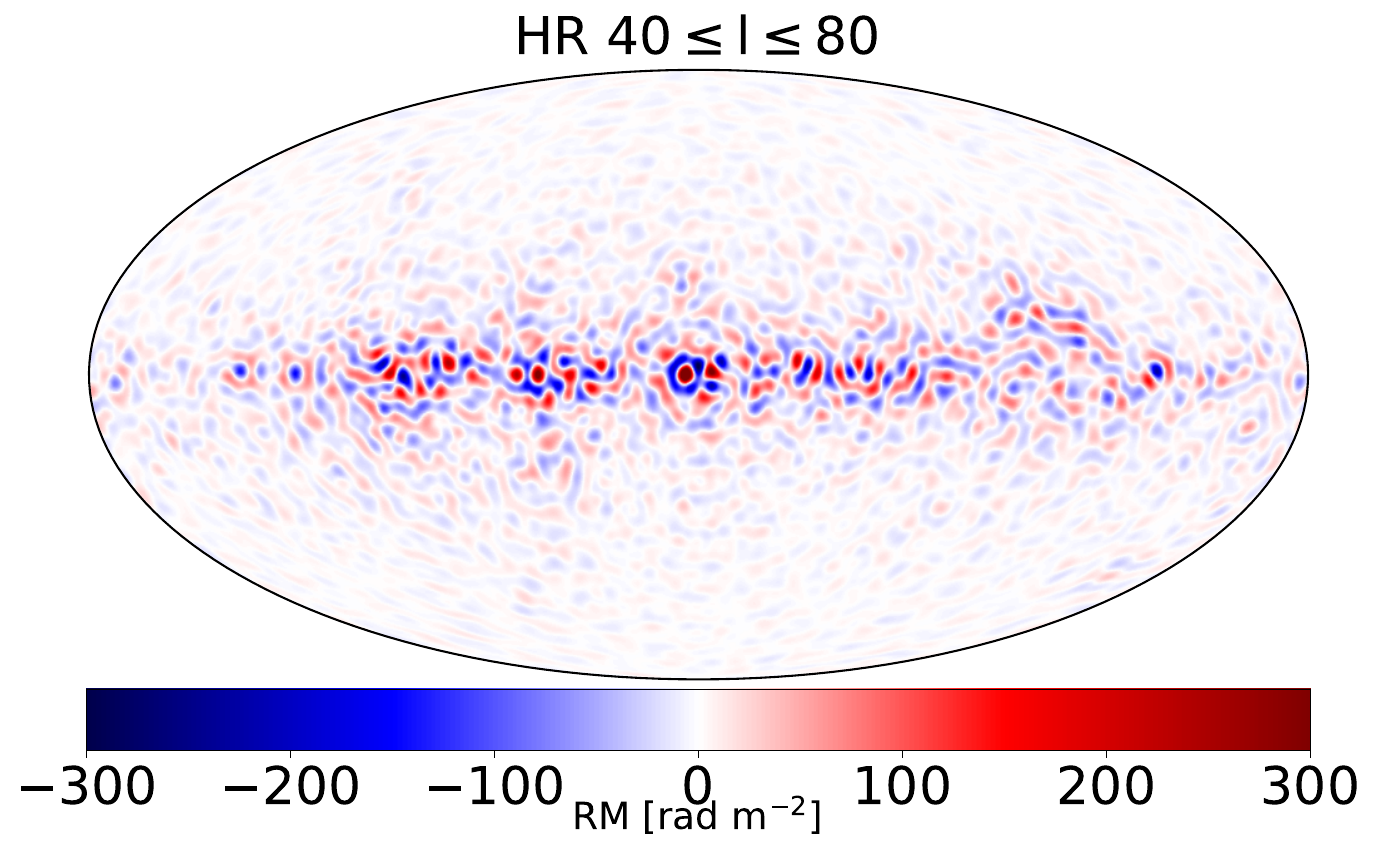}
    \includegraphics[width=0.32\textwidth]{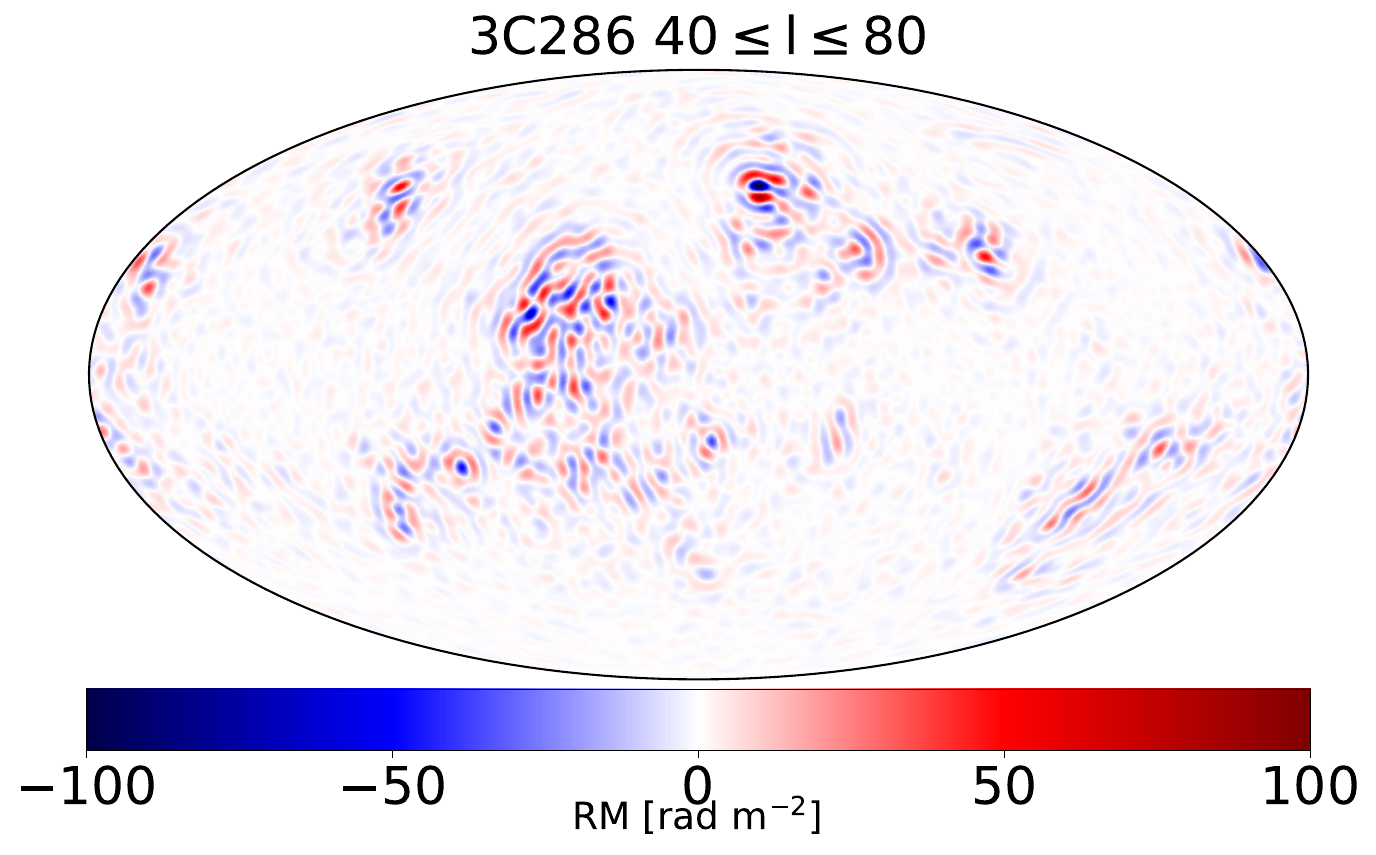}
    \includegraphics[width=0.32\textwidth]{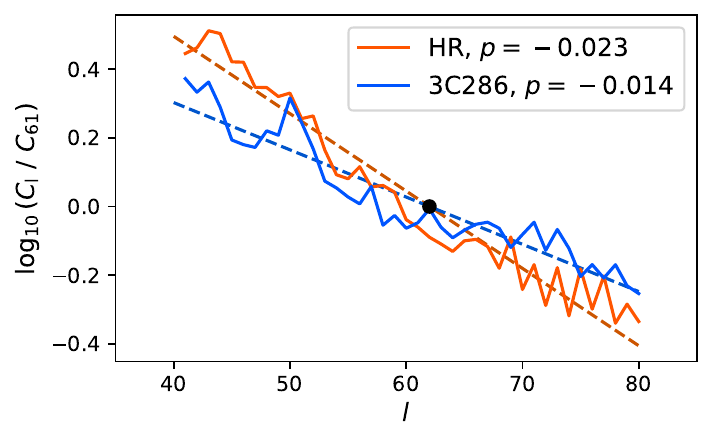}\\

    \includegraphics[width=0.32\textwidth]{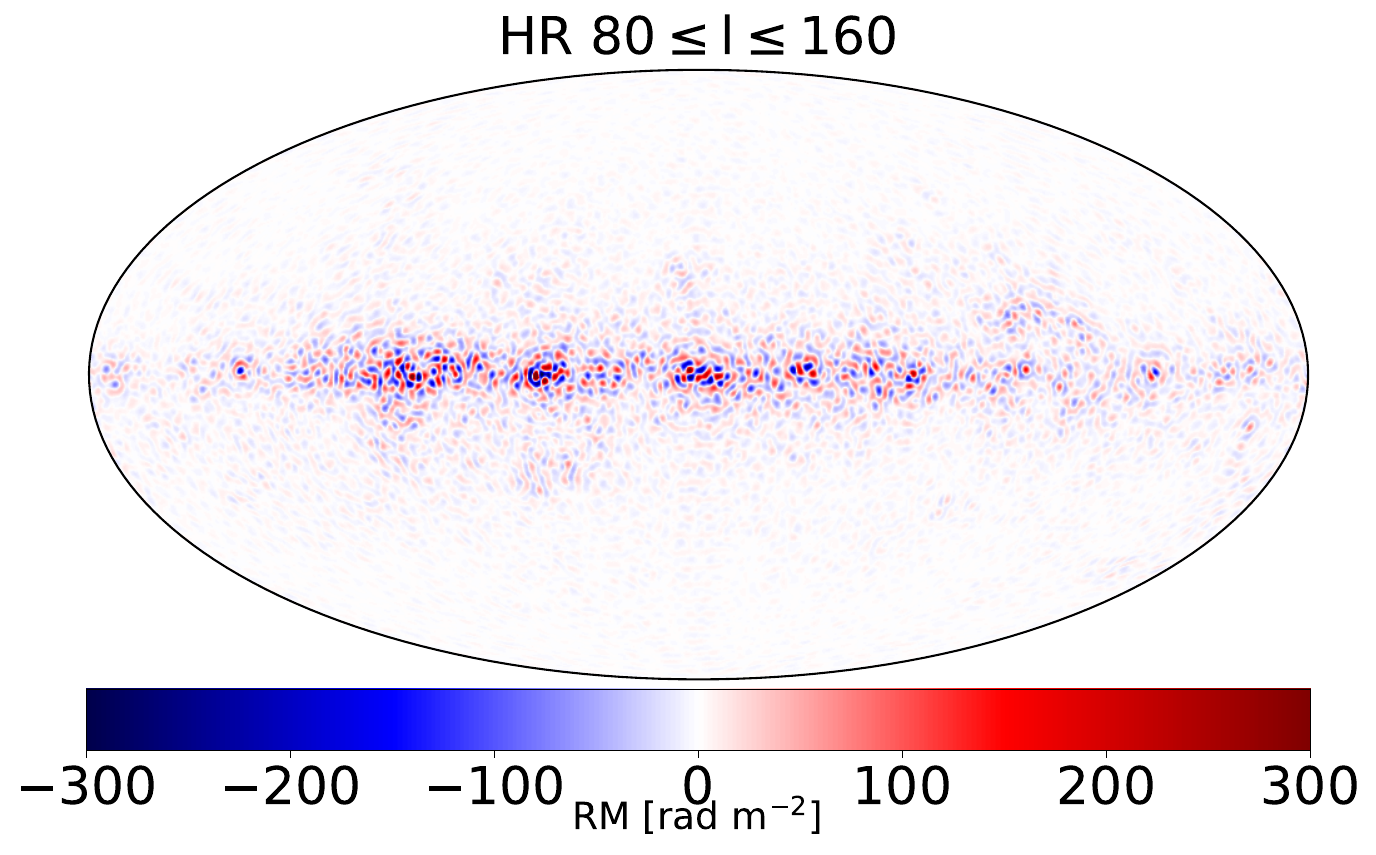}
    \includegraphics[width=0.32\textwidth]{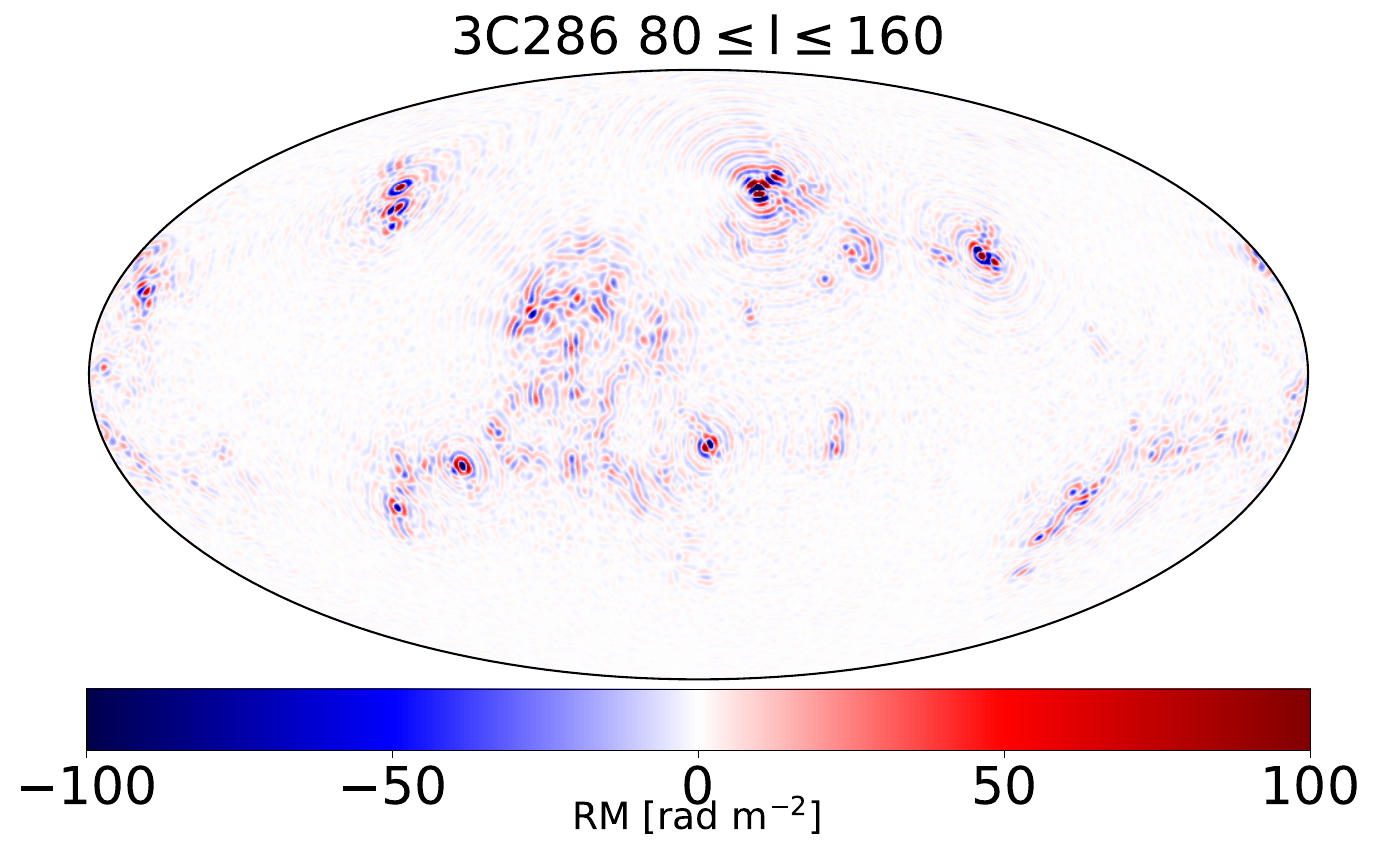}
    \includegraphics[width=0.32\textwidth]{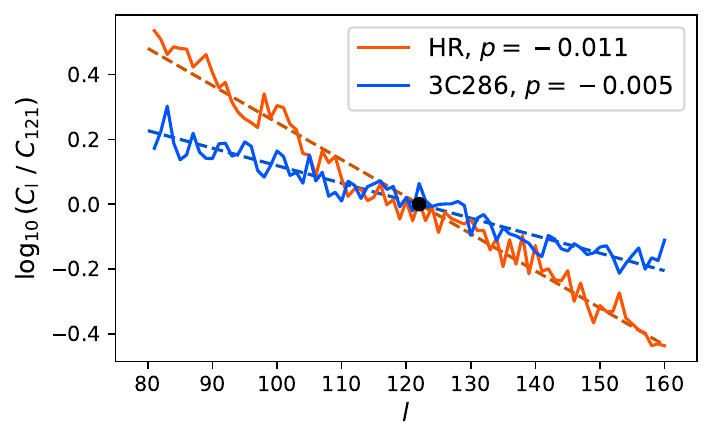}\\

    \includegraphics[width=0.32\textwidth]{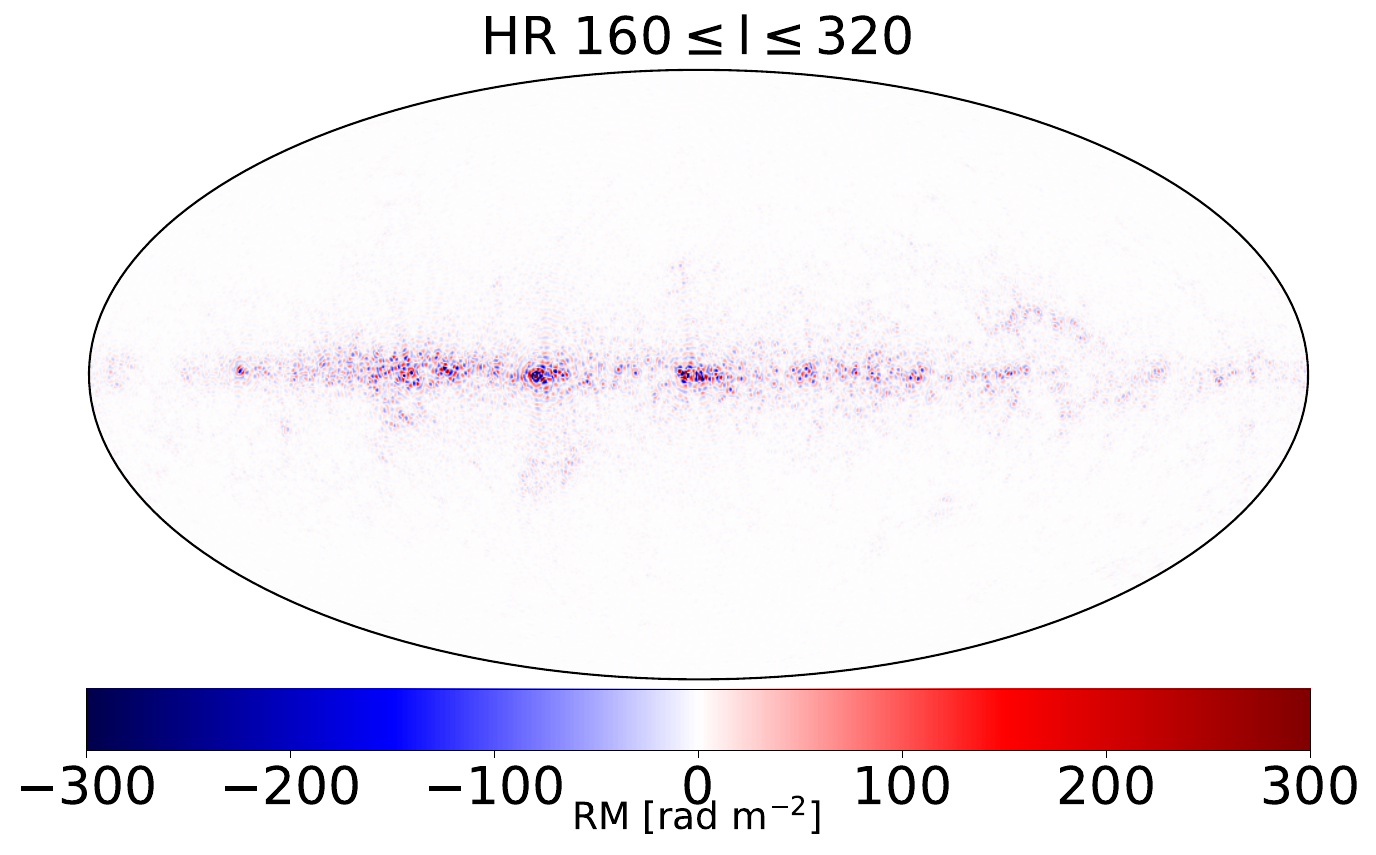}
    \includegraphics[width=0.32\textwidth]{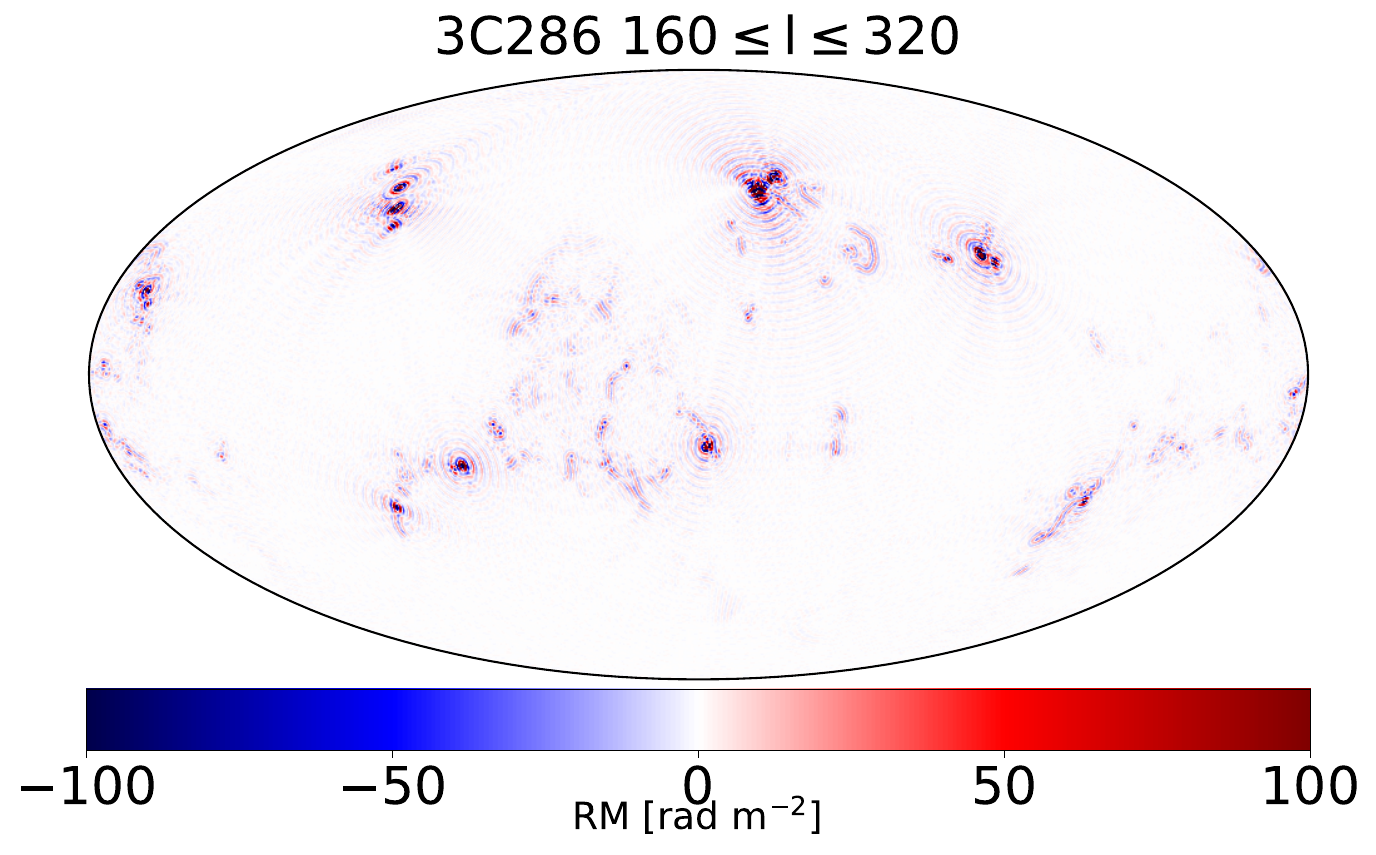}
    \includegraphics[width=0.32\textwidth]{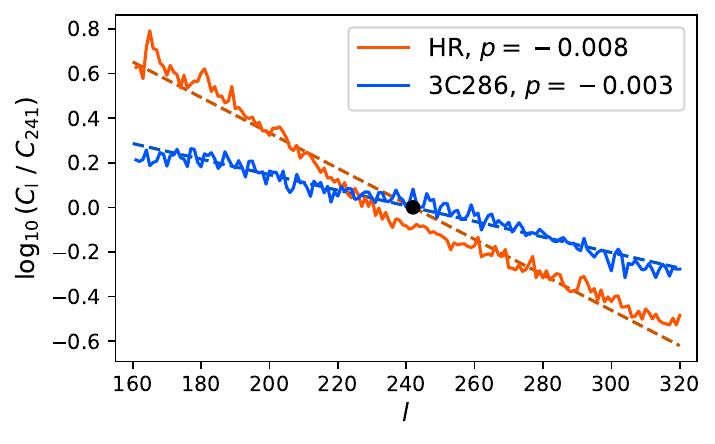}\\
    
	\caption{\correction{Same as Fig.~\ref{fig:FR_fullSkySpectrum} but for the RM map obtained using the source 3C286 as background.}}
	\label{fig:FR_fullSkySpectrum_3C286}
\end{figure*}

In Fig.~\ref{fig:FR_fullSkySpectrum}, we present a comparison of the multipole decomposition between the RM map from \cite{hutschenreuter2022} and the reference map (see Fig.~\ref{fig:FR_fullSkyMaps_reference}) obtained in this work, along with the corresponding power-law spectra for different ranges of $\ell$. Similarly, in Fig.~\ref{fig:FR_fullSkySpectrum_3C286}, the same comparison is shown for the synthetic RM \correctionII{map obtained for the test case of our Faraday RM synthesis routine, as described in Sect.~\ref{app:test_case}.  In this test case, an ideal setup with the standard calibrator 3C286 as the background source is used} (setup E of Fig.~\ref{fig:set_ups_RT_simulations}).
For better comparison, we normalized the spectra to the central multipole moment of each range, with distinct ranges separated by a factor of two. The normalization was determined such that both spectra have the value at 1 at the center of the depicted $\ell$-interval. In reality, the map from the Galaxy has a larger dynamics range due to the influence of the disk midplane.  

The large scale structures reflected in the smallest multipole moments $\ell$ in the Milky Way could be associated with the galactic disk. However, our analysis reveals that a spectrum slope akin to that of a Galactic disk can also be achieved with a much smaller region of a Milky Way-like medium as well (see top panel of Figs.~\ref{fig:FR_fullSkySpectrum} and \ref{fig:FR_fullSkySpectrum_3C286}). We speculate that this occurs because the most prominent structures of nearby SNe driven bubbles are positioned near the plane of the SILCC MHD simulation (compare Fig.~\ref{fig:nth_column} with Fig.~\ref{fig:FR_fullSkyMaps_reference}).

\correction{Large multiple moments can reveal small-scale RM signals that may be associated with far-distant, highly ionized HII regions or SN remnants \citep{Shanahan2019,Reissl2020A}. By decomposing the \correction{maps} into larger ranges of $\ell$, we observe differences in the power spectra behavior between the two analyzed cases. Interestingly, for the reference map (see Fig.\ref{fig:FR_fullSkySpectrum}) the slope of the power spectra aligns closely with that of the RM map from \cite{hutschenreuter2022}, despite the fact that the galactic disk is missing from our MHD simulations. However, we currently lack a definitive explanation for this behavior, which would require testing the full pipeline including ISM, observational, and inference effects. Instead, when 3C286 is used as the background source (see Fig.\ref{fig:FR_fullSkySpectrum_3C286}), the slopes of the two spectra diverge. This divergence could be attributed to additional noise, such as white noise from RM synthesis, which might add sufficient power to raise the spectrum at smaller scales. Nonetheless, as the comparison of power spectra is non-trivial, our discussion remains speculative and further investigation is required. A comprehensive analysis of the power spectrum lies beyond the scope of this work and will be addressed in future studies.
}

\end{document}